# Recent advances in high-contrast metastructures, metasurfaces and photonic crystals

**Pengfei Qiao,[1] Weijian Yang,[1,2] and Connie J. Chang-Hasnain[1,*]**

[1]University of California at Berkeley, Department of Electrical Engineering and Computer Sciences and Tsinghua Berkeley Shenzhen Institute, Berkeley, CA 94720
[2]Columbia University, Department of Biological Sciences, New York, NY 10027
*Corresponding author: cch@berkeley.edu

In the recent decade, the research field using arrays of high-index-contrast near-wavelength dielectric structures on flat surfaces, known as high-contrast metastructures (HCMs) or metasurfaces, has emerged and expanded rapidly. Although the HCMs and metasurfaces share great similarities in physical structures with photonic crystals (PhCs), i.e. periodic nanostructures, many differences exist in their design, analysis, operation conditions, and applications. In this paper, we provide a generalized theoretical understanding of the two subjects and show their intrinsic connections. We further discuss the simulation and design approaches, categorized by their functionalities and applications. The similarity and differences between HCMs, metasurfaces and PhCs are also discussed. New findings are presented regarding the physical connection between the PhC band structures and the 1D and 2D HCM scattering spectra under transverse and longitudinal tilt incidence. Novel designs using HCMs as holograms, spatial light modulators, and surface plasmonic couplers are discussed. Recent advances on HCMs, metasurfaces and PhCs are reviewed and compared for applications such as broadband mirrors, waveguides, couplers, resonators, and reconfigurable optics.

## 1. Introduction

There has been considerable interest, over the past decades, in developing low-cost, compact, high-performance optical components for photonic integration. High-contrast metastructures (HCMs), metasurfaces and photonic crystals (PhCs) prove to be promising candidates. Despite the similarities in physical appearances with periodic subwavelength structures, HCMs and PhCs differ in their design approaches, theoretical analyses and photonic applications, and therefore remain somewhat distinct research topics. Here we present a comprehensive review of their intrinsic connection, leading to intuitive designs and understandings of periodic and quasi-periodic nanostructures.

The study on periodic structures dated back to 1887 by Lord Rayleigh. The term "photonic crystal" came to use after the milestone work by Yablonovitch and John in 1987 [1, 2]. Photonic stop band exists in PhCs in a similar fashion as the bandgap for electrons in crystalline solids and hence the name [3]. 1D PhCs, which was already known as the Bragg gratings and well utilized before 1987, consist of alternating layers that are infinitely wide in the directions normal to the periodicity. The 1D PhCs have been extensively used as mirrors in photonic devices, such as distributed Bragg reflectors (DBRs) in waveguides and lasers. Exploring the periodicity in two and three dimensions, researchers have demonstrated photon trapping and emission in PhC defects [4], PhC fibers [5], supercontinuum generation [6], slow light [7], superlenses [8], invisible cloaking [9], and so on.

For the past decade, there has been extensive research on subwavelength periodic structures and their applications, particularly with their periodicity perpendicular to the incidence direction of light. One example is the ultra-thin high-index-contrast 1D dielectric gratings, known as high-contrast gratings (HCGs) [10], and the 2D variation of such structures, now generalized as high-contrast metastructures (HCMs). At a first glance, 1D HCMs are simply 1D PhCs, except that the dimension perpendicular to the periodicity is ultrathin, and the direction of incident wave may be arbitrary relative to the direction of periodicity. Many extraordinary features were obtained with such simple structures. Notable accomplishments include HCG widely-wavelength-tunable VCSELs [11-16], HCG high-Q resonators [17, 18], HCG hollow-core waveguides [19, 20], HCG surface-normal couplers [21], and HCG high-NA planar lenses [22, 23]. More recent advances include HCG phased arrays [24, 25], surface-normal second harmonic emission from HCGs [26], flexible HCG metasurfaces [27], HCG VCSELs on silicon [28], 2D HCM tunable VCSELs [29], and metasurface holograms [30, 31]. The fast-growing field of metasurfaces [32-36], which emerged from high-contrast nanostructure arrays, also shows the potential of the research area and the necessity of having universal understanding of periodic and, in particular, spatial-varying nanostructures.

In this review, we will focus on the similarities and differences between dielectric-based HCMs and PhCs in theoretical analysis and design methodologies. We will review recent advances in photonic applications of HCMs and PhCs, and discuss the future directions of research. In our last review in 2012 [37], we presented the analytical mode-matching method in details and solved the 1D HCM properties under normal incidence and transverse tilt incidence, where the incidence plane is perpendicular to the gratings. We showed that, most intriguingly, the checker-board-patterned reflection contour plots were observed for 1D HCM under all incidence (normal, transverse tilt and longitudinal tilt).

Such plots are useful tools for designing broadband high reflection, high transmission, or high Q structures [37]. However, the theory was incomplete. In particular, 1D HCMs under longitudinal tilt incidence and 2D HCMs remained unsolved analytically. The longitudinal tilt having propagation constant orthogonal to the direction of periodicity, in fact, has no analogy with PhCs and cannot be predicted by PhC theory. Hence, how it works and connects to PhC remained a puzzle. Here, such problem is solved rigorously for the first time. Strikingly, both transverse and longitudinal tilt plane wave incidence can excite strong resonance distinctive from those commonly observed under normal incidence. In this review, we present a complete analysis with all combinations of tilt planes, polarizations, 1D or 2D HCMs are solved with detailed mathematical formulation and in-depth physical explanation. With the supermode analysis presented in Section 2.2, we are able to identify the few HCM eigenmodes which contribute to many intriguing phenomena, such as high Q resonance and 100% reflection or transmission.

Finally, metasurfaces have gained extensive research interest in the past few years due to their extraordinary optical behavior as artificial material interfaces with ultrahigh compactness. While the earlier research focused on 2D metal nanostructures, there are intense research activities on dielectric structures. Although the HCMs and dielectric metasurfaces came to being from different origins, they are indeed very similar in appearance and applications, and indeed even design methods. In this paper, we will review the evolution and development of metasurfaces in terms of their operation regimes, governing working mechanisms, and the span of functionalities.

This review is organized as follows. Section 2 presents the intrinsic connection between HCMs and PhCs through the theoretical study of periodic structures. Sections 3 to 7 present the detailed comparison of HCMs and PhCs in design and application, grouped by major functionalities of (i) broadband mirrors, (ii) metasurfaces, (iii) waveguides and couplers, (iv) resonators, and (v) reconfigurable optics.

## 2. Theory of Periodic Structures

One-dimensional photonic crystals (1D PhCs), or equivalently Bragg gratings refer to layered structures that are periodic in one direction (chosen as $\hat{x}$) and invariant in the other two directions ($\hat{y}$ and $\hat{z}$), as shown in Fig. 1(a). The one-dimensional high-contrast metastructures (HCM) [10, 37] refers to alternating high-index ($n_b$) and low-index ($n_a$) media with a finite thickness $t_g$ as shown in Fig. 1(b).

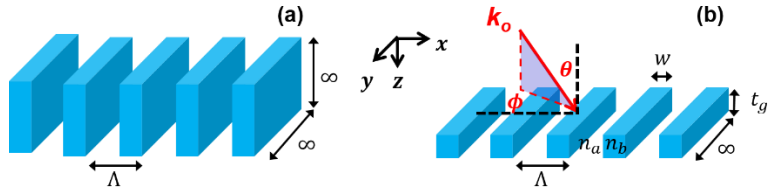

*Fig. 1. Schematics of (a) a 1D photonic crystal and (b) a 1D high-contrast gratings (HCGs) with a finite thickness $t_g$. Physical parameters include period $\Lambda$, thickness $t_g$, bar-width w, high-index material $n_b$ for bars, and low-index material $n_a$ for gaps between bars. The wave number for an incident plane wave is $k_0$. The incidence direction is defined by the polar angle $\theta_0$, and the azimuth angle $\phi_0$. The incidence plane is shaded plane in (b), defined by the surface-normal and the incidence directions. For 1D HCGs, $\hat{x}$ is the periodic direction and $\hat{y}$ is the invariant direction. For normal incidence ($\theta_0 = 0$) on 1D HCGs, TE and TM are defined as the electric field being parallel and perpendicular to the grating bars. For oblique incidence ($\theta_0 \neq 0$), s-polarization and p-polarization refer to the electrical field being perpendicular and parallel to the incidence plane, respectively.*

The 1D HCM design parameters include period $\Lambda$, bar width $w$, bar thickness $t_g$, and grating duty cycle defined as $\eta = w/\Lambda$. The plane wave incidence is characterized by the wave number $k_0$, the polar angle $\theta_0$, and the azimuth angle $\phi_0$.

$$\boldsymbol{k}_0 = \hat{x}k_{0x} + \hat{y}k_{0y} + \hat{z}k_{0z} = n_0\frac{\omega}{c}(\hat{x}\sin\theta_0\cos\varphi_0 + \hat{y}\sin\theta_0\sin\varphi_0 + \hat{z}\cos\theta_0)$$

(1)

For 1D HCMs under normal incidence, it is most convenient to define the polarization relative to the grating direction. In the field of HCGs, transverse-electric (TE) and transverse-magnetic (TM) are defined as the electric field being parallel and perpendicular to the gratings, respectively [37]. Under oblique incidence, the incidence polarization is defined without ambiguity relative to the incidence plane, which is formed by the surface normal and incidence direction, shown as the shaded plane in Fig. 1(b). In this case, s-polarization and p-polarization refer to the electrical field being perpendicular and parallel to the incidence plane, respectively [38].

The unique aspect about HCM/HCG is that it can be designed for incident light with various combinations of the oblique incidence plane and polarization, as shown in Fig. 2. When the incidence plane in perpendicular to the grating direction $\hat{y}$, both the fields and the structure are invariant in $\hat{y}$, as shown in Fig. 2(a) and Fig. 2(b), and the problem is known as a 2D problem. The TE case, which contains only $E_y$, $H_x$, and $H_z$, and the TM case, which contains only $H_y$, $E_x$, and $E_z$ are completely orthogonal to each other, and can be solved independently.

When the incidence plane in parallel to the grating direction, the $\hat{y}$-invariance is only in the structure but not in the fields, as shown in Fig. 2(c) and Fig. 2(d), thus the problem is known as the 2.5D problem [39]. The problem normally requires a 3D solver and hybrid modes in the HCM contains all field components.

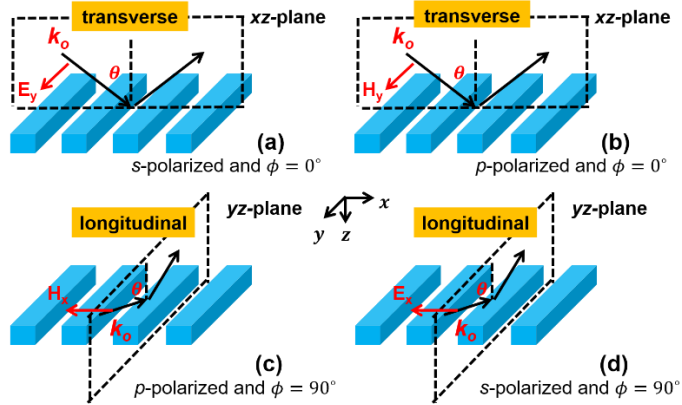

*Fig. 2. Four cases of oblique incidence with combinations of the tilt directions and polarizations. (a) and (b) are referred as the transverse-tilt with the tilt-plane (xz) being perpendicular to gratings. (c) and (d) are referred as the longitudinal-tilt with the tilt-plane (yz) being parallel to gratings. The incidence is s-polarized for (a) and (d), and p-polarized for (b) and (c).*

**2.1. Analytical and Numerical Methods for Periodic Structures**

For the ease of discussion, the meaning and description of math symbols in the review are listed in Table 1. The goal of modeling periodic structures includes solving the complex reflection and transmission coefficients of diffraction orders, the total field distribution, the resonances and quality factors, photonic band structures, and the dispersion of modes.

There are six major analytical and numerical modeling methods for periodic structures and the comparison of their computation methodologies is shown in Table 2

Table 1. Description and meaning of math symbols

| Symbols | Description | Meaning |
|---|---|---|
| $E_x(r), E_y(r), A_i$ | Normal font | **Scalars**: scalar coefficients or field distribution |
| $\mathbf{E}(r), \mathbf{H}(r), \mathbf{A}^+, \mathbf{A}^-$ | Bold font | **Vectors**: total field distribution or state vectors consisting of $A_i$ |
| $\mathcal{E}^{(i)}(x,y), \mathcal{H}^{(i)}(x,y)$ | Script font | **Eigenmodes**: field distribution |
| $\tilde{E}_{x,G}, \tilde{E}_{y,G}, \tilde{H}_{x,G}, \tilde{H}_{y,G}, \tilde{\mathbf{E}}, \tilde{\mathbf{H}}$ | Tilde accent | **Fourier coefficients**: Bloch expansion coefficients or vectors in Fourier domain |
| $\bar{\mathcal{E}}, \bar{\mathcal{H}}, \bar{\tilde{\mathbf{E}}}, \bar{\tilde{\mathbf{H}}}$ | Bar accent | **Matrices**: columns of eigenmodes or state vectors |

The finite-difference time-domain (FDTD) [40, 41], finite-element method (FEM) [41, 42], and the method of moments (MoM) [41, 43] are the three key techniques in computational electromagnetics (CEM). FDTD and FEM are differential equation solvers which discretize either the differential operator or the solution, while MoM is an integral equation solver derived using the Green's function. The entire solution domain is discretized in FDTD and FEM, therefore the computation cost for large problems and the convergence rate have long been challenges. Moreover, since the solution is the total field, separating and extracting modes or diffraction orders from periodic structures are rather inconvenient for FDTD and FEM. MoM can obtain fields throughout the space by calculating only the fields at boundaries, thus is efficient for scattering problems. It can readily produce scattering coefficients for separate diffraction orders, however, the required calculation of periodic Green's function suffers from slow convergence [44].

Despite being the most common techniques in CEM, FDTD, FEM, and MoM do not reveal the eigenmodes and their properties in the periodic region, therefore, much underlying physics is hidden. In the HCM and PhC communities, the rigorous coupled-wave analysis (RCWA) [45, 46] and the plane-wave expansion (PWE) method [3] are two widely used approaches. According to the Floquet-Bloch theorem, the fields in the periodic media can be expanded using the Bloch waves $e^{i\mathbf{G}\cdot\mathbf{r}}$, as shown in Table 2. The expansion coefficients $\tilde{E}_G, \tilde{H}_G$ for these Bloch waves are the unknowns to solve. The difference between RCWA and PWE states as follows. RCWA solves the eigenmodes $\mathcal{E}^{(i)}(r), \mathcal{H}^{(i)}(r)$ through the Bloch waves layer by layer, and is therefore preferred in layered media consisting of periodic structures. PWE expands the entire domain in the Bloch waves and is extensively used in the photonic crystal community for band structure calculation. In RCWA, each region including the periodic layer, is first considered as an infinite waveguide and its eigenmodes are obtained. Therefore, we understand what modes can couple to or be excited by radiative waves. It is also valuable to know their relative propagation velocity and optical phase. This way we can engineer the interference at region interfaces and achieve many extraordinary optical behavior [37, 38, 47]. After matching the boundary conditions at each layer interface, we obtain the interface transfer matrices and solve the eigenmode expansion coefficients $A_i$. With this we know how eigenmodes in each layer form "supermodes", which is the key factor for extraordinary performance in HCMs such as 100% reflection and transmission, and ultrahigh Q resonance [37, 47]. PWE, however, lacks this information since it solves individual eigenmodes in the entire domain.

Alternatively, eigenmodes in each region can be first expressed analytically and used as the eigenmode expansion basis, as shown in Table 2. In this case, fields in each layer can be represented by an array of complex expansion coefficients $a_i$ with eigenmodes being the basis. This method, which we refer as the analytical method or analytical mode-matching (AMM), has been extensively discussed in [37, 47], and proves to be accurate and very efficient for designing 1D HCGs. Similar to RCWA, the supermodes formed by the eigenmodes in each layer can be obtained for studying the resonance and coupling.

The comparison between the AMM and RCWA states as follows. For 1D HCGs, the eigenmodes take the forms of scalar sinusoidal functions, which are analytically rigorous. Therefore, the numerical error in AMM comes only from the finite number of eigenmodes included in calculation. At the interfaces between periodic and homogeneous regions, transfer and scattering matrices are analytically obtained by matching boundary conditions. In addition, it was excellent in revealing the underlying physics, resulting in convenient tools for intuitive designs [37, 47]. However, challenges arise when the grating cross-section is not rectangular, or when the periodicity becomes two-dimensional [38]. Rigorous close-form expressions of the vectorial eigenmodes do not exist in these cases. Even if the eigenmodes are solved numerically, the memory cost for storing the mode profiles, and the computation cost for numerical overlap integrals in the interface matrix elements are both heavy. To overcome these challenges, RCWA projects these eigenmodes further onto the Bloch basis, which consists of exponential functions $e^{i\bm{G}\cdot\bm{r}}$, as shown in Table 2. The expansion coefficients are complex vectors $\widetilde{\bm{E}}_{\bm{G}}^{(i)}$ and $\widetilde{\bm{H}}_{\bm{G}}^{(i)}$. In this case, eigenmodes are solved and represented in the Fourier domain with low computation and memory cost. The drawback, however, of using RCWA without identifying the eigenmodes is that it is exceedingly difficult to differentiate resonances from numerical noise due to the lack of physical interpretation to explain the modes. This is particularly true when there are anomalous dips in broadband reflection or transmission spectra or for high-Q resonator designs. An example of this is shown in the anomalous resonance line for longitudinal tilt plane wave incidence in Fig. 6. This shortcoming is easily overcome with AMM.

Table 2. Comparison among various computation methods for periodic structures

| | How to solve? (Basis functions) | What to solve? (Unknowns) | Results |
|---|---|---|---|
| Finite-Difference Time-Domain (FDTD) [40, 41] | 1. Discretize differential operator into matrix<br>2. Expand fields with delta functions $\delta(x_i, y_i, z_i)$ on finite difference grids $(x_i, y_i, z_i)$<br>3. Build matrix equation and solve | Fields on grids $\mathbf{E}(x_i, y_i, z_i)$ and $\mathbf{H}(x_i, y_i, z_i)$ | Total fields $\mathbf{E}(\mathbf{r})$, $\mathbf{H}(\mathbf{r})$ |
| Finite Element Method (FEM) [41, 42] | 1. Expand fields with basis functions $\psi_i(\mathbf{r})$, predefined commonly on elements in triangular or tetrahedral meshes<br>2. Build matrix from basis functions | Expansion coefficients $a_i$ | Total fields $\mathbf{E}(\mathbf{r})$, $\mathbf{H}(\mathbf{r})$ |
| Method of Moments (MoM) [41, 43] | 1. Express fields integral equations based on known Green's functions $\overline{G}(r, r')$<br>2. Expand surface currents and build matrix equations | Surface electric and magnetic currents $J(r')$, $M(r')$ | Scattered fields $\mathbf{E_s}(\mathbf{r}), \mathbf{H_s}(\mathbf{r})$ |
| Rigorous Coupled-Wave Analysis (RCWA) [45, 46] | 1. Predefined Bloch waves in each layer $e^{iG_x x + iG_y y}$<br>2. Solved eigenmodes in each layer $\mathcal{E}^{(i)}(\mathbf{r})$, $\mathcal{H}^{(i)}(\mathbf{r})$<br>3. Find propagation matrices and interface transfer matrices | 1. Expansion coefficients in each layer $\widetilde{E}_G, \widetilde{H}_G$ to build $i$-th eigenmodes in Eq. (A.8): $\mathcal{E}^{(i)}(\mathbf{r}) = \sum_G \widetilde{E}_G e^{iG \cdot r}$<br>2. Eigenmodes coefficients $A_i$ to build supermodes in Eq. (A.1): $E(\mathbf{r}) = \sum_i A_i \mathcal{E}^{(i)}(\mathbf{r})$ | 1. Eigenmodes in each layer<br>2. Supermodes in each layer<br>3. Total fields |
| Plane Wave Expansion (PWE) [3] | 1. Predefined Bloch waves in entire domain $e^{iG \cdot r}$<br>2. Cast wave equation into matrix equation and solve for expansion coefficients | Expansion coefficients $\widetilde{E}_G$, $\widetilde{H}_G$ to build eigenmodes in the entire domain: $E(\mathbf{r}) = \sum_G \widetilde{E}_G e^{iG \cdot r}$ | Eigenmodes in the entire domain |
| Analytical Mode-Matching (AMM) [37, 47] | 1. Predefined analytical eigenmodes $\mathcal{E}^{(i)}(\mathbf{r})$, $\mathcal{H}^{(i)}(\mathbf{r})$<br>2. Find analytical interface matrices | Eigenmode expansion coefficients $a_i$ to build supermodes in each layer | 1. Supermodes in each layer<br>2. Total fields |

Table 3 shows the comparison among modeling methods in terms of their capabilities. (i) All six methods except PWE can calculate the total fields under excitation since PWE solves only eigenmodes. (ii) The reflection and transmission of each diffraction order can be solved with MoM, RCWA, and AMM, but not with PWE. Post processing of the total fields is required for FDTD and FEM to separate the diffraction orders. (iii) The resonance frequency $\omega_r$ is solved as the eigenfrequencies in PWE, and determined from the scattering spectra in FEM, MOM, RCWA, and AMM. Simple Fourier transform of the FDTD result also gives the resonance frequency. (iv) The Q of a resonance mode is extracted from the time decay in FDTD, and the spectral linewidth in FEM, MOM, RCWA and AMM. (v)

Extracting the band structure from the scattering spectra from FEM, MoM, RCWA, and AMM are non-trivial. However, we can compare and confirm these scattering spectra with the band structure obtained from PWE. With FDTD, careful setup of the excitation sources and observation points are important for producing meaningful band structures, which will be discussed in later sections. (vi) FDTD as a time-domain solver cannot provide eigenmodes and their dispersion. FEM, RCWA, PWE, and AMM as frequency-domain solvers can obtain the eigenmodes by solving the eigenvalue problem with the excitation source being suppressed.

Table 3. Capabilities of various computation methods for periodic structures[a]

|  | Total field under excitation | $|r|e^{i\phi_r}$ and $|t|e^{i\phi_t}$ for each diffraction order | $\omega_r$ | Q | Band structure | Eigenmode & dispersion |
|---|---|---|---|---|---|---|
| FDTD | ✓ |  | ✓ | ✓ | ✓ |  |
| FEM | ✓ |  | ✓ | ✓ |  | ✓ |
| MoM | ✓ | ✓ | ✓ | ✓ |  | ✓ |
| RCWA | ✓ | ✓ | ✓ | ✓ |  | ✓ |
| PWE |  |  | ✓ | ✓ | ✓ | ✓ |
| Analytical | ✓ | ✓ | ✓ | ✓ |  | ✓ |

[a]Check marks indicate directly calculated or simple extraction.

### 2.2. Supermode Analysis in 1D Metastructures

In this section, we provide a complete analysis for 1D HCMs using analytical mode matching method. A review of method and formulation for surface-normal and transverse tilt incidence is included as the lead to the new analysis for longitudinal tilt incidence and 2D HCMs.

To understand cause for the extraordinary optical behaviors of near-wavelength periodic structures, we ought to study the eigenmodes and their resonance and interference. Fig. 3 (a) - Fig. 3(d) show the dispersion of the eigenmodes in coupled periodic waveguides which are infinite in $\hat{z}$. Even and odd polarities about $x = 0$ are defined on the dominant components ($E_y$ for TE and $H_y$ for TM), and labeled as red and blue, respectively. The in-plane wave number from the oblique incidence is embedded in the transverse distribution of the eigenmodes $\mathcal{E}_t^{(i)}(x, y)$ in the RCWA, as shown in in Eq. (**A. 1**). Comparing Fig. 3 (b) to Fig. 3 (a) or Fig. 3 (d) to Fig. 3 (c), the relative propagation speed is different among the eigenmodes under oblique incidence. Therefore their interference at the interfaces is altered, giving rises to angle-dependent new features that are not seen under normal incidence [21, 48].

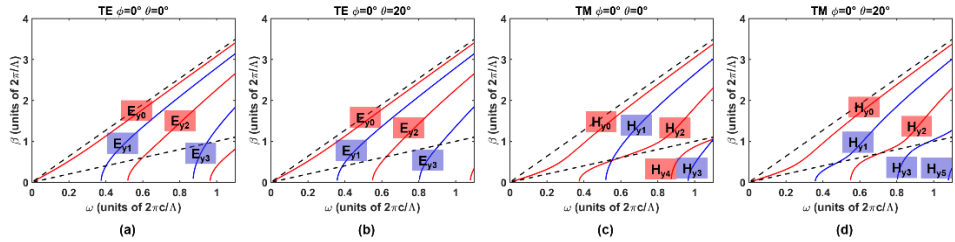

Fig. 3. Dispersion curves for $E_y$-polarized (TE) modes shown in (a) and (b), and $H_y$-polarized (TM) modes shown in (c) and (d) in a $\hat{z}$-infinite waveguide with periodicity in $\hat{x}$. Cases in (a) and (c) have $k_x = 0$, and correspond to normal incidence. Cases (b) and (d) have $k_x \neq 0$, and are matched with the $\theta_0 = 20°$ incidence. Refractive indices are $n_a = 1$ and $n_b = 3.17$. The duty cycle is $\eta = w/\Lambda = 55.556\%$. Red and blue lines indicate the dominant field components being even and odd in $\hat{x}$, respectively.

Resonance conditions in finite-thickness HCMs can be found by building the Fabry-Perot (FP) round-trip (RT) propagation matrix based on the few dominant eigenmodes [37, 38].

For normal incidence cases, choosing the fundamental mode and the 1st high-order even mode are usually sufficient. For this dual-mode analysis, instead of expanding the transverse field in the HCM region on to all eigenmodes as in Eq. (**A. 1**) and Eq. (**A. 2**), we can use a good approximation of

$$\mathbf{E}_t(\mathbf{r}) = A_0 \mathcal{E}_t^{(0)}(x,y) e^{i\beta_0 z} + A_2 \mathcal{E}_t^{(2)}(x,y) e^{i\beta_2 z}$$
$$= \begin{bmatrix} \mathcal{E}_t^{(0)} & \mathcal{E}_t^{(2)} \end{bmatrix} \begin{bmatrix} e^{i\beta_0 z} & 0 \\ 0 & e^{i\beta_2 z} \end{bmatrix} \begin{bmatrix} A_0 \\ A_2 \end{bmatrix}.$$

(2)

And the field can be represented by a state vector $[A_0, A_2]^T$. The RT propagation matrix can be written as [29, 37, 38],

$$\overline{\mathbf{M}}^{\mathrm{RT}}(\lambda, t_g) = \overline{\mathbf{R}}_{\mathrm{II-I}} e^{i\overline{\boldsymbol{\beta}} t_g} \overline{\mathbf{R}}_{\mathrm{II-III}} e^{i\overline{\boldsymbol{\beta}} t_g} \quad (3)$$

where $\overline{\boldsymbol{\beta}}$ is a diagonal propagation matrix composed of the propagation constants $\beta_i$'s of the eigenmodes. Regions I, II, and III refer to the incidence, HCM layer, and the transmission regions, respectively. The reflection matrices at the interfaces are $\overline{\mathbf{R}}_{\mathrm{II-I}}$ and $\overline{\mathbf{R}}_{\mathrm{II-III}}$. Resonance occurs when a supermode composed of the two lowest-order eigenmodes satisfies the FP RT condition, which is expressed as the eigenequation.

$$\overline{\mathbf{M}}_{2\times 2}^{\mathrm{RT}}(\lambda, t_g) \begin{bmatrix} A_0 \\ A_2 \end{bmatrix} = |\Omega| e^{i\phi} \begin{bmatrix} A_0 \\ A_2 \end{bmatrix} \quad (4)$$

Here, $A_0$ is the expansion coefficient of $E_{y0}$-mode for TE cases or $H_{y0}$-mode for TM cases, and $A_2$ is the expansion coefficient of $E_{y2}$-mode for TE cases or $H_{y2}$-mode for TM cases. The combination of $\lambda$ and $t_g$ supports resonance if any one of the two eigenvalues $|\Omega_p|e^{i\phi_p}$ and $|\Omega_q|e^{i\phi_q}$ from Eq. (**4**) satisfies the phase condition

$$\phi_{p,q} = 2m\pi, \quad m = 0,1,2,3 \ldots \quad (5)$$

Here, $p$ and $q$ are the supermode indices. These resonance conditions for TE and TM cases are shown in Fig. 4(a) and Fig. 4(b), respectively. Here, red and blue indicate the resonant supermode is $A_0$ and $A_2$ dominant, respectively. Fig. 4(c) and Fig. 4(d) show the reflectivity contours as functions of the normalized wavelength and the normalized HCM thickness, for TE and TM cases, respectively. The overlaps between the supermode resonances and the reflectivity contours show excellent agreement in Fig. 4(e) and Fig. 4(f). The field profiles of the dominant components in the resonant supermodes in Fig. 4(g) and Fig. 4(h) for TE and TM, respectively.

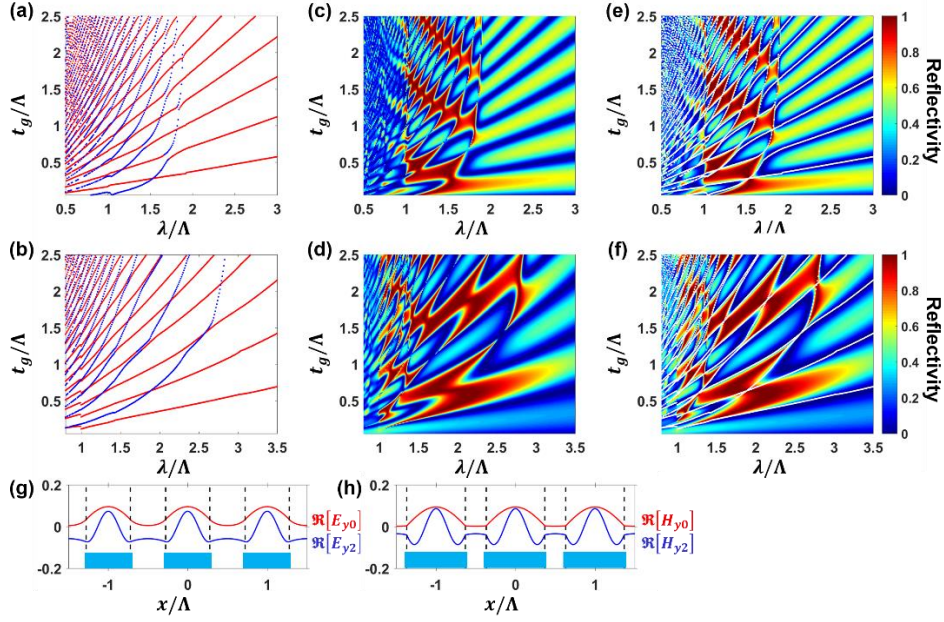

*Fig. 4. 1D-periodic HCM under normal incidence for TE and TM polarization incident light: (a)-(b) Fabry-Perot resonance conditions, (c)-(d) reflectivity contours with the well-known checker-board pattern, and (e)-(f) overlaps between resonance lines and reflectivity contours as functions of the wavelength and the HCM thickness. The polarization is TE for (a), (c), (e), and (g), and TM for (b), (d), (f), and (h). Red and blue dots in (a) indicate $E_{y0}$- and $E_{y2}$-dominant supermode resonances, and the field profiles $\Re[E_{y0}]$ and $\Re[E_{y2}]$ are shown in (g). Red and blue dots in (b) indicate $H_{y0}$- and $H_{y2}$-dominant supermode resonances, and the field profiles $\Re[H_{y0}]$ and $\Re[H_{y2}]$ are shown in (h). The duty cycle $\eta = w/\Lambda$ is 55.556% for (a), (c), (e), and (g), and 73.684% for (b), (d), (f), and (h). The normalized frequency is $\omega\Lambda/2\pi c = 0.65$ in both (g) and (h). Refractive indices are $n_a = 1$ and $n_b = 3.17$.*

For 1D-periodic HCMs under oblique incidence in the transverse plane ($\phi = 0°$), we observe an additional set of resonance lines in the reflectivity contours. Because of the optical phase variation in $\hat{x}$ induced by the incidence light, it is possible to excite an eigenmode with odd parity. The eigenequation of the FP RT condition in Eq. (**4**) is extended onto a three-fold basis,

$$\overline{\mathbf{M}}^{\text{RT}}_{3\times 3}(\lambda, t_g)\begin{bmatrix}A_0\\A_1\\A_2\end{bmatrix} = |\Omega|e^{i\phi}\begin{bmatrix}A_0\\A_1\\A_2\end{bmatrix} \quad (6)$$

Here, $A_1$ is the expansion coefficient of $E_{y1}$-mode for *s*-polarization or $H_{y1}$-mode for *p*-polarization. Resonance occurs if any one of the three eigenvalues $|\Omega_p|e^{i\phi_p}$, $|\Omega_q|e^{i\phi_q}$, and $|\Omega_l|e^{i\phi_l}$, from Eq. (**6**) satisfies the phase condition

$$\phi_n = 2m\pi, \quad m = 0,1,2,3 \ldots \quad (7)$$

Here, $n = p, q$, or $l$ is the supermode index. This is referred as the tri-mode analysis [29]. The red dots in Fig. 5(a) and Fig. 5(b) indicate have the resonant supermodes have dominant components in $A_0$, corresponding to the $E_{y0}$ and $H_{y0}$ eigenmodes for *s*- and *p*-polarizations, respectively. The black dots indicate $A_1$-dominant resonance. The pink dots indicate resonance being mixture of $A_0$ and $A_2$. The tri-mode resonance conditions, the tilt reflectivity contours, and their overlaps are shown in Fig. 5(a)-Fig. 5(f). The excellent agreement proves the well-prediction from the tri-mode analysis. To illustrate how the oblique incidence can excite odd-eigenmodes, we show the field profiles of the three eigenmodes in the HCMs in Fig. 5(g) for *s*-polarization. The real and imaginary parts of the fields are asymmetric across $\hat{x}$, indicating a phase variation along $\hat{x}$ required by the phase matching with the incidence. After excluding this phase variation by multiplying $e^{-ik_{0x}x}$, all three eigenmodes, including the odd-parity one, become symmetric along $\hat{x}$, as shown in Fig. 5(h). This proves that all 3 eigenmodes can be excited. Similar calculation is done for the $p$-polarization, as shown in Fig. 5(i) and Fig. 5(j).

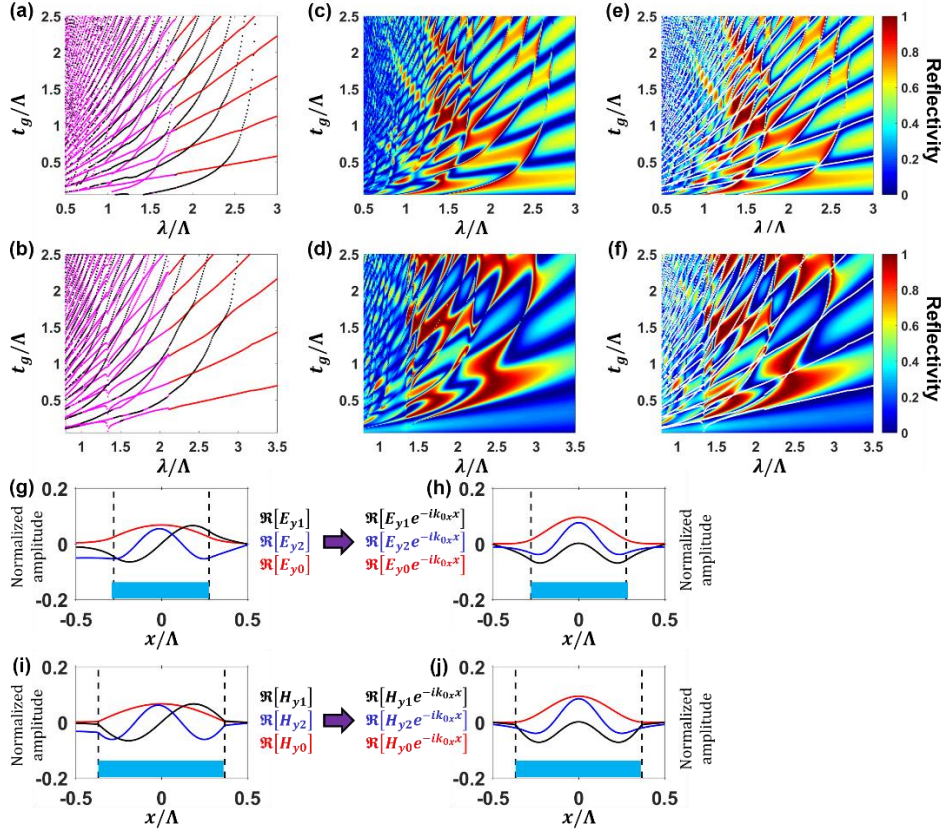

*Fig. 5. 1D-periodic HCM under transverse-tilt ($\phi_0 = 0°$) incidence at $\theta_0 = 20°$ for p- and s-polarized waves: (a)-(b) Fabry-Perot resonance conditions, (c)-(d) reflectivity contours, and (e)-(f) overlaps between resonance lines and reflectivity contours as functions of the wavelength and the HCM thickness. The incidence is s-polarized for (a), (c), (e), and (g), and p-polarized for (b), (d), (f), and (h). Red and black dots in (a) indicate $E_{y0}$- and $E_{y1}$-dominant supermode resonances. Red and black dots in (b) indicate $H_{y0}$- and $H_{y1}$-dominant supermode resonances. Pink dots in (a) and (b) indicate resonant supermodes being a mixture of $E_{y0}$ and $E_{y2}$, and a mixture of $H_{y0}$ and $H_{y2}$, respectively. The duty cycles $\eta = w/\Lambda$ are 55.556% for (a), (c), (e), (g), and (h), and 73.684% for (b), (d), (f), (i), and (j). Refractive indices are $n_a = 1$ and $n_b = 3.17$. S-polarized eigenmode profiles (g) before and (h) after excluding the phase variation $e^{ik_{0x}x}$ matched to the incidence. P-polarized eigenmode profiles (i) before and (j) after excluding the phase variation $e^{ik_{0x}x}$ matched to the incidence. The normalized frequency is $\omega\Lambda/2\pi c = 0.65$ in (g)-(j).*

For 1D-periodic HCMs under oblique incidence in the longitudinal plane ($\phi = 90°$), there is no phase variation in $\hat{x}$ induced by the incidence. Therefore, the odd eigenmodes cannot be excited as in the transverse tilt cases. However, we still see an additional set of resonance lines. We notice this case is a 2.5D problem, and each eigenmode has all six field components instead of only 3 components in previous cases, that is, $E_y$, $H_x$, $H_z$ for TE or $s$-polarization, and $H_y$, $E_x$, $E_z$ for TM or $p$-polarization. For this reason, when we consider $p$-polarization (normally exciting $E_y/H_x$ modes), a $E_x/H_y$-dominant eigenmode can have a $E_y$ component. If the $E_x$-component is odd along $\hat{x}$, then the associated $E_y$-component will be even along $\hat{x}$, which can be excited by the incidence. We now build the tri-mode FP resonance condition as in Eq. (6) and Eq. (7). But now, $A_0$, $A_1$ and $A_2$ correspond to $E_{y0}$, $E_{x1}$, $E_{y2}$ eigenmodes, respectively, for *p*-polarized incidence. For *s*-polarization, $A_0$, $A_1$ and $A_2$ correspond to $H_{y0}$, $H_{x1}$, $H_{y2}$ eigenmodes, respectively. Red and black dots indicate $A_0$- and $A_1$-dominant resonances, and pink dots indicate $A_0$-$A_2$-mixed resonance. The tri-mode resonance conditions, the tilt reflectivity contours, and their overlaps are shown in Fig. 6(a), Fig. 6(c) and Fig. 6(e) for *p*-polarization and Fig. 6(b), Fig. 6(d) and Fig. 6(f) for *s*-polarization. It should be noted that, the extra resonance lines for *p*-polarized longitudinal tilt incidence (corresponding to black lines in Fig. 6(a)), though being weak, are observable in Fig. 6(c). For example, one of such lines crosses through $\lambda/\Lambda = 1.5$ and $t_g/\Lambda = 0.37$ in Fig. 6(c). One may easily miss this resonance if only looking at the reflection contour. This further shows the usefulness of the supermode analysis. Excellent agreement means the chosen eigenmodes are causes of resonance. To illustrate the excitation of odd eigenmodes, we show for *p*-polarized incidence in Fig. 6(g), the $E_x$-polarized odd eigenmode contains an even $E_y$-component in Fig. 6(h), which matches with the incidence light. Similar illustration is done for the *s*-polarized incidence in Fig. 6(i) and Fig. 6(j).

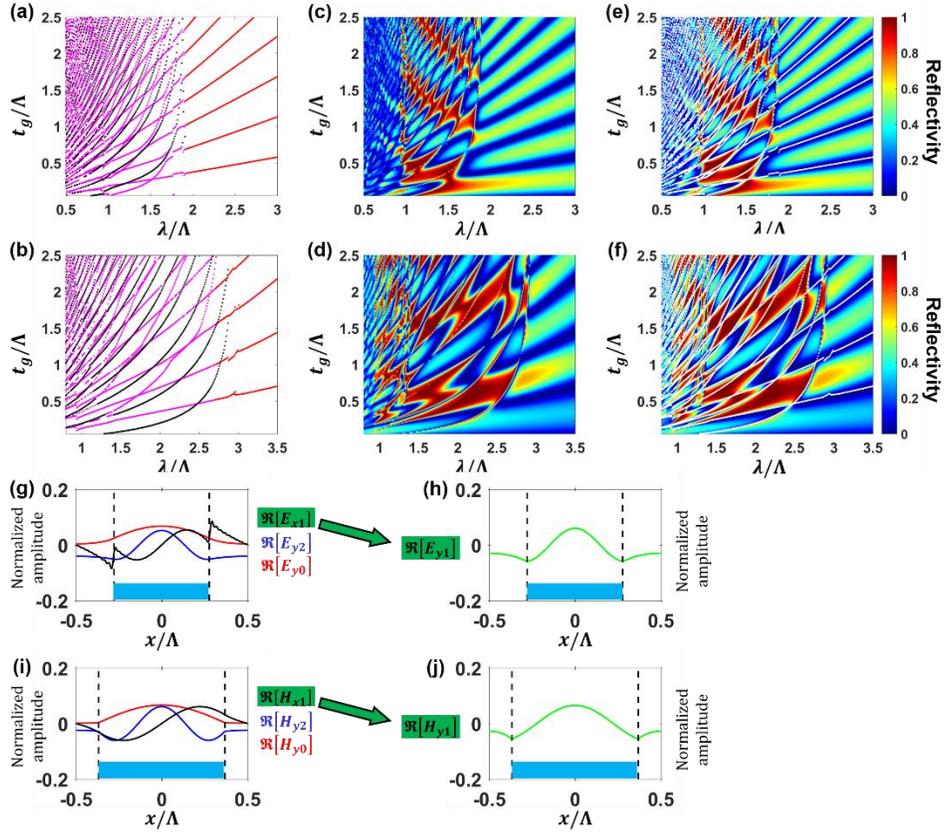

*Fig. 6. 1D-periodic HCM under longitudinal-tilt ($\phi_0 = 0°$) incidence at $\theta_0 = 20°$ for p- and s-polarized waves: (a)-(b) Fabry-Perot resonance conditions, (c)-(d) reflectivity contours, and (e)-(f) overlaps between resonance lines and reflectivity contours as functions of the wavelength and the HCM thickness. The incidence is p-polarized for (a), (c), (e), and (g), and s-polarized for (b), (d), (f), and (h). Red and blue dots in (a) indicate $E_{y0}$- and $E_{x1}$-dominant supermode resonances. Red and blue dots in (b) indicate $H_{y0}$- and $H_{x1}$-dominant supermode resonances. Pink dots in (a) and (b) indicate the resonances of supermodes being a mixture of $E_{y0}$ and $E_{y2}$, and a mixture of $H_{y0}$ and $H_{y2}$, respectively. The duty cycles $\eta = w/\Lambda$ are 55.556% for (a), (c), (e), (g), and (h), and 73.684% for (b), (d), (f), (i), and (j). Refractive indices are $n_a = 1$ and $n_b = 3.17$. (g) Field profiles of the 3 eigenmodes contributing to the resonance conditions in (a), including two $E_y$-dominant even modes, and one $E_x$-dominant odd mode. (h) The $E_x$-dominant odd mode in (g) has an even $E_y$-component which couples with the p-polarized incidence. (i) Field profiles of the 3 eigenmodes*

*contributing to the resonance conditions in (b), including two $H_y$-dominant even modes, and one $H_x$-dominant odd mode. (j) The $H_x$-dominant odd mode in (i) has an even $H_y$-component which couples with the s-polarized incidence. The normalized frequency is $\omega\Lambda/2\pi c = 0.65$ in (g)-(j).*

In Section 2.2, resonance conditions of 1D HCMs are studied using the supermode analysis, under normal, transverse tilt, and longitudinal tilt plane wave incidence, with combinations of both *s*- and *p*-polarizations. For the normal incidence case, two polarizations are decoupled, and we can refer as TE and TM (relative to the grating direction). The resonance is dominant by two-fold supermodes formed with the fundamental HCM eigenmode and the first higher-order even mode. For transverse tilt incidence, two polarizations are still decoupled, which we refer as *s*- and *p*-polarizations to remove confusion. The resonance is dominant by three-fold supermodes formed with the fundamental eigenmode, the first higher-order even mode and the first higher-order odd mode. For the longitudinal tilt incidence, the two polarizations are coupled, which means both TE-like and TM-like eigenmodes can be excited. Indeed the resonance is dominant by three-fold supermodes formed with the fundamental eigenmode, the first higher-order even mode, and the first higher-order odd mode with a nearly orthogonal polarization.

### 2.3. Band Structures and Oblique Incidence for 1D Metastructures

The photonic band structure is a key property to analyze when studying PhCs. One can obtain the frequencies of discrete modes allowed to propagate in a PhC at a given periodic phase shift, known as the Bloch wave number. The frequency stop band (band gap), resulting from the structural periodicity, gives rise to a broadband reflection and engineerable group velocities.d. In the HCM community, the broadband reflection and transmission are frequently analyzedand the connection to PhC is clear for the transverse tilt and, to a lesser extent, surface-normal incidence. We find when the light momentum from the oblique incidence matches with the Bloch wave number, the reflection and transmission spectra of a periodic structure are intrinsically connected to its band structure. However, the connection with longitudinal tilt was never established since the propagation constant is orthogonal to periodicity. Here we attempt to address the connection and comparison. .

We start with the band structure of a 1D PhC, as shown in Fig. 7(a), calculated using the PWE method and the 2D FDTD method, where we use a cloud dipoles as excitation and monitor the spectra with different Bloch phase shift at the periodic boundaries in $\hat{x}$. Photonic band gaps are opened as a result of the periodic index contrast. The Γ-point and X-point refer to $k_x = 0$ and $k_x = \pi/\Lambda$ in the *k*-space, respectively.

For transverse 1D-periodic HCM *s*-polarized case in Fig. 7(a) with a finite thickness $t_g = 0.5\Lambda$, we orient ensembles of electric dipoles along $\hat{y}$ in FDTD, and corresponding frequency bands are shown in Fig. 7(b). One can observe clear resemblance between Fig. 7(a) and Fig. 7(b), particularly on band-edge frequencies. The reason is obvious; Fig. 3(b) and Fig. 3(d) show the dispersion curves of the resonance modes in the structure with group velocity in the $\hat{x}$ direction, which is the same as modes for PhC. The only difference being Fabry-Perot effect for a finite thickness HCM in $\hat{z}$ direction.

We further calculate the *s*-polarized reflectivity spectra at various incidence angle $\theta_0$ using RCWA with 1D-periodicity. The contour reflectivity plot as a function of the normalized incidence wave number in $\hat{x}$, and the normalized frequency is shown in Fig. 7(c). The extracted peaks from the contour spectra in Fig. 7(b) are overlapped with the contour reflectivity in Fig. 7(c), and excellent agreement is shown in Fig. 7(d). However, the reflectivity values provided by RCWA cannot be obtained by the FDTD calculations nor PWE approximations in a PhC. FDTD is inconvenient for extracting the complex reflection and transmission coefficients, especially when more than one diffraction orders exist.

*The more peculiar case is for longitudinal HCMs. Similar comparisons are done for cases in Fig. 2(c) and Fig. 2(d). For the longitudinal tilt, neither the incidence wave number nor the Bloch phase shift exists in $\hat{y}$ direction. The first Brillouin zone is shown in Fig. 8(a). P-polarized frequency bands in Fig. 8(c) are excited with ensembles of $\hat{y}$-oriented magnetic dipoles in 2D FDTD, which match excellently with the reflectivity contour from RCWA, as shown in Fig. 8(d).*

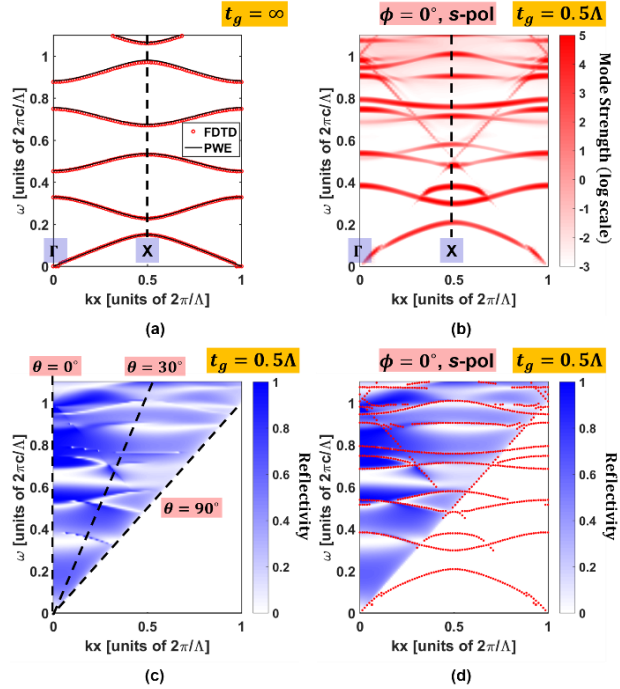

*Fig. 7. (a) Band structure of a 1D photonic crystal $t_g = \infty$ calculated using the FDTD and PWE methods, shown in red circles and black lines respectively. (b) Transverse-tilt s-polarized band structure of a 1D-periodic HCM with a finite thickness $t_g = 0.5\Lambda$ calculated using FDTD. (c) Transverse-tilt s-polarized angle-dependent reflectivity contour of the same structure, calculated using RCWA. (d) Overlap between (c) and the extracted photonic bands (red dots) from (b). Parameters: $n_a = 1$, $n_b = 3.48$, $\eta = 0.6$.*

To solve the band structures for the longitudinal-tilt cases, we use 3D FDTD by forcing a phase shift in $\hat{y}$ on the artificial boundaries, which matches with the incidence wave number in $\hat{y}$. The first Brillouin zone is shown in Fig. 8(b). The *s*-polarized and *p*-polarized frequency bands are excited with ensembles of $\hat{x}$-oriented electric and magnetic line sources, respectively. The reflectivity contours are still calculated by the RCWA with 1D-periodicity. Excellent matches for the longitudinal-tilt cases are shown in Fig. 8(e)-Fig. 8(h).

Most critically, it is important to note that for the longitudinal tilt, the propagation constant does not have any projection in $\hat{x}$ direction. Theoretically there should be no zone folding in $\hat{k}_y$ direction, since there is no periodicity along $\hat{y}$ direction. However, to numerically solve (e.g. in FDTD) the band structure along $\hat{y}$ direction, one typically has to impose artificial periodic *y*-boundaries with phase shift, as shown in Fig. 8(b). As long as the artificial period in $\hat{y}$ is small compared to the period $\Lambda$ (e.g. $0.1\Lambda$), the artificial folding boundary will be far away from our region of interest, as shown in Fig. 8(e) and Fig. 8(g), with negligible influence on the band structure calculation. This treatment is not necessary in RCWA reflection contours in Fig. 8(f) and Fig. 8(h), because one can still use 1D RCWA (i.e. 1D periodic in $\hat{x}$) and the $k_y$ propagation is induced from the oblique incidence. The above analysis is in sharp contrast with the case for transverse tilt in Fig. 8(c) and Fig. 8(d).

The photonic band structure provides the PhC community with most of the information required to predict the PhC optical properties [3]. For the HCM community, it is of great interest how the incoming light is scattered, confined, or guided. For instance, there are frequency regimes in between the bands that provide extraordinary features such ultrabroadband high reflection [10] and high-Q resonance [17, 49]. In this sense, invaluable information is revealed from the reflectivity and transmissivity contours, which proves to be efficient design tools [37].

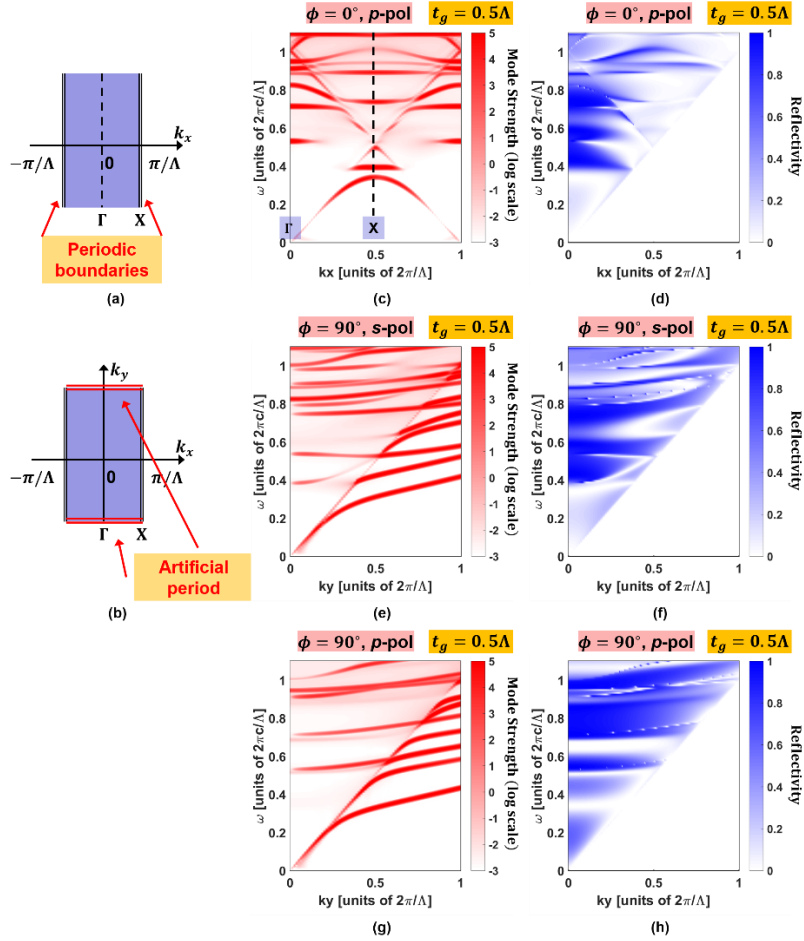

*Fig. 8. Brillouin zones of metastructures with 1D periodicity in $\hat{x}$ with (a) $k_y = 0$, and (b) $k_y \neq 0$. (c) and (d) correspond to the transverse-tilt p-polarized case. (e) and (f) correspond to the longitudinal-tilt s-polarized case. (g) and (h) correspond to the longitudinal-tilt p-polarized case. The red and blue plots are band structures and tilt reflectivity contours obtained from FDTD and RCWA, respectively. The HCM parameters are the same as in Fig. 7.*

In Section 2.3, we compare the band structures of 1D HCMs and the reflectivity contour under oblique incidence. Their overlaps show excellent agreement for four cases of tilt planes and polarizations. These two analyses commonly used in PhC and HCM communities, respectively, are intrinsically connected. The photonic bands and the dips in the reflectivity contour are both results of the resonance of supermodes, which are superpositions of one or a few dominant eigenmodes in the periodic region. Both the HCM lateral dimensions and the thickness affect the relative optical phases between the dominant eigenmodes when they interfere at the exiting interface, thus may lead to very different optical behaviors. There is, however, no clear analogy of longitudinal HCM with PhC.

### 2.4. 2D Metastructures

Metastructures with 2D periodicity are more challenging to design and analyze due to the large parameter space and complexity of the optical modes [38]. The thin layer of 2D-periodic HCM is also referred as the 2D PhC slab [3]. The 2D lattice structure can be rectangular, hexagonal, or even quasiperiodic [50, 51]. The building blocks can be dielectric cubes, nanobeams, circular rods, air-holes, V-antennas, and so on. Here we use HCMs on a square lattice to illustrate, as in Fig. 9. Two oblique incidence planes are chosen, aligning with the two periodic directions. Four cases in Fig. 9(a)-Fig. 9(d) with combinations of the incidence planes and polarizations are studied. The 2D-periodic HCMs are referred as the island-type when high-index building block is surrounded by low-index materials ($n_a < n_b$), as shown in Fig. 9(e). The opposite case with $n_a > n_b$ is referred as the island type, as shown in Fig. 9(f). The design parameters include the thickness $t_g$, periodicities ($\Lambda_x$ and $\Lambda_y$) and duty cycles ($\eta_x = w_x/\Lambda_x$ and $\eta_y = w_y/\Lambda_y$) in both $\hat{x}$ and $\hat{y}$ directions.

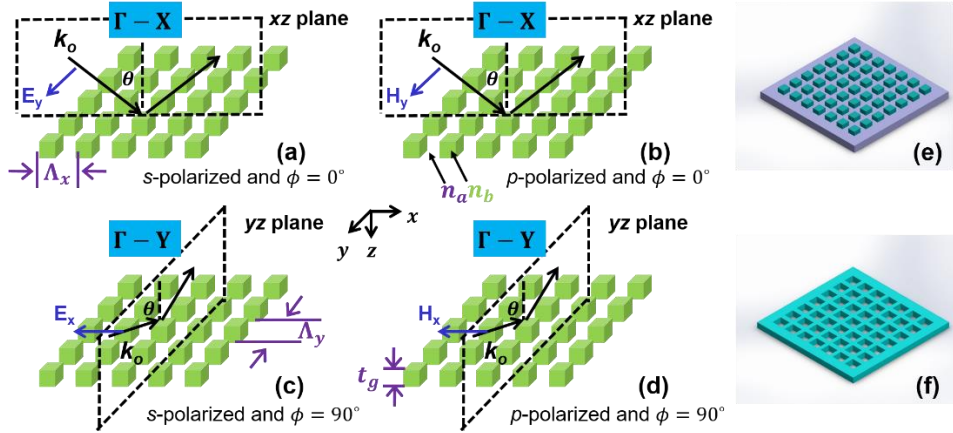

*Fig. 9. (a)-(d) Schematics of 2D-periodic HCMs under various combinations of incidence planes and polarizations. The refractive index for the building block is $n_b$, and the surrounding index is $n_a$. (e) An island-type 2D-periodic HCM with $n_a < n_b$. (f) A mesh-type 2D-periodic HCM with $n_a > n_b$. The periodicities in $\hat{x}$ and $\hat{y}$ are $\Lambda_x$ and $\Lambda_y$, respectively. The HCM thickness is $t_g$. The widths and the duty cycles of the building block are $w_x$, $w_y$, $\eta_x = w_x/\Lambda_x$, and $\eta_y = w_y/\Lambda_y$.*

The band structures of 2D-periodic HCMs or 2D PhCs are closely connected to their optical properties under oblique incidence, as shown in Fig. 10, given that the Bloch wave number in the *k*-space matches with the oblique incidence wave number. Here, we use the mesh-type HCMs on a square lattice, with rectangular holes as the building blocks. The 1st Brillouin zone for a square lattice is shown in Fig. 11(a). We calculate the band structures in 3D FDTD and the reflectivity contours in 2D-periodic RCWA, from Γ-point ($k_x = k_y = 0$) to X-point ($k_x = \pi/\Lambda_x$, $k_y = 0$) in Fig. 10(a)-Fig. 10(d). *S*- and *p*-polarized simulations are isolated by setting up ensembles of $\hat{y}$-oriented electric and magnetic infinite line sources, respectively. Fig. 10(e)-Fig. 10 (h) correspond to simulations from Γ-point ($k_x = k_y = 0$) to Y-point ($k_x = 0$, $k_y = \pi/\Lambda_y$), and ensembles of $\hat{x}$-oriented electric and magnetic infinite line sources are used for *s*- and *p*-polarizations in 3D FDTD, respectively.

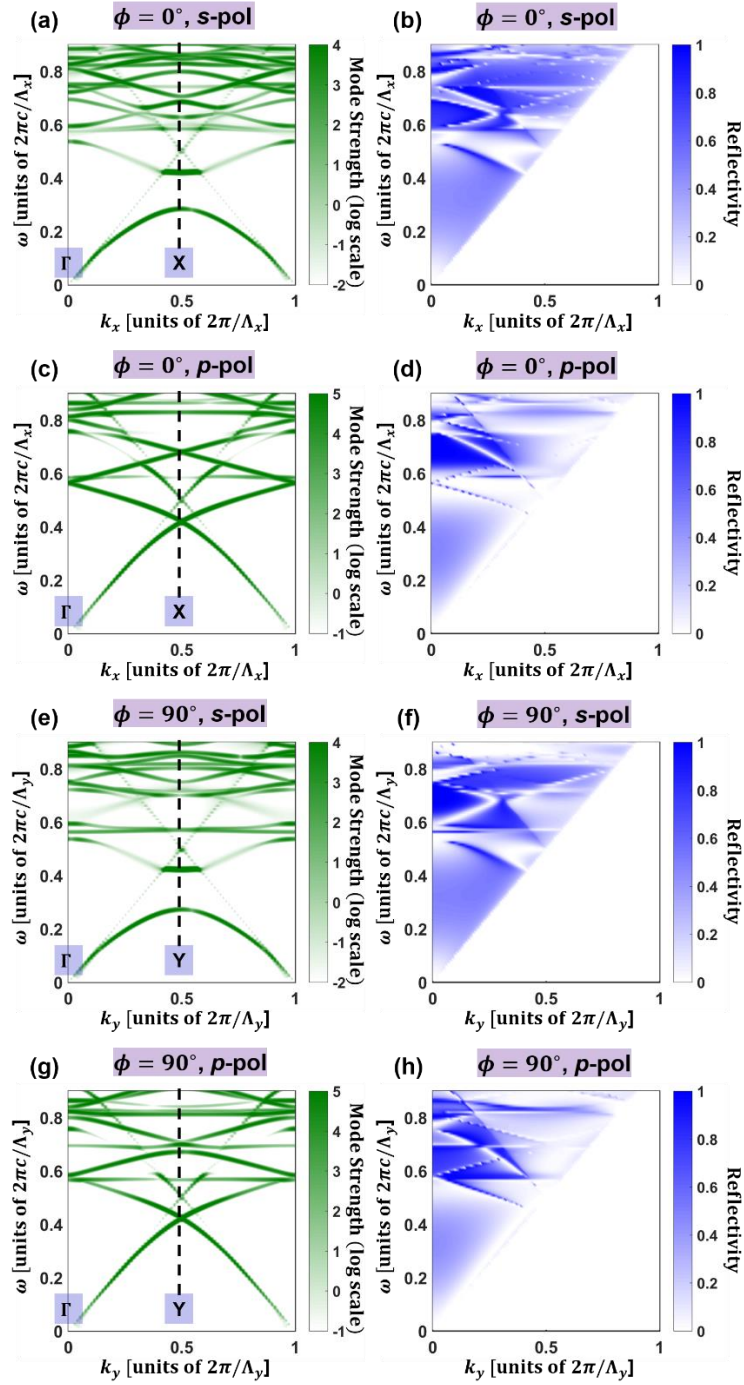

*Fig. 10. Comparisons between the band structures in (a), (c), (e), and (g) calculated using FDTD, and the angle-dependent reflectivity contours in (b), (d), (f), and (h) calculated using RCWA, of a mesh-type 2D-periodic HCM. (a)-(d) correspond to the xz-incidence plane ($\phi = 0°$). (e)-(h) correspond to the yz-incidence plane ($\phi = 90°$). (a), (b), (e), and (f) correspond to s-polarization with regard to their incidence planes. (c), (d), (g), and (h) correspond to p-polarization with regard to their incidence planes. Parameters: duty cycles $\eta_x = w_x/\Lambda_x = 73.333\%$ and $\eta_y = w_y/\Lambda_x = 66.667\%$, thickness $t_g = 0.4\Lambda_x = 0.4\Lambda_y$, refractive indices $n_a = 3.141$ and $n_b = 1$.*

Sweeping in the vast dimensions of parameter space for 2D HCMs or PhCs is rather impractical. Understanding the way modes interference and propagate in 2D-periodic structures greatly facilitate the design and optimization. Here, we consider the $E_x/H_y$-dominant eigenmodes in $\hat{z}$-infinite periodic coupled waveguides. We choose three lowest-order eigenmodes, namely EH$_{00}$, EH$_{20}$, and EH$_{22}$ modes, which are symmetric in both $\hat{x}$ and $\hat{y}$. Dispersion curves are shown in Fig. 11(b), The cutoff frequencies for the EH$_{20}$ and EH$_{22}$ modes are $\omega_{c2} = 0.397 \frac{2\pi c}{\Lambda_x}$ and $\omega_{c4} = 0.518 \frac{2\pi c}{\Lambda_x}$, respectively. The field profiles $\Re[H_y]$ are shown in Fig. 11(c)-Fig. 11(e). All three modes are possible to couple with $E_x/H_y$-polarized normal incidence. The elements in the state vector $[A_0, A_2, A_4]^T$ for the supermode correspond to the three eigenmodes.

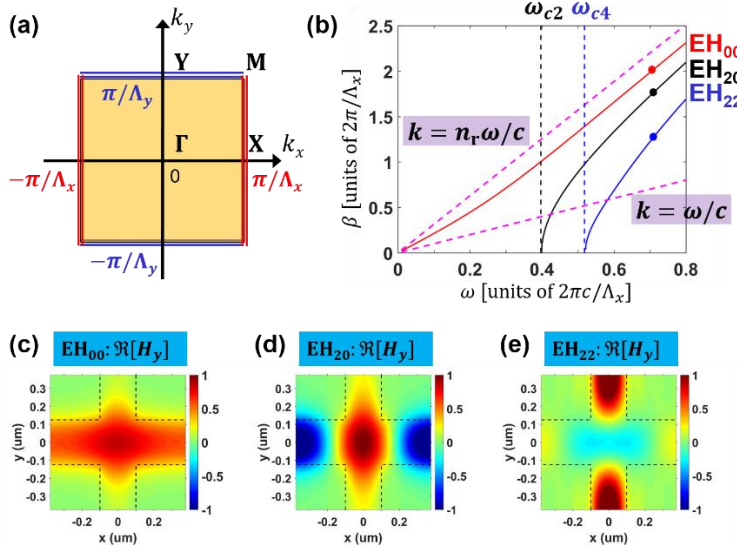

*Fig. 11. (a) Fourier space showing the 1st Brillouin zone of 2D-periodic HCMs on a square lattice. (b) Dispersion curves of the three lowest-order $E_x/H_y$-dominant eigenmodes. The light lines in the high- and low-index materials are shown as the pink dotted lines. The cutoff frequencies for the EH$_{20}$ and EH$_{22}$ modes are $\omega_{c2}$ and $\omega_{c4}$, respectively. (c)-(e) Magnetic field profiles $\Re[H_y]$ for EH$_{00}$, EH$_{20}$, EH$_{22}$ modes. The refractive indices are $n_a = 3.141$, and $n_b = 1$. The periods are $\Lambda_x = \Lambda_y = 750 \ nm$. The length and width of the air holes are $w_x = 550 \ nm$ and $w_y = 500 \ nm$. The incidence is surface-normal and $E_x/H_y$-polarized. The normalized frequency is $\omega \Lambda_x/2\pi c = 0.7$ in (c)-(e). [29]*

Using the tri-mode analysis, the resonance conditions as functions of the HCMs thickness and the wavelength are shown in Fig. 12(a). Red and black dots indicate the resonant supermodes being EH$_{00}$- and EH$_{20}$-dominant. Pink and blue dots indicate the resonant supermodes being mixture of EH$_{00}$ and EH$_{22}$ modes. The reflectivity contour under normal incidence and its overlap with the tri-mode resonance are shown in Fig. 12(b) and Fig. 12(c), respectively.

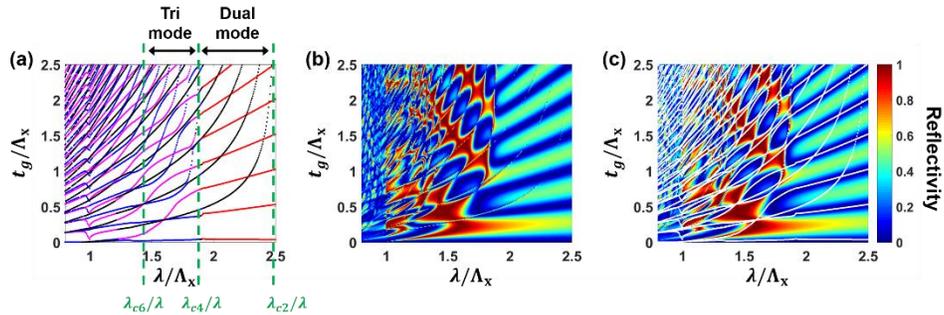

*Fig. 12. 2D-periodic mesh-type HCM under $E_x/H_y$-polarized normal incidence. (a) Fabry-Perot resonance conditions, (b) reflectivity contour, and (c) overlaps between resonance lines and the reflectivity contour as functions of the wavelength and the HCM thickness. Red and black dots in (a) indicate the resonant supermodes dominated by the $EH_{00}$ and $EH_{20}$ modes, respectively. Pink and blue dots in (a) both indicate resonant supermodes being a mixture of $EH_{00}$ and $EH_{22}$ modes. HCM parameters are the same as in Fig. 11. Adapted from [29].*

Fig. 13 shows a map of main operation regimes for HCMs, PhCs, and diffraction gratings, overlapped with the FDTD-calculated band structure of a 2D HCM (or 2D PhC slab). PhCs typically operate below the light line [52], where the modes are guided in the *x-y* plane within the slab. HCMs mainly operate above the light line, which is the radiation regime for PhCs, because HCM applications typically involve light impinging [18, 53], scattering [10], and radiative coupling [21]. Occasionally, HCMs are also optimized for a higher diffraction order [27]. The band analysis reveals the intrinsic physical connection between HCMs and PhCs. They are periodic in nature and support discrete resonance modes (eigenmodes) in the periodic region, and this is the same theoretical foundation for both HCMs and PhCs. For HCMs, radiative waves from outside interacts strongly with the periodic region, in which the eigenmodes largely determines the way HCMs respond to incoming waves. This is the reason for the excellent agreement between the reflection contours and the band structures, as shown in Fig. 7, Fig. 8, and Fig. 10. However, for the HCM community, the band structure alone does not provide quantitative information regarding the reflectivity and transmissivity. Moreover, instead of solving the band structures directly (as with the PWE method), the HCM community analyze the periodic layer as coupled waveguides. With the knowledge of these waveguide modes, along with their propagation constants, relative optical phases, we can precisely engineer their interference, which results in extraordinary optical phenomena, such as 100% reflection [47] and extremely high-Q resonator [17]. The excellent agreement shown in Fig. 4, Fig. 5, Fig. 6, and Fig. 12 proves that we are indeed working with the correct eigenmodes which are responsible for the extraordinary phenomena we observe.

As a summary of Section 2, periodic structures exhibit intriguing optical behaviors in the near-wavelength regime. Both the PhC and HCM communities study behaviors of periodic structures with different approaches. In particular, PWE is widely used to calculate band structures and engineer the band gap. However, it does not enable design for different functions, such as ultrahigh broadband reflection, perfect transmission, high Q resonance, reflection and transmission phase control, etc. With the help of supermode analysis, in combination with the reflection and transmission properties calculated by AMM or RCWA, very wide range of applications in Section 3 to Section 7 are made possible with engineered HCMs.

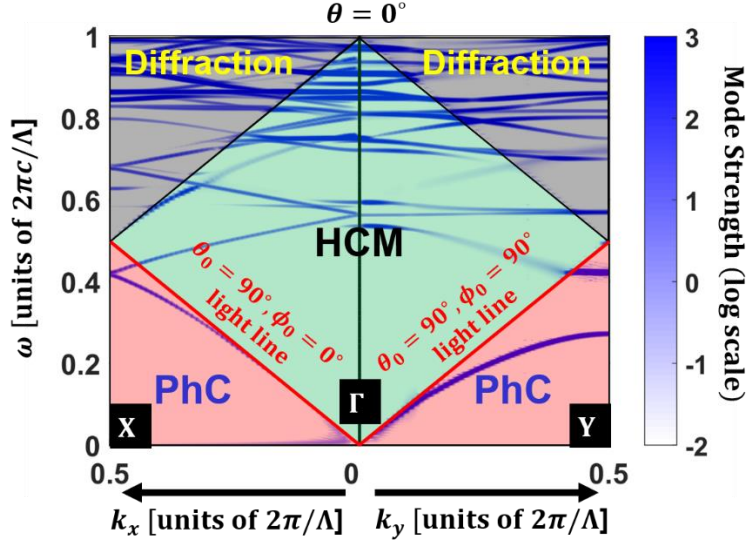

*Fig. 13. The FDTD simulated band diagram of a 2D HCM or a 2D PhC slab on a square lattice ($\Lambda_x = \Lambda_y = \Lambda$), indicating their operation regimes. The incidence light is $E_x/H_y$-polarized. The left half is the $\Gamma - X$ band structure, corresponding to Fig. 9(b), Fig. 10(c) and Fig. 10(d). The right half is the $\Gamma - Y$ band structure, corresponding to Fig. 9(c), Fig. 10(e) and Fig. 10(f). The operation regimes for HCMs and PhCs are separated by the $\theta_0 = 90°$ light lines (red lines), and are shaded in red and green, respectively. The HCM and diffraction regimes are separated by folded light lines from neighboring Brillouin zones. The HCM parameters are the same as in Fig. 11.*

## 3. Broadband Mirrors

### 3.1. High-Contrast Metastructure Mirrors

HCM mirrors are referred as planar arrays of metastructures of which the periodicity is not in-plane with the incidence wave vector. As shown in the previous section, the HCMs can be considered as coupled waveguides in the *z*-direction. For a wide wavelength range, there could be only a few modes propagating, as shown in the dispersion relations in Fig. 3 and Fig. 11(b). When these few modes are excited at the HCM input *z*-plane, and propagate for a finite distance $t_g$, the accumulated optical phase for each mode is different. With a correct choice of HCM thickness $t_g$, optical field intensity may perfectly cancel through destructive interference. This is the underlying mechanism for broadband high reflection and even 100% reflection [37, 47].

In 2004, Mateus et al. numerically simulated and experimentally demonstrated HCMs, which then was termed "high-contrast gratings (HCGs)", with ultrabroadband reflection ($\Delta\lambda/\lambda_0 \sim 30\%$) with very high reflectivity ($R > 99\%$) [10], as shown in Fig. 14(a). Magnusson complemented the work on HCGs by showing broadband reflection even when the gratings are index-matched with homogeneous layer directly below them, as shown in Fig. 14(b). This structure was then referred as "zero-contrast gratings (ZCGs)." [54] [55]. Subwavelength periodic structures similar to ZGCs were, in fact, used to provide antireflection [56], spatial variation in polarization state [57, 58] and optical phase [59]. The most significant difference is that a HCG has high index gratings fully surrounded by index contrast, whereas the a ZGC is typically a deep grating made on a homogeneous layer without the full contrast from the bottom exiting plane.

Viktorovitch et al. discussed the combination of a 1D/2D high-index-contrast lateral structure with a 1D high-index-contrast vertical structure, using multilayer membrane stacks including 1D/2D PhC membranes, thus resulting in so-called 2.5D PhC [60]. One example is the double grating membrane structure, as shown in Fig. 14(c) forming a Fabry-Perot cavity. Here, 2.5D is referring to the structure only, which is not to be confused with the 2.5D electromagnetic problem in Fig. 9(c)-Fig. 9(d). Despite the name of 2.5D PhC, we categorize these structures as HCM mirrors because the incidence light is out-of-plane from the periodic direction.

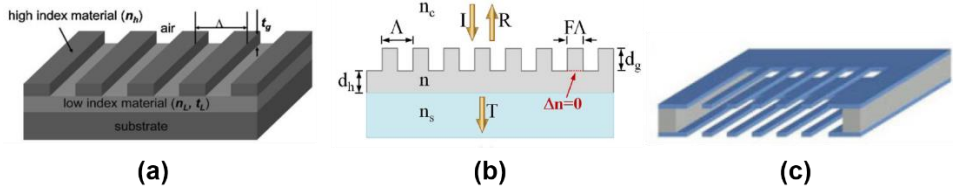

*Fig. 14. Examples of HCM mirrors: (a) 1D high-contrast gratings [10]; (2) 1D zero-contrast gratings [54]; (c) Combination of lateral periodic structures with 1D high-contrast vertical structure, using multilayer membrane stacks [60].*

Due to the extraordinary performance as broadband mirrors, HCMs have been widely used during the past decade in widely-wavelength-tunable or multi-wavelength VCSELs operating at 850-nm [11], 980-nm [14], 1060-nm [15], 1310-nm [12], and 1550-nm [13] wavelengths. Examples of 1550-nm proton-implanted and 1060-nm oxide-confined HCM-VCSELs are shown in Fig. 15(a) and Fig. 15(b), respectively. 1D HCMs have also been used for VCSEL polarization control. Zhou et al. measured polarization-resolved emission of 850-nm HCM-VCSELs with more than 28 dB selectivity between two orthogonal polarizations [61]. Rao et al. measured above 20 dB selectivity for 1550-nm HCM-VCSELs [13].

Using the same thickness by changing the lateral dimensions, 2D HCMs can function as polarization-dependent or polarization-independent mirrors. Polarization-insensitive (or bistable) VCSELs [29, 62-64] are made possible for all-optical flip-flops and buffers with the use of 2D HCMs [65]. Fig. 15(a) and Fig. 15(b) show the schematics of widely-tunable VCSELs using microelectromechanical system (MEMS) controlled 1D and 2D HCM mirrors, respectively. Notably, 2D HCM is able to provide [29] above $\Delta\lambda/\lambda_0 > 10\%$ for 99.5% reflection bandwidth, which is much beyond the record wavelength swept range (~6.9%), which is limited by the cavity free spectral range, for existing electrically-pumped tunable VCSELs [16]. As a result, 2D HCMs can readily be used as broadband mirrors for tunable VCSELs, as shown in Fig. 15(b) while opening up broad functionalities as integrated spatial phase or amplitude modulators [29].

It should be emphasized that with a single thin layer of HCM, either 100% reflection or 100% transmission can be achieved, which is proven rigorously by the theoretical analysis of supermodes [37, 47] in 1D HCMs. To understand physically, when the incidence light couples to the only two existing HCM modes, it is possible for the field DC components (i.e. the zeroth-order Fourier component) of these two modes to cancel perfectly at either the input or output plane. Such conditions can even be engineered, because the lateral HCM dimensions determine the relative propagation speed between the two modes, and the HCM thickness determines their relative optical phase accumulated. With such perfect cancellation, 100% reflection or transmission will naturally be achieved.

Such analysis of 100% reflection or 100% transmission is extended to 2D HCMs [38]. Although the incidence light can couple to a few HCM modes (more than two), as indicated in Fig. 11(b) and Fig. 12(a), their field DC components can still cancel perfectly. Fig. 16(a) shows the reflection spectrum of a 2D HCM and 100% reflectivity occur at wavelengths of 1024 nm and 1093 nm. When the reflection is perfect, the transmission goes zero and the transmission phase spectrum will experience a $\pi$-discontinuity. With the help of RCWA, we can isolate the contribution from each HCM modes to the output transmission. As shown in Fig. 11(c)-Fig. 11(e), $EH_{00}$ and $EH_{22}$ have the largest field DC components.

They are both strongly excited by the *x*-polarized incidence and strongly interfere with each other. When the $EH_{00}$ and $EH_{22}$ modes at the exit *z*-plane are similar in DC magnitude but out-of-phase ($\pi$-phase difference), as shown in Fig. 16(b) and Fig. 16(c), we expect large cancelation of the transmitted wave.

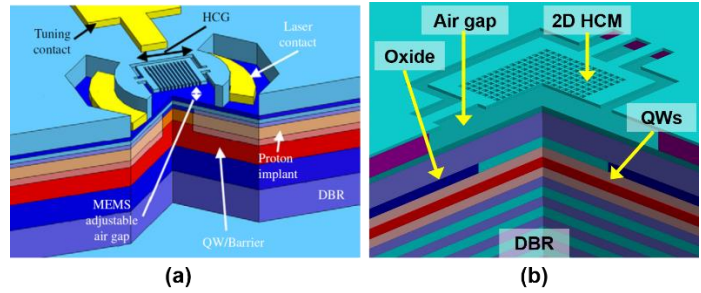

(a)            (b)

*Fig. 15. (a) Schematic of a proton-implanted tunable VCSEL using MEMS-controlled 1D HCM mirror. [13] (b) Magnified view of the top mirror of an oxide-confined MEMS-tunable VCSEL using 2D HCM. [29]*

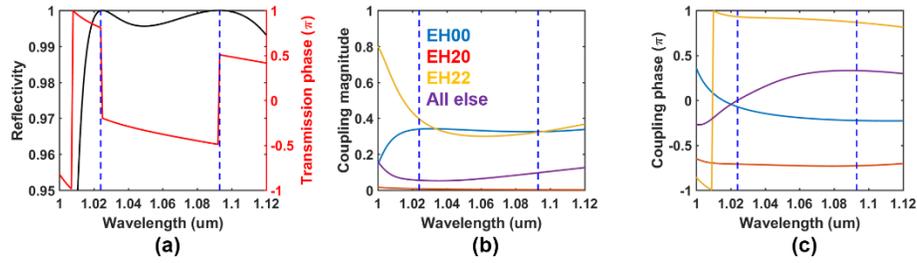

*Fig. 16. (a) Reflectivity spectrum (black) of a 2D HCM showing 100% reflection at wavelengths of 1024 nm and 1093 nm, indicated by the blue dashed lines. When perfect reflection occurs, the transmission phase spectrum (red) has a $\pi$ discontinuity. (b) Magnitudes and (c) phases of the transmission coefficients out-coupled from individual HCM modes, with $EH_{00}$, $EH_{20}$, $EH_{22}$, and all the remaining evanescent modes shown in blue, red, orange, and purple, respectively. The 2D HCM dimensions are the same as in Fig. 11.*

To further utilize the ultra-compactness of HCMs, double 1D-HCM VCSEL has been demonstrated at long-wavelength (1550-nm) under optical injection by Sciancalepore et al. [66]. The schematic is shown in Fig. 17(a). Single-mode operation is achieved without gain- or index-guiding mechanisms, and the integrated device is compatible with CMOS technology. The top and bottom reflectors are referred as "photonic crystal mirrors (PCMs)". However, in order to have surface normal emission, the periodic membrane indeed operates above the light line, corresponding to the HCM regime in Fig. 13. Another example of double periodic membrane surface emitting laser is shown in Fig. 17(b). The integration of the III-V gain medium and 2D-periodic silicon nanomembranes is achieved by the stamp-assisted transfer-printing process [67]. Optically-pumped room-temperature lasing at 1550-nm wavelength is achieved. Similar to Fig. 17(a) where 1D-periodic membrane is used, here the 2D-periodic membrane reflects out-of-plane light to form a vertical Fabry-Perot cavity. The light transmission is also out-of-plane from the periodic directions to enable surface normal emission. Therefore, these two examples are categorized as HCM mirrors.

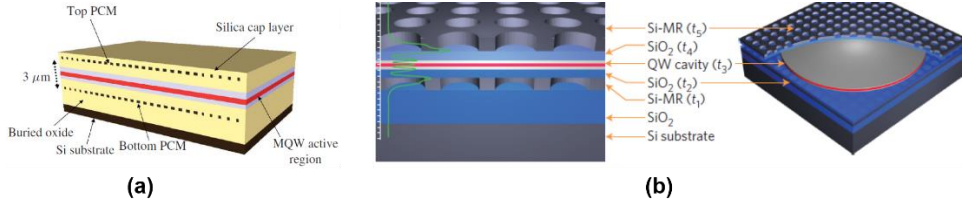

*Fig. 17. (a) Schematic of an optically-pumped CMOS-compatible double-HCM VCSEL. Adapted from [66]. (b) Schematic of an optically-pumped silicon membrane reflector VCSEL fabricated by transfer printing. Adapted from [67]*

One key advantage of HCM mirrors is their ultralow thickness required to maintain broadband high reflection. For a reflectivity above 99% over a $\Delta\lambda/\lambda_0 \sim 30\%$ wavelength range, the HCM thickness can be as low as $\sim 0.15\lambda_0$. As a comparison, DBR typically requires 20-40 pairs of alternating layers for a similar reflectivity, where the total thickness would be 10-20 times larger. With the combination of two lateral dimensions, the HCM mass could be ~10,000 times lighter than a DBR, resulting in ~100 times faster tuning speed. A 3µm×3µm 1D HCM with only four periods as the tunable mirror of a 850 nm VCSEL has shown a 27 MHz mechanical tuning speed [68]. With this advantage, arrays of MEMS-controlled HCM reflectors have been fabricated on a SOI wafer and demonstrated with ultrafast beam steering (response time ~µs) with low actuation voltage [24]. Moreover, each pixel of the phased array can be designed as a Fabry-Perot etalon using suspended HCMs on a DBR, as shown in Fig. 18(a)-(b). The air gap thickness is actuated by MEMS voltage, and each pixel behaves as an all-pass filter [25]. The 8x8 phased array with individually electrically addressed pixels have been fabricated on a GaAs-based epitaxial wafer, as shown in Fig. 18(c). Fast beam steering ($>$ 0.5 MHz) with low actuation voltage (10 V) and 9.14° total field-of-view.

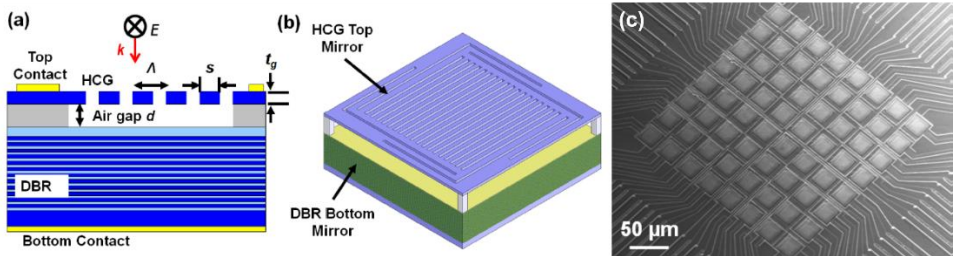

*Fig. 18. (a) Cross-sectional and (b) tilted-view schematics of an individual pixel of an optical phased array. Each pixel is a HCM all-pass filter (APF) individually controlled by MEMS voltage. The device is fabricated on a GaAs-based epitaxial wafer. GaAs HCMs are suspended on GaAs/AlGaAs DBR, forming the Fabry-Perot etalon. (c) Scanning electron micrograph of the 8x8 optical phased array using MEMS HCM-APFs. Adapted from [25].*

### 3.2. Photonic Crystal Mirrors

PhC mirrors are referred as arrays of subwavelength structures, of which the periodicity is in-plane with the incidence light wave vector. By this definition, traditional DBR mirrors fall into this category, which indeed are 1D PhCs. DBR (1D PhC) high reflectivity mirrors have been extensively used in semiconductors lasers, for both surface-emission (e.g. VCSELs) and edge-emission (e.g. DBR lasers) technologies. Moreover, PhC mirrors with 2D periodicity have also been used in in-plane devices such as ridge waveguide lasers, as shown in Fig. 19. Due to the high reflectivity from PhC mirrors, both the lasing threshold and the required cavity length has been significantly reduced, leading to large Fabry-Perot mode spacing and single-mode operation [69]. Since the PhC in-plane mirrors are well-understood with the photonic band theory [1, 3] and extensively applied in commercialized photonic devices, lengthy discussion will be omitted in this review.

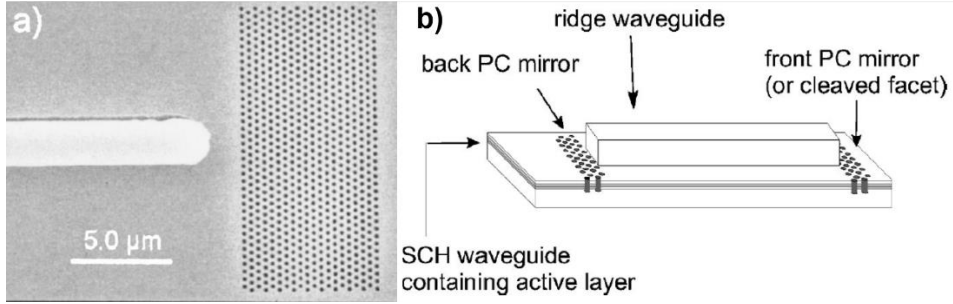

*Fig. 19. (a) SEM of an in-plane 20-row 2D PhC mirror next to a ridge waveguide. (b) Schematics of a short-cavity ridge waveguide laser using 2D PhC mirrors. Adapted from [69].*

## 4. Metasurfaces

Metasurfaces have gained extensive research interest in the past few years due to their extraordinary optical behavior as artificial material interfaces with ultrahigh compactness. The early research focused on 2D metal nanostructures and, recently, there are intense research activities on dielectric structures. Although the HCMs and dielectric metasurfaces came to being from different origins, they are indeed very similar in appearance and applications, and indeed even design methods. In this section, we will review the evolution and development of metasurfaces in terms of their operation regimes, governing working mechanisms, and the span of functionalities.

### 4.1. Historical review

The use of periodic structures (e.g. diffraction gratings) as artificial interfaces for diffracting the impinging light into wavelength-dependent directions dated back to 17th century, following the discovery of diffraction pattern through a bird feather by James Gregory in 1673. Up to now diffraction gratings remain the fundamental elements in optical systems such as monochromators and spectrometers. The diffraction angle $\theta_n$ of the $n$-th diffraction mode (also known as the Floquet-Bloch mode) is well predicted by the constructive and destructive interference conditions, or equivalently the grating equation [70]

$$\sin \theta_n = \sin \theta_0 + \frac{n\lambda}{\Lambda}, \quad n = 0, \pm 1, \pm 2, \ldots \tag{8}$$

The grating dimensions are typically larger than the wavelength of light, especially when high diffraction orders are used. The individual resonance behavior of the building blocks (e.g. stripes or grooves), the field interaction among them, and the near field distribution have been less of concern, while the spatial wavelength separation, spectral resolution, and diffraction efficiency in the far-field are main engineering targets. As a result, classical diffraction theory [70] is considered sufficient, and the operation regime is thus referred as the diffraction regime, as shown in Fig. 13.

In 1902, surprising phenomena were observed by Wood from the diffraction spectra of metallic grating [71]. Sharp changes of diffraction amplitude occurred within narrow spectral ranges near the cutoff wavelengths of certain diffraction orders (i.e. $\theta_n \approx 90°$), and were highly polarization dependent, i.e. only for E-field perpendicular to the grating. These phenomena were later known as the Wood's anomaly, which drew significant attention back to optical gratings, and opened up investigations beyond classical diffraction theory. Initial explanation by Rayleigh attributed the anomalies to the emergence of diffraction orders [72], despite the discrepancy with measurements. Fano complemented Rayleigh's theory by [73] explaining the spectral shape of the anomalies, and noticed a resonance-like and clearly different type of anomaly associated with the leaky waves supported by the grating. Using rigorous numerical tools, Hessel and Oliner proved in 1960s the resonance-type anomalies, and through understanding of surface impedance explained the possibility of anomalies for E-field parallel to the gratings with deep grooves [74]. Because of the extensive use of metallic surfaces, the study of Wood's anomalies eventually led to the discovery of SPP. Interested readers may refer to [75] for further details. In short, researchers throughout the 20[th] century realized that besides the periods and duty cycles, the thicknesses, shapes and material types of the grating could strongly alter the diffraction behavior via various resonance and coupling effects.

Before the appearance of metasurfaces, a closely-related structure known as the frequency selective surface (FSS) [76, 77] has been intensively studied since 1960s. FSSs were initially intended for spectral filtering, with diverse applications ranging from microwave ovens to antenna and radomes. Arrays of resonant structures are modelled with equivalent RLC circuits, which provide physical insight of the resonance frequency, reflection and transmission. The inaccuracy of circuit models to capture the geometry and material of FSS has been later overcome by advanced computation methods such as MoM, FEM and FDTD.

In 1979, Agrawal and Imbriale [78] demonstrated a dual-frequency (2.0-2.3 GHz and 13-15 GHz) microwave reflector comprising arrays of copper scatterers referred as "cross dipoles", printed on a dielectric sheet, as shown in Fig. 20(a). The surface parameters including the length, width and spacing of the scatterers are theoretically studied via MoM, which expands the current distribution on the scatterers. Both the simulation and measurements show that the resonance frequency is dependent on the size of the scatterer, but insensitive to the spacing. The surface is highly reflective near the resonance of the "dipoles", while nearly transparent at off-resonance. The scatterer spacing in [78] is smaller than the free-space wavelength so that higher-order Floquet-Bloch modes are made evanescent.

In 1989, Bertoni et al. [79] used alternating dielectric strips, as shown in Fig. 20(b), to demonstrate the possibility to have multiple frequencies for total reflection or total transmission to take place. Examples in [79] indicate both total reflection and total transmission can occur either below or above the cutoff wavelengths of higher diffraction orders. It was stated that total transmission would occur when the layer thickness was one half the effective wavelength along surface normal direction, and the total reflection was associated with the leaky waves. However, bases on the analysis in this review, we can tell the conclusion in [79] was rather incomplete. In fact, total reflection or transmission can happen at multiple layer thicknesses. Further, they can occur with or without all higher-order Floquet-Bloch-modes being evanescent. In 1993, Wang and Magnusson presented the theory of guided mode resonance [80], which refined Bertoni's study of dielectric waveguide gratings, and confirmed the two different types of resonance identified by Hessel and Oliner: a Rayleigh type due to the appearing or disappearing of a spectral order, and a guided mode resonance type supported by the waveguide gratings. The resonance regimes associated with diffraction orders are provided. In 1997, Lalanne and Morris [56] reported the antireflection behavior at visible wavelengths for 2D subwavelength gratings fabricated on silicon wafers. The performance was comparable to quarter-wavelength antireflection coating over 60° field of view.

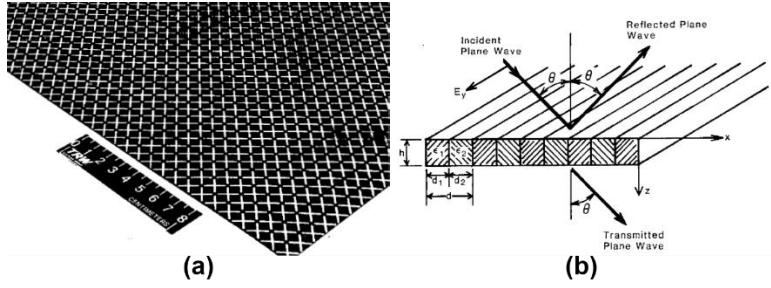

*Fig. 20. Examples of frequency selective surfaces. (a) Copper scatterers on a dielectric sheet as a dual-frequency reflector. Adapted from [78]. (b) A periodic array of dielectric strips for total reflection or total transmission. Adapted from [79].*

Kuester, Holloway, and co-workers first coined the term "metafilm" in 2003, referring to a surface distribution of electrically small scatterers [81], as shown in Fig. 21(a). The term later evolved into "metasurface", which may comprise scatterers of arbitrary shape, as long as the physical dimensions are small compared to the wavelength in the surrounding medium [34]. Metafilms and metasurfaces are regarded as the 2D equivalence of metamaterials in manipulating electromagnetic waves. One possible feature of this medium is to have the effective permittivity and permeability being simultaneously negative, also known as the double negative (DNG) material or left-handed material (LHM) [82], in which light can be "bent" in an anomalous fashion. Kuester et al. proposed the generalized sheet transition conditions (GSTCs) to characterize metasurfaces, as shown in Fig. 21(b), in contrast to the effectively medium theory for metamaterials. The surface distribution of scatterers were replaced by an infinitely thin interface with generalized boundary conditions to incorporate scatterer characteristics. In 2011, Yu, Capasso, and co-workers proposed and experimentally demonstrated the generalized laws of reflection and refraction using metallic metasurfaces with abrupt optical phase shift Φ at the interface [83]. The phase shift Φ is a continuous function of position (*x,y*) along the interface. As shown in Fig. 21(c), the conventional Snell's law is modified by the gradient of the phase shift Φ,

$$n_t \sin\theta_t - n_i \sin\theta_i = \frac{\lambda_0}{2\pi}\frac{d\Phi}{dx} \quad (9)$$

Here, the phase gradient can be controlled through metasurface design.

The work on 2D arrays of dielectric metastructures also started around two decades ago. In 2002, Hasman and co-workers used dielectric gratings with spatially-varying orientation for polarization conversion and vortex beam generation [58], as shown in Fig. 21(d). The dimensions of the building blocks are the same. In 2010, both Lu et al. [22] and Fattal et al. [23] used 1D dielectric gratings with varying sizes in both lateral directions to function as 2D lenses, as shown in Fig. 21(e) and Fig. 21(f), respectively. The metastructure arrangement for Fig. 21(e) and Fig. 21(f) belongs to a different type than Fig. 21(d), which will be discussed in the following sections.

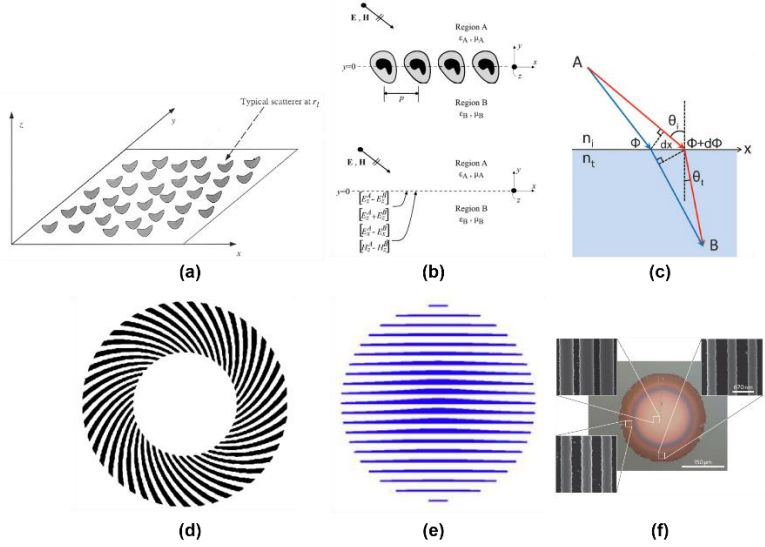

*Fig. 21. (a) "Metafilms", first coined in 2003, refer to a 2D distribution of electrically small scatterers, which are characterized by electrical and magnetic polarization densities. Adapted from [81]. (b) Generalized sheet transition conditions for characterizing metasurfaces. Adapted from [84]. (c) Generalized laws of reflection and refraction at metasurfaces with interfacial phase gradients. Adapted from [83]. (d) A 2D metasurface consisting of orientation-varying dielectric gratings [58]. (e) and (f) 2D metasurfaces consisting of dimension-varying dielectric gratings. [22, 23]*

### 4.2. Regions of operation

In order to differentiate metasurfaces from the traditional FSSs, authors in [34] extensively discussed the types of metasurfaces based on their regions of operation. According to [34], the electromagnetic behavior of periodic structures is categorized into three regions based on their resonance behaviors. Region I, is a quasistatic region where the wavelength is much larger than the periodicity of the scatterers, and the composite structure is equivalent to a uniform medium with effective permittivity and permeability. In Region II, the resonances of the surface are associated with the resonances of individual scatterers. Region III, is where the surface resonances are associated with the lattice periodicity, or higher order Floquet-Bloch modes need to be considered. Bragg scattering and GMR both belong to this region. Authors in [34] distinguished metasurfaces, which operate in Region II, from FSSs and photonic bandgap structures, which operate in the Region III. However, it is mentioned that the Region II may not occur and sometimes FSS operates in the third region. In fact, the two examples of FSSs in Fig. 20(a) and Fig. 20(b) clearly fall into Region II and Region III, respectively.

However, there are limitations for categorizing HCMs as in [34]. Regions I, II, and III were associated with frequency ranges from low to high, which is not strictly correct. In fact, both scatterer-type and lattice-type resonances can occur at similar wavelengths below period, and Regions II and III typically have a significant overlap of operation wavelengths. For example, PhC slabs belong to Region III, and rely on lattice-associated interference. The guided resonance may occur at $\Lambda/\lambda_0 = 0.31$ and above according to [85]. For another example, the FSSs in [78] and [82] both belong to Region II and rely on the resonance of the scatterers. However, they operate at $\Lambda/\lambda_0 \sim 0.46$ ($f_0 \sim 15$ GHz and $\Lambda \sim 0.92$ cm), and at $\Lambda/\lambda_0 = 0.17$ ($\lambda_0 = 3$ cm and $\Lambda = 5$ mm), respectively. Therefore, we see Region II and Region III in [34] are not necessarily distinguished by the effective wavelength. Moreover, periodic dielectric resonators may have contributions from both the scatterer-type and the lattice-type resonance, since the resonant frequency is sensitive to both the dimension and the period [37].

In the HCM community, an unambiguous approach for categorizing the metasurfaces is shown in Fig. 22. Depending on the ratio of $\lambda_0/\Lambda$, there are three regimes: diffranction, near-wavelength, and effective medium. Firstly, the effective-medium regime with $\lambda_0 \gg \Lambda$ is where the individual resonance or mutual coupling are weak, corresponding to Region I in [34]. In this regime, only the fundamental eigenmode is traveling within the coupled metastructures. Secondly, the diffraction regime with $\lambda_0 < \Lambda$ is where multiple scattering (Floquet-Bloch) orders exist, and correspond to gray areas in Fig. 13. Optical behaviors are well-explained with the traditional diffraction theory. Finally, most intriguing phenomena occur in the near-wavelength regime with $\Lambda < \lambda_0 < \lambda_{c2}$, where $\lambda_{c2}$ is the cutoff wavelength of the 1st higher order propagating eigenmode in the coupled metastructures. In this regime, a few eigenmodes travel within the coupled metastructures, with propagation constants shown in Fig. 3 and Fig. 11(b), and interference occurs at the discontinuous $z$-boundaries.

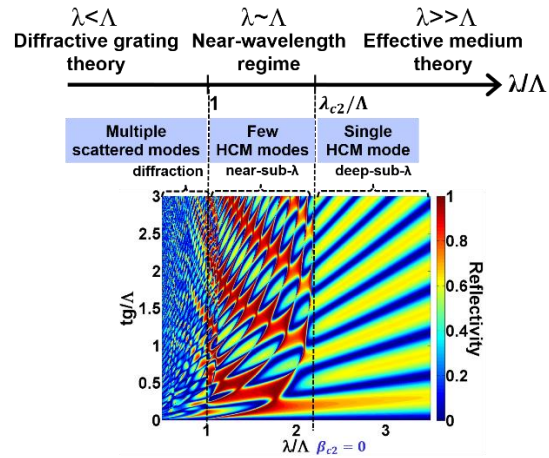

*Fig. 22. Regions of operation for metasurfaces classified based on the effective wavelength and mode interference.*

### 4.3. Properties and arrangement of building blocks

In the microwave regime, resonant metallic structures have long been used to form antenna arrays, and to provide high directivity or beam steering, which would be difficult to achieve for a single resonator. With the rapid growth of the field of plasmonics, researchers grasp the opportunity of using noble metals (e.g. gold and silver) in a shorter-wavelength (e.g. infrared and visible) regime. Large collective oscillation of electrons can be excited in these metals or at the metal-dielectric interfaces, even with subwavelength structures. Strong localized optical field allows precise manipulation of the optical behaviors. A few examples of metallic subwavelength structures as building blocks of metasurfaces are shown in Fig. 23(a)-(d). Notably, Yu et al. used V-shaped subwavelength gold antennas to realize the generalized Snell's law, as shown in Fig. 21(c) [83]. The arm length, the angle, and the orientation of the V structures combine to determine the optical phase difference between the incidence and scattered fields. Similar approach using H-shaped antennas with varying arm lengths was shown in [86]. These nano-antennas can be considered as LC resonators where the inductance and capacitance are controlled by the arm lengths and spacings (or angles). As the resonant wavelength is tuned, so is the phase response. Moreover, due to the presence of symmetric and anti-symmetric modes, these arrays typically exhibit different response to two orthogonal polarizations [83], and thus available for dual functionalities. According to the Bibinet's principle, which is proven to be also valid for optical nanoantennas [87], the reflection and transmission are to interchange for complementary structures illuminated with complementary polarizations of light. Thus, one can design complementary metallic structures (e.g. slot antennas), as shown in Fig.

23(c) [88], for phase gradient metasurfaces. Researchers in [89] used gold nanorod antennas as building blocks for their metasurface. However, the orientation rather than the dimensions of the antennas is spatially-varying, as shown in Fig. 23(d), which belongs to the Pancharatnam-Berry type of metasurface arrangement to be discussed later in this review.

Despite the ability of strong field localization, metallic structures suffer severely from high losses. Meanwhile, strong scattering resonance is not unique to metallic or plasmonic structures. Dielectric nanostructures were demonstrated with strong optical resonance [17, 36], and are capable of being building blocks for metasurfaces. As shown in Fig. 23(e)-(h), possible dielectric nanostructures include dimension- or orientation-varying nanoposts [30], orientation-varying nanobeams [90], double rectangular resonators with varying spacings and widths [91], and rectangular mesh [29].

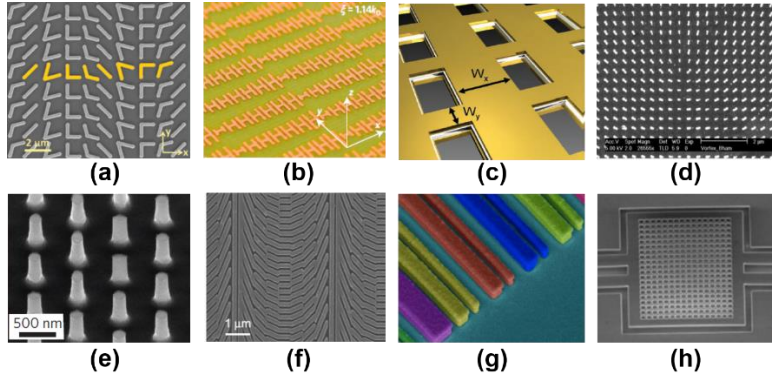

*Fig. 23. Examples of building blocks on metasurfaces. Metallic nanostructures: (a) V-antennas [83], (b) H-antennas [86], (c) slot antennas [88], and (d) nanorod antennas [89]. Dielectric nanostructures: (e) nanoposts [30], (f) nanobeams [90], (g) double rectangular resonators [91], and (h) dielectric mesh [29].*

High index contrast building blocks have strong localized effect of altering optical phase front. Arrays of these building blocks are used to form metasurfaces. Due to the spatial dependent phase change, metasurfaces can behave as phase plates. There are two major types of arrangement of building blocks for metasurfaces, as shown in Fig. 24, which we refer as dimension-varying and orientation-varying metasurfaces. For the dimension-varying metasurfaces, the sizes or shapes of the HCMs are spatial-dependent, and determine the local optical phase. This type of arrangement was referred as "resonance tuning" in a recent review paper by Genevet et al. [33]. Here, we avoid the use of "resonance" for this type of metasurface, because HCMs may not need to be at resonance in order to tuning the output phase. With plasmonic antenna arrays (e.g. [83]), the shapes of the HCMs affects the resonance frequencies, similar to LC resonators. With dielectric HCMs (e.g. [22]), the sizes and shapes affects the effective indices for modes propagating in these coupled structures, thus affecting the optical phase. HCMs do not necessarily operate at resonance conditions indicated by the dotted lines in Fig. 12(a). The dimension-varying metasurfaces does not include the arrangement with varying only the orientation of the HCMs, which belongs to the second type in Fig. 24. This second type is also known to utilize the Pancharatnam-Berry (PB) phase [57, 92] or geometric phase, where the term "geometric" should not be confused with the dimension-varying designs. With the PB phase design, the local optical phase is completely controlled by the orientation of the HCMs with uniform size and shape.

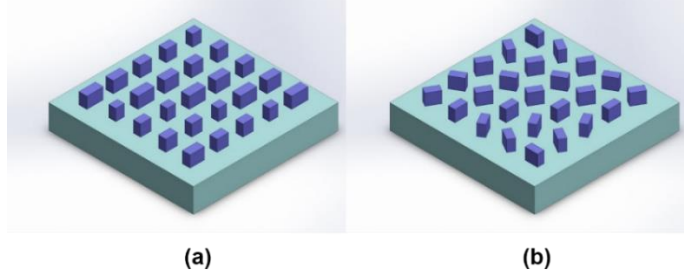

*Fig. 24. Two types of arrangement for building blocks on metasurfaces: (a) dimension-varying and (b) orientation-varying metastructures. Adapted from [31].*

With the dimension-varying approach, one needs we need to optimize a range of geometry variation which provides high transmission (or reflection), and a full $2\pi$ phase variation. Here we use 2D RCWA to generate the contour plots of transmission intensity and phase as functions of the width and length of the rectangular HCMs, as shown in Fig. 25(a)-(b). One optimized dimension-varying range for $\lambda = 633$ nm operation is shown as the black dashed lines. From the transmission and phase spectra in Fig. 25(c), we confirm as a full coverage of $2\pi$ transmission phase and high power efficiency (>95%). It should be noted the design is based on a single wavelength. Since the effective geometry is strongly dependent on wavelength, multiwavelength or broadband operation with this approach is challenging. Researcher have used composite HCMs in each unit cell, such as double rectangular resonators [91] and "meta-atoms" (circular posts with two different diameters) [93], to overcome the bandwidth limitation.

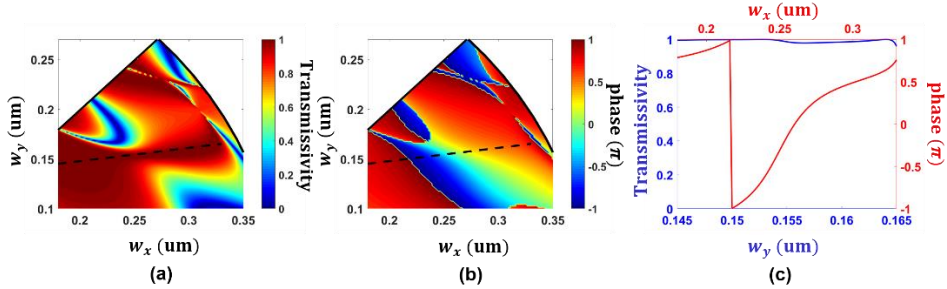

*Fig. 25. Design strategy for dimension-varying metasurfaces. (a) Transmissivity contour as a function of the lateral dimensions ($w_x$ and $w_y$) of the rectangular building blocks. (b) Transmission phase $\phi_{xx}$ contour as a function of $w_x$ and $w_y$. The dashed lines in (a) and (b) show the varying ranges of $w_x$ and $w_y$ that provides high-transmission (>95%) and covers a full $2\pi$ transmission phase range. (c) Transmissivity (blue) and phase (red) variations as functions of the lateral dimensions. Both the incidence and the calculated transmission are x-polarized at $\lambda = 633$ nm. The HCM parameters are $\Lambda = 510$ nm, $t_g = 240$ nm, $n_a = 1$, and $n_b = 3.85$.*

To understand the orientation-varying HCMs, the Jones matrix describing the transmission of anisotropic scatterers is commonly used [94],

$$W_0 = \begin{pmatrix} t_{xx} & 0 \\ 0 & t_{yy} \end{pmatrix} \quad (10)$$

Here, $t_{xx}$ is the transmission coefficient for *x*-polarized output under *x*-polarized input, and $t_{yy}$ is the transmission coefficient for *y*-polarized output under *y*-polarized input. The *x*-polarization, *y*-polarization, left-hand circular polarization (LHCP), and right-hand circular polarization (RHCP) are represented with state vectors $(1,0)^T$, $(0,1)^T$, $(1,i)^T$, and $(1,-i)^T$, respectively. When the scatterer is rotated around the optical axis by $\psi$, then the Jones matrix becomes

$$W(\psi) = \begin{pmatrix} \cos\psi & -\sin\psi \\ \sin\psi & \cos\psi \end{pmatrix} W_0 \begin{pmatrix} \cos\psi & \sin\psi \\ \sin\psi & \cos\psi \end{pmatrix} \tag{11}$$

If the incidence wave is circularly polarized, the transmission will contain two circularly polarized components, one having the same, other having the opposite handedness. The transmitted wave can be written as

$$E_t^{L/R} = W(\psi)E_i^{L/R} = \frac{t_{xx}+t_{yy}}{2}E_i^{L/R} + \frac{t_{xx}-t_{yy}}{2}\exp(\mp 2i\psi)E_i^{R/L} \tag{12}$$

Here, $L$ and $R$ indicate LHCP and RHCP, and there is conversion of handedness if the scatterers are anisotropic ($t_{xx} \neq t_{yy}$). The cross-polarized transmission experiences an additional phase change $\exp(\mp 2i\psi)$, which is determined solely by the rotation angle. Therefore, we only need to find one optimal geometry of the HCMs, and a full $2\pi$ phase range is already covered by 0-to-180-degree rotation.

In order to maximize the conversion from LHCP to RHCP, or vice versa, we desire $t_{xx}$ and $t_{yy}$ to have equally high magnitudes but completely out-of-phase. Fig. 26(a) shows the contour plot of the cross-polarized transmissivity $|t_{cross}|^2 = |t_{xx} - t_{yy}|^2/4$ as a function of the width and length of the rectangular HCMs. The HCM period is 1.1 μm, designed for $\lambda = 1550$ nm, and our chosen lateral design is indicated as the star in Fig. 26(a). The intensity and phase spectra for $x$- and $y$-polarization are shown in Fig. 26(b). We observe broadband high transmission for both $t_{xx}$ and $t_{yy}$, and their phase difference is $\pi$ at 1550 nm. As a result, the cross-polarized transmission, which determines the power efficiency, exhibits a 3dB-bandwidth as large as 12.9%. Moreover, even when operated away from target wavelength, the phase distribution across the metasurface still retains. Only the efficiency will drop but the optical behaviors will not be distorted.

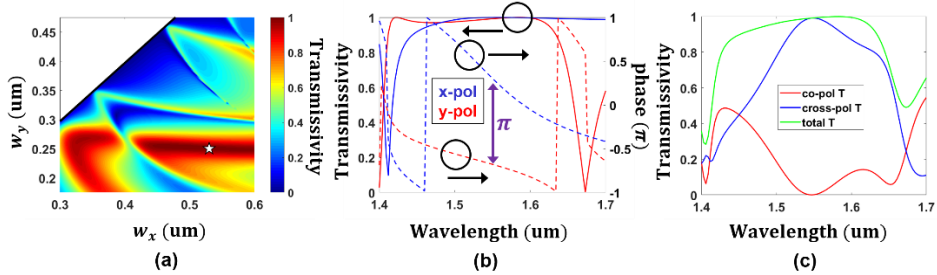

*Fig. 26. Design strategy for orientation-varying metasurfaces. (a) Cross-polarized transmissivity $|t_{cross}|^2$ contour at $\lambda = 1550\ nm$ as a function of the lateral dimensions of the rectangular building blocks. The white star indicates the chosen design. (b) Transmissivity spectra (solid) and phase (dashed) spectra for x-polarization (blue) and y-polarization (red). At 1550 nm, the difference between $\phi_{xx}$ and $\phi_{yy}$ is π. (c) Co-polarized (red), cross-polarized (blue), and total transmissivity spectra (green) showing the broadband operation of orientation-varying metasurfaces. The 3dB-bandwidth for cross polarization is $\Delta\lambda/\lambda_0 = 12.9\%$. The HCM parameters are $\Lambda = 1.1\ \mu m$, $t_g = 1\ \mu m$, $w_x = 540\ nm$, $w_y = 250\ nm$, $n_a = 1$, and $n_b = 3.48$.*

### 4.4. Applications of phase metasurfaces

As mentioned in 3.1, periodic subwavelength structures (ZGCs) have been used alter the optical phase locally. What is most amazing about the HCMs, on the other hand, is the fact that the optical phase can be altered locally by one single subwavelength structure, as discussed in 4.3. Hence, many devices can be designed with very high spatial resolution, such as planar beam deflectors, lenses, axicons, vortex plates and so on [32, 38].

The metasurfaces which mainly provide spatial phase modulation, which we refer as phase metasurfaces, are categorized by their functionalities as following:

- Anomalous reflection and refraction: scattered beam is deflected anomalously at designed angles [9, 27, 82, 83, 86, 95-97]
- Lenses and collimators: beam propagation properties (e.g. beam waists and divergence angle) are engineered [22, 23, 30, 59, 90, 91, 98]
- Beam shaping: the spatial beam profiles (e.g. Gaussian or vortex) are modified, [48, 51, 89, 99, 100]
- Polarization control: stabilization or conversion of the beam polarization [57, 58, 101]
- Holography: phase or amplitude hologram generation [30, 88, 102-104]

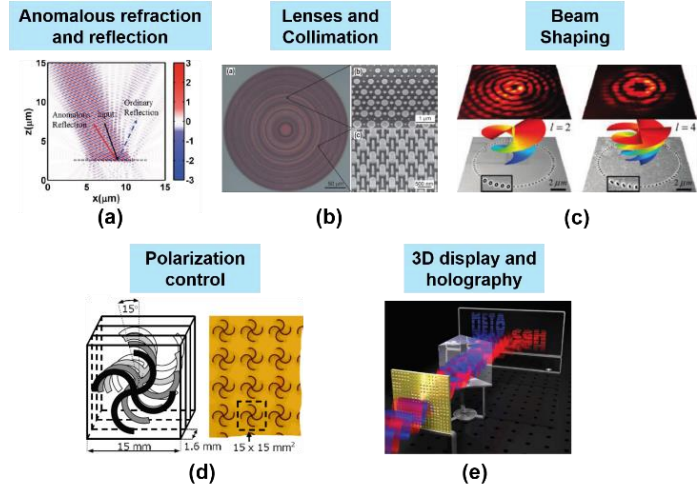

*Fig. 27. Major applications for metasurfaces. (a) Reflection and refraction of light at anomalous angles. (b) Focusing, diverging, and collimation of light. (c) Shaping of the spatial beam profiles. E.g. Generation of vortex beams. (d) Polarization control or conversion of light. (e) 3D display and holography. Figures in (a)-(e) are adapted from [27], [93], [99], [105], and [88]. respectively.*

In this review, we will investigate particularly metasurfaces utilized as holograms [30, 31, 102, 104] where optical near-field is rapidly varying across the 2D surface. Here, we discuss two algorithms to construct holograms using 2D HCMs. Based on either the dimension-varying or the orientation-varying 2D metastructures, the output optical phase distribution across the metasurface can be engineered while maintaining nearly uniform intensity distribution. Therefore, the Gerchbery-Saxton (GS) algorithm is ideal for producing the phase holograms. The iteration process of the algorithm is shown in Fig. 28(a). The target image to be reconstructed is chosen as Fig. 28(b). After the algorithm converges, a desired near-field phase distribution with uniform intensity is obtained. The near-field is pixelated at 300-by-300, with the pixel spacing being the period (1.1 um) of the 2D HCM. Each pixel contains one rectangular HCM optimized through the orientation-varying approach in Fig. 24(b) and Fig. 26. The design is verified by the FDTD simulation using the LHCP incidence light at $\lambda = 1550$ nm. The metasurface output near-field is monitored at the same wavelength, and decomposed into both LHCP and RHCP components. The LHCP components is filtered and the Fourier transform is performed on the RHCP component to produce the far-field. The simulated reconstruction is shown in Fig. 28(c) [31].

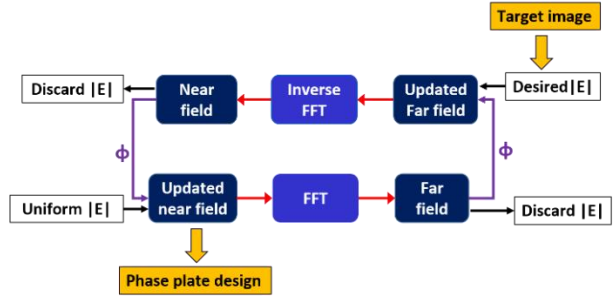

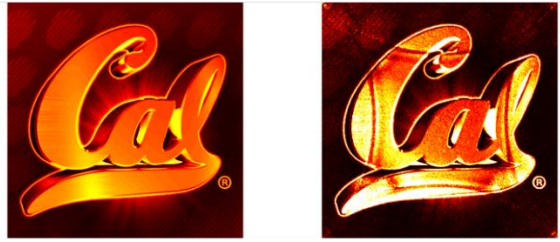

*Fig. 28. (a) Gerchberg-Saxton (GS) algorithm for generating the near-field phase distribution and the phase hologram from a target image. (b) Target image with pseudo-color representation of the image brightness. (c) FDTD-simulation of the reconstructed image from the transmitted near-field through an orientation-varying metasurface. The incidence is LHCP at $\lambda = 1550$ nm and the RHCP transmitted field is captured. Field intensity is represented via pseudo coloration. The hologram is generated by the GS algorithm and the HCM design is the same as in Fig. 26.*

Although the GS algorithm ensures desired far-field patterns to be generated, the designed phase plates are not strictly 3D hologram due to the lack of depth information and viewing angle dependence. Here we use the point-source propagation algorithm [31, 104, 106] to capture both the amplitude (brightness) and phase (depth) information. The hologram construction process is shown in Fig. 29(a). Each point $(x_i, y_i, z_i)$ on the object is considered as a point source, which generates a complex near-field at $(x_j, y_j)$ on the hologram as

$$E(x_j, y_j) = \sum_i \frac{A_i}{r_{ij}} \exp\left(i \frac{2\pi}{\lambda} r_{ij} + i\phi_i\right) \tag{13}$$

Here, $r_{ij} = \sqrt{(x_i - x_j)^2 + (y_i - y_j)^2 + z_i^2}$ is the distance, $A_i$ is the amplitude, and $\phi_i$ is a random phase which reduces large amplitude variation on the hologram and behaves as an optical diffuser. We first draw a 3D object (a DNA in this case) in a computer-aided-design (CAD) software, and extract the amplitude and distance information of each point on the object. Using Eq. (**13**) we produce a hologram plate pixelated with a period of 530 nm, which is the same as the HCM period as in Fig. 26 for $\lambda = 633$ nm. The hologram plate is located at $z = 0$, and the depth of the object spans from $z = -4.3$ mm to $z = -2.8$ mm. The reconstruction of the 3D object is shown in Fig. 29(b). The HCM is designed with the orientation-varying approach. We observe the transmitted RHCP component when the incidence is a LHCP wave. The RHCP component should reconstruct a virtual object within $(-4.3 \text{ mm}, -2.8 \text{ mm})$ if designed properly. To confirm that the image is 3D, we use a lens with a focal length of 2.795 mm located at $z = 2.8$ mm to project the image onto a movable object plane. From Fig. 29(c) to Fig. 29(e), we move the object plane farther away from the lens. The projected images show the focused part of the object moves closer to the lens, as indicated by the arrows. The calculated object distances for Fig. 29(c) to Fig. 29(e) are 6.848 mm, 6.554 mm, and 6.3 mm, which correspond to $z = -4.048$ mm, $z = -3.754$ mm, and $z = -3.5$ mm, respectively. Indeed, a virtual 3D object is reconstructed at our designed location within $(-4.3 \text{ mm}, -2.8 \text{ mm})$.

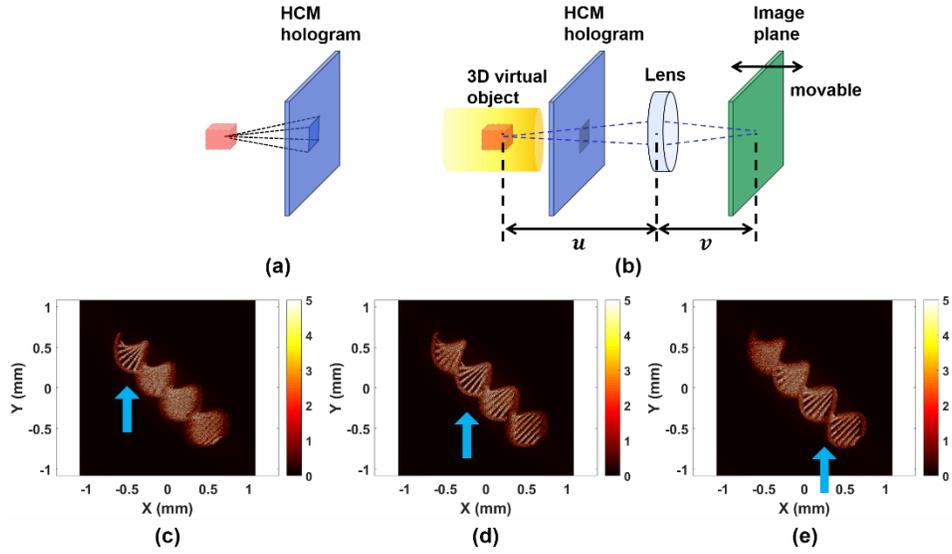

*Fig. 29. (a) Generation of hologram by the point-source algorithm. (b) Reconstruction of the 3D object with the virtual object and image distances at u and v, respectively. Simulation of the transmitted near-field through a metasurface being projected onto an image plane at distances of (c) $v = 4.722$ mm, (d) $v = 4.873$ mm, and (e) $v = 5$ mm. The focal length is $f = 2.795$ mm. The distance between the metasurface and the lens is 2.8 mm. The virtual object distances are (c) $u = 6.848$ mm, (d) $u = 6.554$ mm, and (e) $u = 6.3$ mm, with blue arrows indicating the focused parts of the virtual object.*

### 4.5. Novel amplitude metasurfaces

Metasurfaces that provide spatial-dependent or angle-dependent amplitude modulation are referred as amplitude metasurfaces. Many intriguing applications have been demonstrated in recent years, among which a notable example is the glass-free 3D display using the multi-direction backlight technology by Fattal et al. [103]. Instead of generating a hologram phase plate, the authors control individually the scattering amplitudes in 64 different viewing angles using the commercial liquid-crystal technology. Each unit contains a set of static directional gratings to scatter in-plane polarized light into various angles through first-order diffraction. Multi-color operation is achieved by including RGB grating pixels in each cell. The active display (30 frames per second) is achieved using the liquid-crystal display shutter plane to modulate the angle-dependent scattering amplitude.

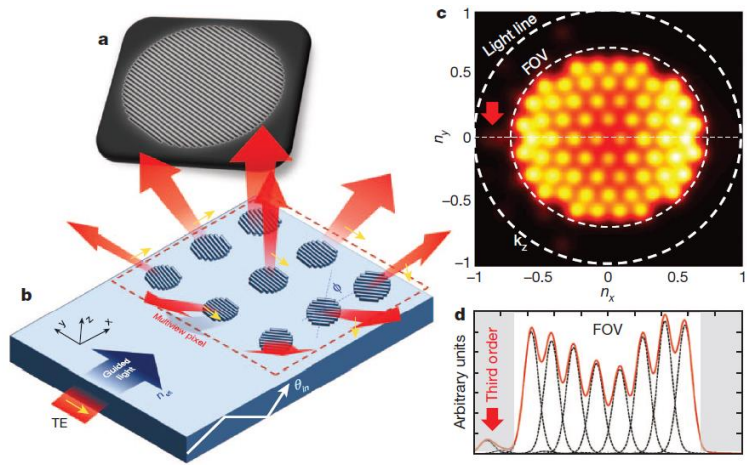

*Fig. 30. SEM, schematic and full-wave simulation of the multi-directional backlight technology for glass-free 3D display. Each unit cell contains a group of directional gratings to cover the 64 views within the field-of-view. Input polarized and guided light is scattered from in-plane direction into the viewing angle using the first-order diffraction. Adapted from [103].*

Another notable example is the beam shaping of single-mode VCSELs using HCMs [48]. While high reflectivity is maintained for single-mode operation, which is proved by the emission spectra with side-mode suppression above 45 dB, the transmission far-field pattern can be tailored through the lateral design of HCMs. The emission from the bottom DBR mirror is a Gaussian beam, while the emission from top HCM mirror can be single-lobed, double-lobed, or donut beam. The underlying mechanism is the angle-dependent transmission amplitude from the HCM. Such angular dependence of the transmitted near-field generates various distributions of far-field intensities after Fourier transform. Excellent agreement between theory and experiments has been demonstrated [48].

Actively-controlled HCMs can be used as color filters or spatial light modulators. Complex-shape 2D HCMs can be made tunable by incorporating MEMS. Here we present a novel design of HCM which consists a wave-shape grating bar (s-bar) and a line grating bar (l-bar) as the fundamental element with periodic repetition in the lateral direction, illustrated in Fig. 31(a). The surface-normal optical reflection can be greatly changed with a minute change in the spacing between the s- and l- bars. Fig. 31(b) shows the displacement of the s-bars when the tuning is on. Because the sharp bending of the s-bar creates significant light leaking in the in-plane direction (*x*-*y* plane), strong mode coupling between the s-bar and l-bar can be achieved. Fig. 32(a) and Fig. 32(b) show the reflection spectra for our designed 2D HCM for *x*- and *y*-polarized incidence, respectively. For *x*- polarized light, the tunable 2D HCM behaves as a light switch, with high transmissivity ($R < 0.15\%$ at 633 nm) without displacement ("off") and extremely low transmissivity ($R > 99.8\%$ at 633 nm) with shifting ("on") the s-bars for 50 nm, with a 3-dB bandwidth of 34.4 nm. Fig. 32(c) and Fig. 32(d) show the magnetic field profiles normalized by the incidence amplitude across the middle $z$-plane of the HCM. With the tuning off, the intensity is weak and majority of the incidence light transmits through the HCM. With the tuning on, the intensity is high, and the internal resonance causes the incidence light to be mostly reflected. The coupling between the two bars is strongly affected by their distance when the electric field is normal to the bars, similar to a capacitor. Therefore, such broadband intensity modulation is achievable with only 50-nm displacement! Such displacement can be mechanically actuated by the MEMS structure with a high mechanical speed and low actuation voltage.

The same device operating under $y$-polarization does not exhibit such behavior. The device remains as a broadband reflector with or without displacement. As shown in Fig. 32(b), the reflectivity at 633 nm remains above 99.6%. Fig. 32(e) and Fig. 32(f) confirms that the field profiles exhibit only small changes upon the tuning. However, the device exhibits a narrow pass bandwidth with high reflectivity for the rest of the spectrum. The displacement can effectively tune the pass wavelength which correspondingly control the color of the device.

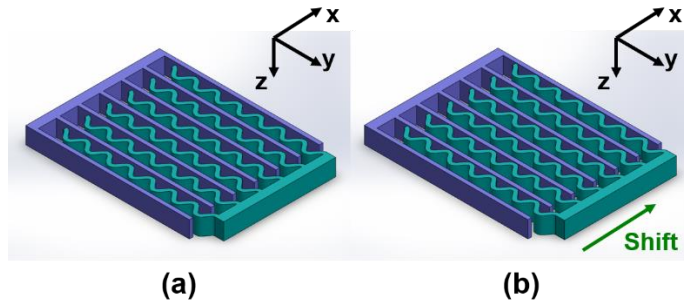

*Fig. 31. Schematics of laterally-tunable 2D metasurfaces. Each period comprises a wave-shape grating bar (w-bar) and a straight grating bar (s-bar). (a) No displacement of w-bars when tuning is off. (b) Displacement of w-bars in $+x$-direction when tuning is on.*

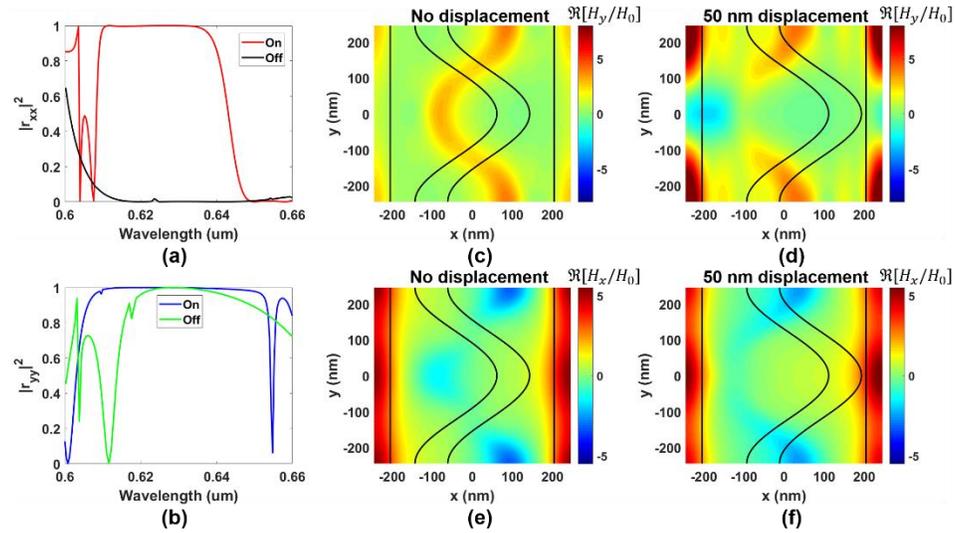

*Fig. 32. (a) Reflection spectra for x-polarized light with the tuning on (red) and off (blue). (b) Reflection spectra for y-polarized light with the tuning on (blue) and off (green). Tuning on and off*

*correspond to zero and 50-nm displacement, respectively. (c) and (d) correspond to the field profiles $\Re[H_y/H_0]$ across the middle z-plane ($z = t_g/2$) under $E_x/H_y$-polarized incidence. (e) and (f) correspond to the field profiles $\Re[H_x/H_0]$ across the middle z-plane ($z = t_g/2$) under $E_y/H_x$-polarized incidence. The incidence wavelength is $\lambda = 633\ nm$. Here, $\Lambda_x = \Lambda_y = 490\ nm$ and $t_g = 380\ nm$. Widths of both bars are 81.7 nm. The index of the bars is 3.48.*

## 5. Waveguides and Couplers

Waveguides and couplers are perhaps the most fundamental elements in integrated optoelectronics. The fundamental properties of waveguides are to confine light in one or two dimensions and guide it in the third dimension. As mentioned in earlier chapters, both HCMs and PhCs are good reflectors for light, rendering them excellent building blocks to confine lights and thus construct waveguides. The HCMs and PhCs can be engineered to have a designated response during their interaction with light; this leads to a whole realm design of light-couplers, such as couplers from free-space to waveguides, optical splitter, and optical switch.

### 5.1 HCM waveguides and couplers

HCM can offer a superiorly broadband high reflection for light incident from a low refractive index media, making it an ideal guiding reflector in chip-scale hollow-core waveguides (HCWs). HCWs guide light through a low-index core surrounded by high-index cladding layers. They are desirable for compact sensors and gas-based nonlinear optics because of the increased lengths for light-matter interaction.

Light propagation direction can be either parallel or perpendicular to the grating bars. One-dimensional waveguide can be achieved with two parallel planes of highly reflective HCGs. The optical beam is guided by zig-zag reflections from the guiding walls – an intuitive ray optics model for HCW. For practical application, lateral confinement is critical. Previously, we demonstrated a lateral confinement scheme by varying the HCG dimension laterally such that the effective index of the core is higher than that of the cladding [19]. This is a novel use of the effective index method in HCW. Fig. 33 shows the HCW schematic and the experimentally measured mode profile. The HCW is constructed by two planar, parallel, silicon-on-insulator wafers with HCGs running in parallel with the light propagation direction. The waveguide is designed to have a height of 9 μm, with an effective refractive index difference of $4\times10^{-4}$ between the core and cladding. In between the core and cladding, a transition region with chirped HCG dimensions is introduced to reduce the loss. The net propagation loss and coupling loss are extracted from waveguides with different lengths. The extracted propagation loss for a 9-μm high waveguide is 0.37 dB/cm at 1535 nm, the lowest reported loss for an HCW with such a small core.

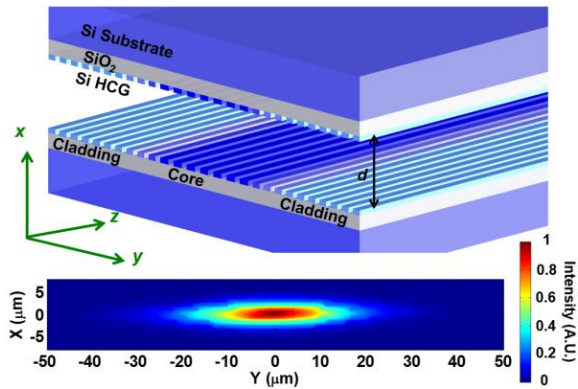

*Fig. 33. Schematic of a 2D HCG HCW. In the lateral direction, the core and cladding are defined by different HCG parameters to provide lateral confinement. In between the core and cladding region is the chirped HCG to provide a graded index profile. The waveguide height is d. [19]*

Given the high reflectivity of the HCG, two HCWs can run in parallel and share the same HCG sidewall with little cross-talk. This enables compact coupler, splitter and switch designs based on multi-mode interference (MMI) [20] (Fig. 34) In an MMI design, multiple modes are excited at the transition of a single-mode input waveguide to a multi-mode waveguide. Due to the difference in their propagation constants, these guided modes interfere along the length of the waveguide and form single/multiple images of the incident field. One specialty of HCG HCW is that the two waveguides can be literally adjacent to each other thanks to the high reflectivity of the HCG. As a result, the width of the MMI region $w_{MMI}$ can be as small as $2d$, where $d$ is the width of each waveguide. This is in contrast with the conventional dielectric waveguide, where $w_{MMI}$ is typically $3d\sim4d$, as the two waveguides are typically separated by at least $d\sim2d$ so as to avoid undesirable coupling between various components. As the length of the MMI region scales with $w_{MMI}^2$, the splitters and couplers built on the HCW platform can be relatively compact.

The rich property of HCG enables its many interesting applications based on the HCW platform. One of them is the optical switch, shown in Fig. 34(d)- Fig. 34 (f). This HCG layer has two-fold of functionalities. It completely isolates the two HCWs, serving as a highly reflective mirror at OFF state; by applying a very small change of the refractive index of this HCG layer (-7×10$^{-3}$), either by optical induced carrier or electrically injected carrier, the HCG becomes transparent and allows light switching between the two HCWs [20]. The rich property of HCG enables its many interesting applications based on the HCW platform. One of them is the optical switch, shown in Fig. 34(d)- Fig. 34 (f). This HCG layer has two-fold of functionalities. It completely isolates the two HCWs, serving as a highly reflective mirror at OFF state; by applying a very small change of the refractive index of this HCG layer (-7×10$^{-3}$), either by optical induced carrier or electrically injected carrier, the HCG becomes transparent and allows light switching between the two HCWs applications [107].

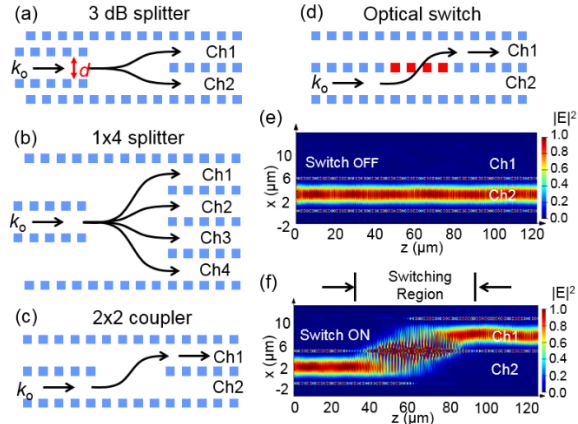

*Fig. 34. Compact optical components based on HCG hollow-core waveguide. (a) 3dB MMI based splitter (b) 1×4 splitter (c) 2×2 MMI coupler (d) An optical switch where the output can be switched from CH1 to CH2 by changing the refractive index of HCG in the MMI region. (e) and (f) shows the FDTD simulation of the switch. The switching length is ~60 μm, and the refractive index change is -7x10$^{-3}$ at this switching region. [20]*

HCG can be designed to operate in other regimes to serve as a coupler for light from free space in surface-normal direction to couple into an in-plane dielectric waveguide. One straightforward example is the second order grating, where the incident light picks up a wave-vector provided by the grating, and gets coupled into a guided mode in the waveguide underneath the grating. Conventional second order gratings have limited efficiency, often significantly below 25% in each in-plane directions. Reflection DBRs and slanted grating can be employed to enhance the efficiency [108, 109]. Motivated by the resonance effect of HCG, we recently reported using HCG in a vertical-to-in-plane (VIP) coupler [21], as shown in Fig. 35(a), which could achieve a close-to-100% coupling efficiency between free-space wave and in-plane waveguide mode. In this design, the HCG is placed on top of the waveguide, with a low refractive index material in between to create the operation environment for HCG. Light comes from the surface normal direction to the HCG and excites the supermodes as combinations of waveguide eigenmodes. If one of these modes is at resonance, light could get stored in the HCG. This greatly increases the photon lifetime within the waveguide. Light can thus be coupled into the dielectric waveguide before escaping to the zeroth-order reflection or transmission. In other words, it is the interaction between the HCG resonance mode and the propagation mode of the dielectric waveguide that leads to the close-to-100% coupling efficiency, as shown in Fig. 35(b).

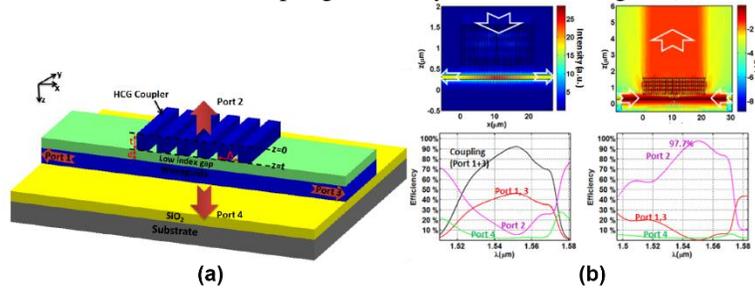

Fig. 35. (a) Schematic of a HCM vertical coupler. (b) Field profiles of vertical-to-in-plane coupling (top left) and in-plane-to-vertical coupling (top right) with corresponding coupling efficiency spectra (bottom left and bottom right). Adapted from [21]

Another novel application of HCM VIP coupler is to couple surface-normal propagating plane wave with the surface plasmon polariton (SPP) propagation mode in a plasmonic-dielectric waveguide, as shown in Fig. 36(a). This device can be attractive for bio-sensing applications. The TM-polarized normal-incident light $\mathbf{E}_{inc} = \hat{x} E_x e^{i n_{super} k_0 z}$ enters the superstrate. The high-index HCMs are embedded in the low-index spacer. A 50-nm gold layer is coated on the spacer and the substrate is typically the low-index sample under test. Strong resonant coupling occurs when the in-plane wave number of the +1-diffraction order $k_x = 2\pi/\Lambda$ from the grating matches the surface plasmon polariton (SPP) wave number $k_x(\omega) = \frac{\omega}{c}\sqrt{\frac{\varepsilon_d \varepsilon'_m(\omega)}{\varepsilon_d + \varepsilon'_m(\omega)}}$, as shown in Fig. 36(b). Here, $\varepsilon'_m$ is the real part of the metal permittivity and $\varepsilon_d$ is the dielectric permittivity. The SPP can occur at both the spacer-gold and the substrate-gold interfaces. The $k_x$ dispersion curves are shown as the blue and green curves, respectively, and the matching conditions are satisfied at their intersections with the red line of $k_x = 2\pi/\Lambda$, which are at $\lambda = 854$ nm, and $\lambda = 759$ nm, respectively. Most strikingly, the resonant coupling of incidence to the SPP mode occurs at the same wavelength, as shown in Fig. 36(b), and is independent of the spacer thickness! The cutoff wavelength of the $\pm 1$-diffraction order in the superstrate is indicated by the black line. Fig. 36(d) and Fig. 36(e) show maximum field intensity enhancement $|E_z/E_{inc}|^2$ of 296 and 297 at the substrate-gold interface with spacer thicknesses of 2260 nm and 400 nm, respectively. Fig. 36(f) shows field intensity enhancement of 173 at the spacer-gold interface with a spacer thickness of 500 nm.

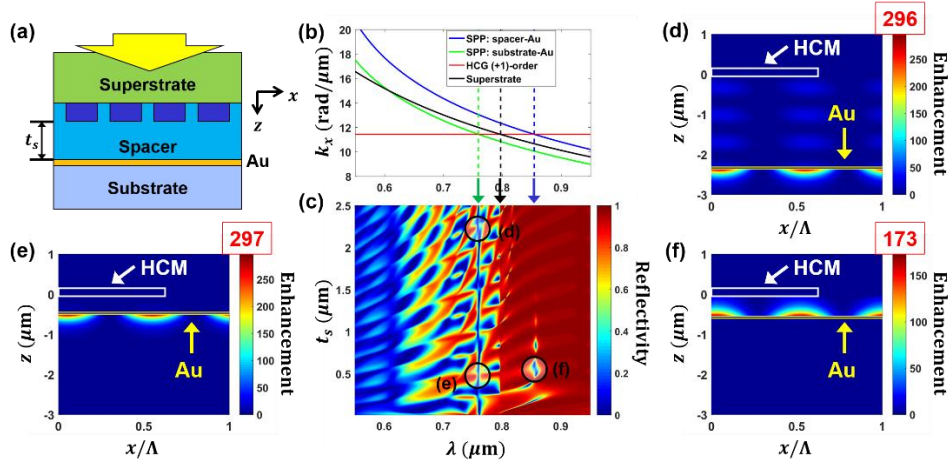

*Fig. 36. (a) Schematic of a HCM-plasmonic sensor under normal x-polarized (TM) incidence. The refractive indices for the superstrate, substrate, spacer, and HCM are $n_{super} = 1.45$, $n_{sub} = 1.33$, and $n_{spacer} = 1.5$, $n_{HCM} = 3.84$, respectively. The HCM dimensions are $\Lambda = 549.35\ nm$, $\eta = 62.4\%$, and $t_g = 200\ nm$. The gold layer thickness is 50 nm, and the permittivity is interpolated from the CRC Handbook (b) In-plane propagation constants $k_x$ as functions of the wavelength for SPP at spacer-Au interface (blue), SPP at substrate-Au interface (green), and the (+1)-diffraction order from the HCM (red) with $k_x = 2\pi/\Lambda$. The total propagation constant $k = n_{super}\omega/c$ in the superstrate is shown as the black line. (c) RCWA-calculated reflectivity contour as a function of the wavelength and the spacer thickness $t_s$. Strong SPP resonance is expected along the coupling lines indicated by the green and blue arrows. Three designs indicated by the black circles are analyzed for field enhancement. The FDTD-calculated intensity distributions $|E_z/E_{inc}|^2$ for (d) $t_s = 2260\ nm$, $\lambda = 762.66\ nm$, (e) $t_s = 400\ nm$, $\lambda = 765.46\ nm$, (f) $t_s = 500\ nm$, $\lambda = 848.39\ nm$ show enhancement of 296, 297, and 173, respectively.*

### 5.2 PhC waveguides and couplers

Just like HCM, the rich properties of PhC also enable its wide applications in waveguiding, optical couplers, and optical switches. A typical PhC waveguide is composed of a line defect of missing air holes in a PhC slab, which is high-index thin film with a two-dimensional array of air holes surrounded by air cladding, as shown in Fig. 37. Light propagates along the line defect, confined by total internal reflection in the vertical direction and Bragg reflection in the lateral direction, where the PhC operates in its stop band regime. This is largely different from the HCM hollow-core waveguide, as the PhC waveguide mode operates below the light line (propagation constant larger than that in the surrounding free-space low-index material), whereas the HCM hollow-core waveguide operates right at the opposite regime.

One particular application of PhC waveguide is slow light [7, 110]. Due to the zone-folding of the guided-mode band, the ω-k diagram appears to be a flat line at the cut-off point called the band edge. The group index diverges to infinity, leading to slow light (or stopped light) effect. At this band edge, the second-order dispersion (group-velocity dispersion, GVD) can become very large, which could strongly disperse the pulse and thus limit the capacity of the PhC waveguide. By designing the geometry of the PhC, and thus the photonic band structure, the GVD can be suppressed. Fig. 37(c)- Fig. 37(d) show one example where the diameter of the innermost air holes adjacent to the line defect is reduced compared to the others. By tuning the air hole diameters, a nearly flat spectrum of the group index can be obtained [111].

The slow light effect effectively increase the optical path inside the waveguide while maintaining a small physical footprint. This leads to compact directional coupler or optical switch designs based on PhC waveguide. Fig. 38(a) shows one such example. The geometry of the holes in the PhC waveguide are tuned to modify the mode dispersion properties and hence the coupling characteristics. By changing the temperature of the PhC waveguide (leading to a change of refractive index of $4.2\times10^{-3}$), light can switch between these two waveguides over a switching length of 5.2 μm [112].

Another realization of optical switch is through a PhC nanocavity, as shown in Fig. 38(b)-Fig. 38(c), which shares conceptual similarity with the HCM switch shown in Fig. 34(d)-Fig. 34(f). In this example, a point defect creates a PhC cavity in between the input and output PhC waveguides. This cavity leads to a low transmission of the signal light from input to output. When a pump light is injected to the input waveguide, free carriers are generated, which induces a wavelength shift in the resonant transmission spectrum of the cavity. This allowed the simultaneously injected signal light to couple into the output waveguide. Switching energies with a contrast of 3 dB and 10 dB of 0.42 fJ and 0.66 fJ respectively, were obtained [113].

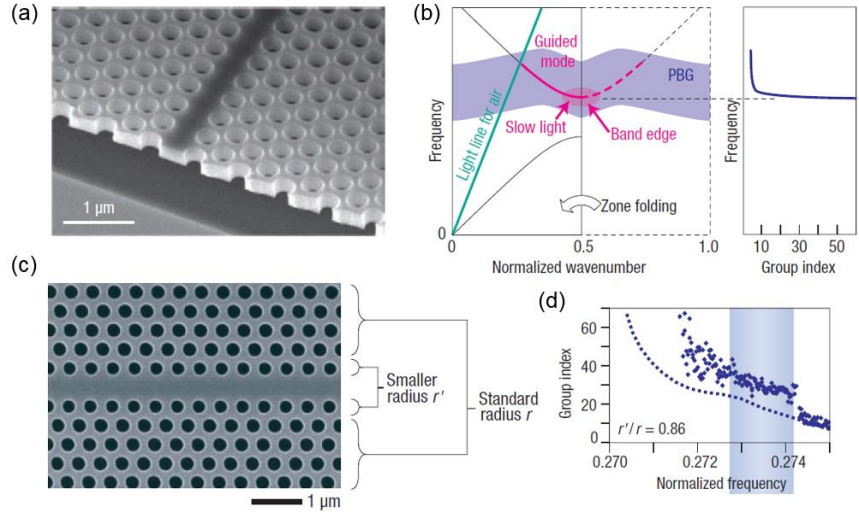

*Fig. 37. Slow light in PhC waveguide. (a) Scanning electron microscope image of a PhC waveguide. (b) ω-k diagram and the group index spectrum of a silicon PhC waveguide. (c) Scanning electron microscope image of a PhC waveguide designed to have zero GVD slow light. (d) The group index of the zero GVD PhC waveguide. (a) and (b) reprinted with permission from [7], (c) and (d) reprinted and adapted with permission from [7, 111].*

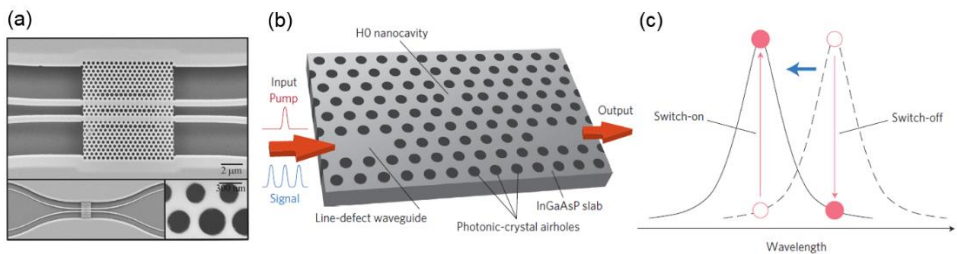

*Fig. 38. Optical switch based on PhC waveguide. (a) Scanning electron microscope image of a directional coupler switch. (b) Schematic of an all-optical switch based on PhC waveguide and an H0 cavity. (c) Operating principle of all-optical switching. A pump carrier induces a wavelength shift in the resonant transmission spectrum of the signal. (a) reprinted with permission from [112]; (b) and (c) reprinted with permission from [113].*

While the discussions so far have been focused on PhC waveguides hosted by a two-dimensional membrane, which are relevant to planar integrated photonics, waveguide surrounded by three-dimensional PhC existed, and they are more commonly termed as photonic crystal fibers (PCF). A whole assortment of PCF have been developed, ranging from solid-core to hollow-core [114], as shown in Fig. 39. Thanks to its air-hole nature, PCF has been widely applied to gas-based nonlinear optics such as ultralow-threshold stimulated Raman scattering in molecular gases and high harmonic generation. PCF with extremely small solid cores and high air-filling fractions could yield very high optical intensities in the core. Together with a delicate control of the chromatic dispersion, it is an ideal candidate for ultrahigh nonlinearity applications, such as supercontinuum generations. Other application of PCF includes gas sensing, high power laser delivery, atom and particle guidance, etc.

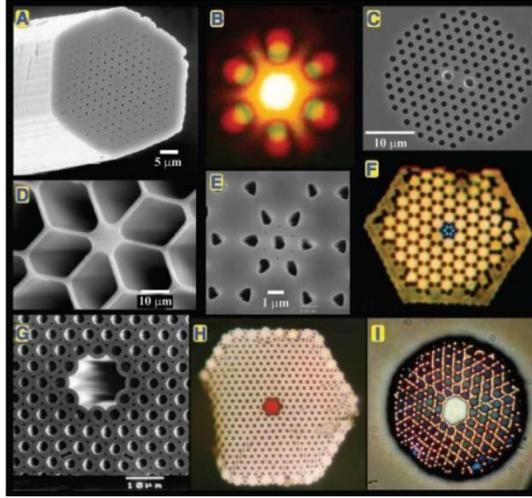

*Fig. 39. An assortment of optical and scanning electron microscope imaging of PCF structures. (A) Single-mode solid core PCF. (B) Far-field pattern produced by (A) when excited by red and green laser light. (C) Birefringent PCF. (D) Small core PCF. (E) First photonic band gap PCF. (F) Near-field of the six-leaved blue mode that appears when (E) is excited by white light. (G) Hollow-core photonic band gap fiber. (H) Near-field of a red mode in a hollow-core PCF. (I) A hollow-core PCF with a Kagome cladding lattice, guiding white light. Reprinted with permission from [114].*

## 6. Resonators

### 6.1. High-Contrast Metastructure Resonators

HCMs can be designed as high-Q resonators without the need of a physical void or defect to form a cavity. Zhou et al. demonstrated for the first time a resonator using a near-wavelength 1D HCM [17]. The surface-normal emission had an ultra-narrow linewidth (Q>14,000) due to the in-plane resonance, as shown in Fig. 40(a). $In_{0.2}Ga_{0.8}As$ multiple quantum wells were embedded in the $Al_{0.6}Ga_{0.4}As$ grating layers. To prevent lateral light leakage due to finite-sized HCM, DBR gratings are fabricated as in-plane mirrors. Soon afterwards, the theoretical work by Karagodsky et al. [115] showed even without DBRs, high-Q resonance was achievable with a single-layer HCM. Fabry-Perot (FP) resonance along the surface-normal direction can occur within the HCM layer in between the two high-contrast interfaces. There are remarkable differences between the HCM resonance and the traditional FP resonance. Neither gain medium nor highly-reflective mirrors is required to satisfy the resonance condition [115]. The supermode which formed with two or a few HCM modes operating in the "near-wavelength" regime in Fig. 22, can perfectly resonate with itself, as indicated by the resonance lines in Fig. 4, Fig. 5, Fig. 6, and Fig. 12. As shown in [37], when the resonance lines "anti-cross" with (or repel) each other, indicating strong coupling among the HCM modes, very strong resonance will occur, with intensity enhancement above $10^7$.

With such remarkable phenomena, single-layer HCM resonators fabricated on silicon-on-insulator (SOI) wafers, as shown in Fig. 40(b), can be used as label-free biosensors [18] and nonlinear optical frequency conversion devices [116]. When operated at strong resonance, HCM reflection spectra is highly sensitive to the surrounding refractive index. A refractive index sensitivity (RIS) of 418 nm/RIU (refractive index unit) was demonstrated for a HCM biosensor, as shown in Fig. 40(d). On-chip HCM resonator has also shown enhanced four-wave mixing in silicon platform. Under a surface-normal pump at the 1538.9-nm wavelength, and a signal-pump detuning of 70-240 pm, a conversion efficiency of −19.5 dB was demonstrated at a low pump power of 900 µW, as shown in Fig. 40(e). When integrating a single atomic layer of graphene on the HCM resonator, the device can be used as a COMS-compatible electro-optic spatial light modulator, as shown in Fig. 40(c). With a change in the gate voltage from 0 V to −2 V, the graphene layer is tuned from charge neutral to hole-doped state, and an 11-dB modulation depth in reflectivity is observed, as shown in Fig. 40(f).

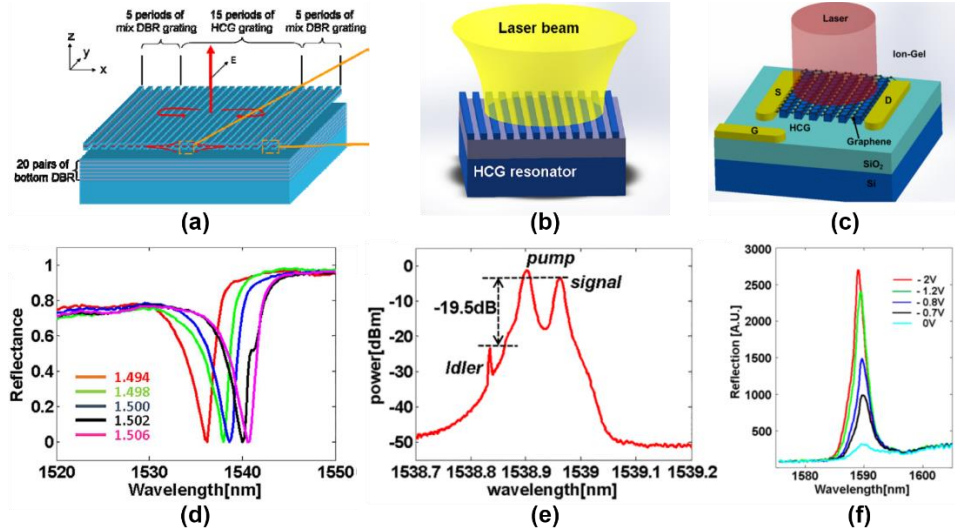

*Fig. 40. (a) Schematic of a high-Q resonator with 15-period middle 1D HCM grating and two 5-period mix DBR grating as the side mirror. (b) Schematic of surface-normal coupled Si-based 1D HCM resonator. (c) Schematic of surface-normal spatial light modulator using graphene on a HCM resonator. (d) Reflection spectra of a HCM resonator as a biosensor immersed in liquids with various refractive indices. (e) HCM resonator for enhancing the four-wave mixing of the normal incidence. (f) Voltage-tunable reflectivity spectra of a graphene-integrated HCM resonator showing electro-optic spatial light modulation. (a) is adapted from [17]. (b) and (e) are adapted from [116]. (c) and (f) are adapted from [117]. (d) is adapted from [18].*

### 6.2 Photonic Crystal Resonators

Very similar to HCM resonators, PhC membrane can also be designed and operated at resonance. While HCM resonators operates along the resonance curves above the light line and in the non-diffraction region (Fig. 13), the PhC membrane resonators can operate at the rest of the region shown in Fig. 13. To facilitate light coupling from PhC into free-space, it typically operates in the diffraction region.

One exemplary design is to have the resonance condition locates at the high order Γ point, where the group velocity of light is zero. The lattice constant can be designed to be the same as the light wavelength. Lightwave propagating along the lattice period direction can get reflected backwards because of the second-order Bragg diffraction effect. In addition, the lightwave is also diffracted to the out-of-plane direction because of the first-order Bragg diffraction effect. Consequently, lightwaves propagating in these directions are coupled with each other, and a standing wave state can be formed in the PhC plane (Fig. 41). While this provides an intuitive explanation of the PhC resonator, the rigorous analysis can be implemented following the supermode concept in Section 2.3.

When combining the PhC resonator with an active material, a PhC laser can be realized. In the example shown in Fig. 41 [118], a PhC layer is fabricated below the MQW layer on an epitaxial wafer, followed by a regrowth to form a cladding layer. Light confined in the PhC layer is evanescently coupled to MQW layer. Due to a very broad coherent-oscillation area, a very high-power output (1.5 W) and high beam quality (<3º divergence angle) could be simultaneously obtained with electrical pump. The experimentally measured band structure in Fig. 41(c) confirmed its single-mode lasing condition at the band-edge. By engineering the lattice geometry and thus shifting the zero-group-velocity point away from Γ, lasing conditions with different free-space output coupling angle can be obtained [119].

While the above PhC resonator features a resonance mode occupying a broad area, another type of PhC resonator – the point defect type PhC resonator [4] – has a completely opposite feature. The point defect type PhC resonator is implemented in a PhC membrane. The operating principle shares similarity with the PhC waveguide discussed in Section 5.2. Whereas a line defect makes a waveguide, the point defect makes a resonator. Light can be coupled out through an adjacent waveguide, as shown in Fig. 38(b); or through scattering into free space. By locating a gain material within the point defect, a PhC laser can be realized. To facilitate electrical pump, a lateral p-i-n junction structure could be implemented, shown in Fig. 42 [120]. In this example, the intrinsic region is the PhC cavity (modified three-hole defect), and the *p*-doped and *n*-doped region are formed by ion implantation. Lasing thresholds of 181 nA at 50 K and 287 nA at 150 K are experimentally demonstrated.

Besides lasing application, PhC resonator can also be used in bio-sensing application. The device concept is very similar to the HCM resonator sensing in Section 6.1, except that light is coupled to a high Q PhC resonator through an adjacent PhC waveguide, instead of through free-space. An example can be found at [121].

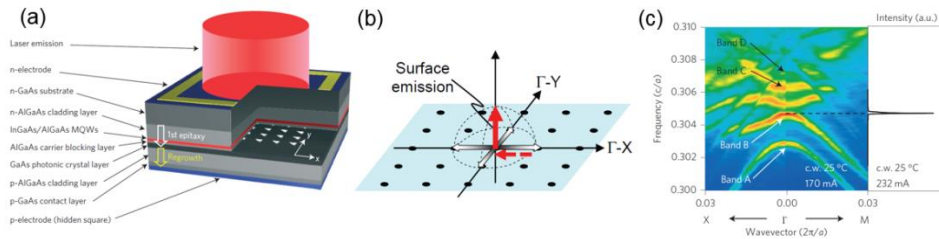

*Fig. 41. PhC laser operating at band-edge. (a) Schematic of the PhC structure. Arrows indicate the growth direction of the first epitaxial and regrowth structures. (b) Coupling diagram shown in the reciprocal space of the square-lattice PhC. Out-of-plane coupling occurs due to first-order Bragg diffraction. (c) Left panel, band structure of the PhC laser measured well below threshold current; right panel, the lasing spectrum measured in the surface-normal direction by injecting current above threshold. Reprinted with permission from [118].*

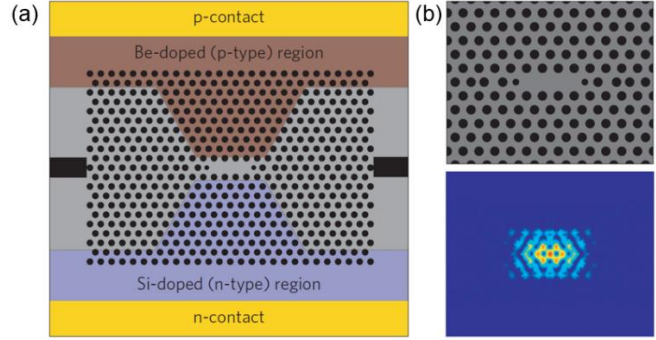

*Fig. 42. Electrically pump point-defect type PhC laser. (a) Schematic of the electrically pumped PhC laser made of GaAs with InAs quantum dots as the active layer. The width of the intrinsic region is narrow in the cavity region to direct current flow to the active region of the laser. (b) Modified three-hole defect PhC cavity design (top), and a FDTD simulation of the electrical field of the cavity mode (bottom). Reprinted with permission from [120]*

## 7. Reconfigurable Optics

With all the extraordinary performance exhibited through thin, compact arrays of metastructures, it is greatly desirable if these optical components can become reconfigurable, either switchable among operation states, or continuously tunable in certain optical properties. In this section, we review recent advances of metastructures in reconfigurable optics.

As an overview, recent approaches for building reconfigurable metastructures fall into the following major categories:
  A. Mechanical reconfiguration of structures [27]
  B. Optical reconfiguration of structures [53, 122]
  C. Thermal reconfiguration of structures [123]
  D. Electrostatic and magnetic reconfiguration of structures [124, 125]
  E. Electrical tuning of material properties (e.g. refractive index [126], electric and magnetic dipole resonance [127])
  F. Thermal tuning of material properties (e.g. refractive index [128], material phase transition [129])

By implementing HCMs on flexible materials, researchers are able to mechanically deform the arrangement of the metastructures, such as the period. As an example in [27], by stretching the HCM period, the diffraction angle of the $-1^{st}$ order can be steered by 5.5° with 5.45% deformation. On the other hand, at fixed incidence and observation angles, the peak wavelength $-1^{st}$ diffraction shifts by 39 nm with 4.9% deformation, causing a significant change in display color, as shown in Fig. 43(a). In other examples [130, 131], stretching of the period changes the optical phase distribution, and thus the light refraction angle and the focal length of the HCM lens, as shown in Fig. 43(b).

When the radiation force induced by optical pumps acts on these light-weight metastructures, it is possible to change their spatial arrangement and thus the optical behavior. As shown in Fig. 43(b), resonance modes in the metastructures are excited by the pump light, which then generates strong optical forces [122]. The optical forces are sufficient to produce movements of the nanostructures, thus the modulation of the probe light transmission, and even large nonlinear optical response.
Suspended HCM mirrors can be electrostatically actuated as tunable mirrors in VCSELs, which is well understood as discussed in Section 3. However, it has been recently discovered in [53] that the light emission can sustain self-oscillation of the HCM mirror, as shown in the SEM pictures in Fig. 43(d), with the laser being turned off (left) and on (right). By correctly designing the mirror reflection spectrum and the response delay between the optical force relative to the mirror movement, either large mechanical oscillation or cooling of the mechanical modes can be achieved.

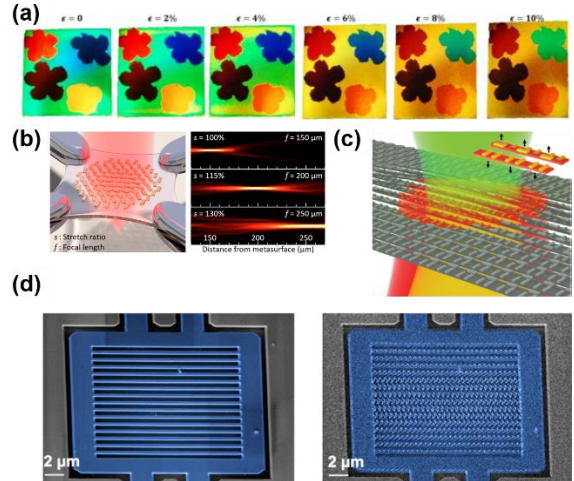

*Fig. 43. Examples of mechanical and optical reconfigurable metastructures. (a) Tunable coloration using flexible HCMs under various stretching deformation ϵ. (b) Stretchable HCM lens showing continuously tunable focal length. (c) Optically actuated HCMs giving rise to a large optical nonlinearity (d) Optomechanical oscillation of the HCM suspended mirror on a VCSEL. (a), (b), (c), and (d) are adapted from [27], [130], [122], and [53], respectively.*

Making use of the material thermal expansion, the deformation of the metastructures can be controlled by temperature. As shown in Fig. 44(a), the metastructure is a two-layer membrane consisting of 50-nm thick gold layer and a 100-nm thick $SiN_x$ layer [123]. Due to the difference of the thermal expansion coefficients (5x larger for gold than $SiN_x$), the membrane bends toward opposite directions with increasing and decreasing temperatures. With a temperature change from 76 K to 270 K, 51% increase of the transmission increase was observed at transmission resonance near 1735 nm.

With the use of MEMS, the positions of metastructures can be electrostatically actuated by the Coulomb force. As an example in Fig. 44(b), alternating straight and meander gold plasmonic wires are supported by ion-milled 50-nm thick $SiN_x$ membranes [124]. The initial gap size between the wires is 125 nm. Optical reflectance and transmission modulation up to 8% can be obtained at up to megahertz rates. Alternatively, plasmonic nanowires with conducting current experience Lorentz force under external magnetic field, thus can be actuated for the modulation of light transmission or reflection. As an example in [125], under 112 mT static magnetic field and ~50 mV electric voltage for generating ~160 µA current in the nanowires, ~2.5% transmission modulation is achieved at ~200 kHz rates.

Other than structural actuation, the material properties can be modulated electrically or thermally. As shown in Fig. 44(c) [126], a thin conducting ITO and a thin $Al_2O_3$ film are sandwiched between gold stripes and a gold back plane. Charge accumulation occurs at the $Al_2O_3$/ITO interface under applied voltage. Tunability of the metastructures result from the field-effect modulation of the ITO refractive index. Reflectance change of ~30% has been observed with 2.5 V gate bias with modulation speed up to 10 MHz. A beam steering angle of 40° has also been demonstrated. Authors in [127] introduce free charge carriers in InSb metastructures and the engineer the electric and magnetic dipole resonances by applied voltage. The transmission phase can be tuned continuously between 0 and $2\pi$ with less than 3 dB intensity loss. Thus the structures can be used as phase metasurfaces.

Material properties of the metastructures can also be tuned by temperature. Fig. 44(d) [128] shows a $8 \times 8$ tunable optical nanoantenna phased array integrated on silicon. Each pixel consists of a nanoantenna to couple optical power to free space and optical phase delay lines to control the emitting phase. The silicon waveguide in each pixel is partly doped to form a resistive heater for the thermal phase tuning. Dynamic far-field patterns have been demonstrated by applying different voltage combinations to the pixels.

In recent years, significant interest in room-temperature phase-transition materials has emerged [132]. One of such materials is vanadium dioxide ($VO_2$), which exhibits reversible metal-insulator transition at ~67°C, leading to drastic changes in resistivity and refractive index. With this material, authors in [129] has built metastructures to behave as tunable absorbers and tunable polarizers, as shown in Fig. 44(e). The $VO_2$ HCM is laterally patterned by selective defect engineering, which shifts the local critical temperature. With temperature above 75°C, above alternating regions are metallic, and the polarization dependence vanishes. Below 75°C the reflection shows large polarization selectivity.

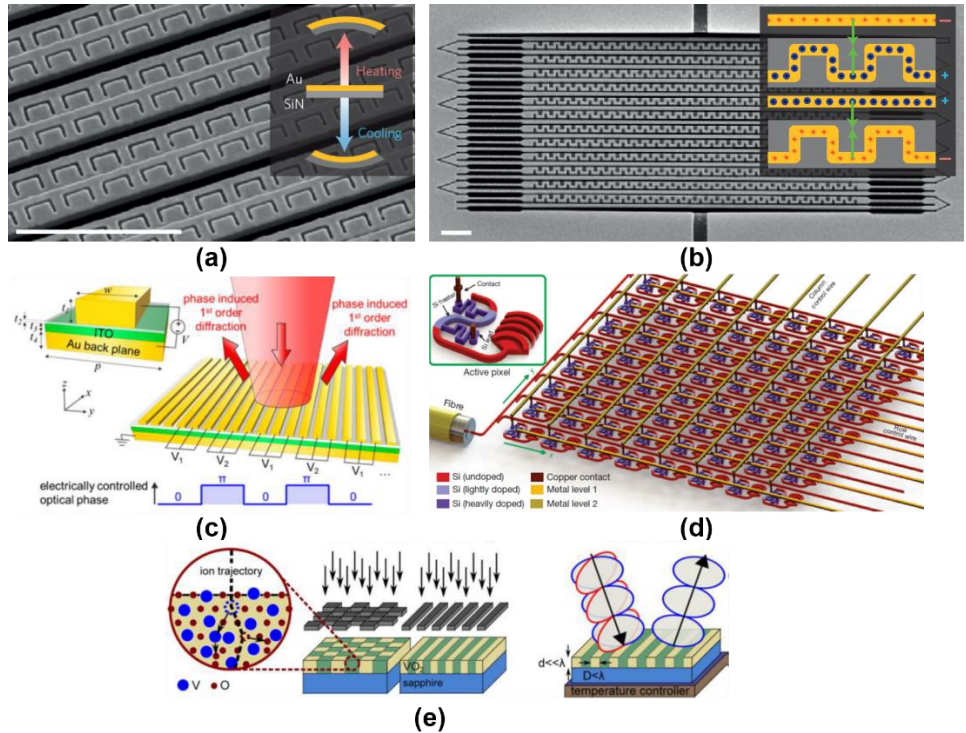

*Fig. 44. Examples of thermal and electrical reconfigurable metastructures. (a) Thermal control of the structural deformation. (b) Electrostatic actuation of the structural position. (c) Gate voltage control of the permittivity of conducting oxides. (d) Thermal control of waveguide optical delay. (e) Thermal control of material phase transition. (a), (b), (c), (d), and (e) are adapted from [123], [124], [126], [128], and [129], respectively.*

## 8. Summary

Similarities and differences between HCMs and PhCs are extensively discussed. Examples of extraordinary optical behaviors are shown with simple, compact arrays of subwavelength structures. Detailed analysis has brought light on the intrinsic connection between HCMs and PhCs arising from their periodic nature, as well as resonance behavior. Although the commonly-used design tools differ among HCM and PhC communities, a single diagram is presented in this review to identify difference of their operation conditions. HCMs typically apply to light impinging on the periodic structures, while PhC apply to light confined and propagating within the periodic structures. The major region of interest associated with the band structure is above and below the "light line" for HCMs and PhCs, meaning the optical field is propagating and evanescent outside of the periodic region, respectively.

The design HCMs concerns mainly the scattering properties when used as artificial structures. The design of PhCs concerns mainly of the properties of modes within the structure, their photonic band structures, and the stop bands. They both can be designed as mirrors and resonators. We differentiate HCMs and PhC mirrors from their directions of reflection, being out-of-plane or in-plane with their periodic directions. HCMs and PhCs both can support high-Q resonance. HCM itself acts as a Fabry-Perot cavity for the coupled modes within the periodic structure. PhC typically relies on artificial defects within the periodic structure. Most intriguingly, HCMs can behave as metasurfaces or artificial surfaces to tailor the scattering of light in an extraordinary fashion, similar to PhCs acting as metamaterials or artificial media, in which light propagate in an unusual manner. With much underlying physics being revealed in this article, we believe period and quasi-periodic subwavelength structures will continue to have major impact in the fields of optical communication, light detection and ranging, optical sensing, and illumination and display.

**Appendix A**

In this appendix, we present a generalized formulation for calculating layered periodic media. In the HCG/HCM community, RCWA and AMM are two commonly used simulation methods, and the following serves as a common ground for the two methods. The total transverse field can be expressed through eigenmode expansion as

$$\mathbf{E}_t(\mathbf{r}) = \hat{x}E_x(\mathbf{r}) + \hat{y}E_y(\mathbf{r}) = \sum_i^L A_i \boldsymbol{\mathcal{E}}_t^{(i)}(x,y)e^{i\beta_i z}$$

$$\mathbf{H}_t(\mathbf{r}) = \hat{x}H_x(\mathbf{r}) + \hat{y}H_y(\mathbf{r}) = \sum_i^L A_i \boldsymbol{\mathcal{H}}_t^{(i)}(x,y)e^{i\beta_i z}$$

(A. 1)

where $A_i$ is the scalar coefficients and $\boldsymbol{\mathcal{E}}_t^{(i)}(x,y)$ and $\boldsymbol{\mathcal{H}}_t^{(i)}(x,y)$ are the transverse vector fields of the $i$-th eigenmode. Considering the forward and backward propagation, the above equation can be written in the matrix form,

$$\begin{bmatrix} \mathbf{E}_t(\mathbf{r}) \\ \mathbf{H}_t(\mathbf{r}) \end{bmatrix} = \begin{bmatrix} \boldsymbol{\mathcal{E}}_t^{(1)} & \cdots & \boldsymbol{\mathcal{E}}_t^{(L)} & \boldsymbol{\mathcal{E}}_t^{(1)} & \cdots & \boldsymbol{\mathcal{E}}_t^{(L)} \\ \boldsymbol{\mathcal{H}}_t^{(1)} & \cdots & \boldsymbol{\mathcal{H}}_t^{(L)} & -\boldsymbol{\mathcal{H}}_t^{(1)} & \cdots & -\boldsymbol{\mathcal{H}}_t^{(L)} \end{bmatrix} \begin{bmatrix} e^{i\bar{\beta}z} & \overline{0} \\ \overline{0} & e^{-i\bar{\beta}z} \end{bmatrix} \begin{bmatrix} A_1^+ \\ \vdots \\ A_L^+ \\ A_1^- \\ \vdots \\ A_L^- \end{bmatrix}$$

(A. 2)

Here, $L$ is the number of modes to keep in simulation, and $\overline{\boldsymbol{\beta}}$ is a diagonal matrix consisting of the propagation constants $\beta_i$'s for eigenmodes. Casting the vector eigenmodes into matrices $\overline{\mathcal{E}}_t$ and $\overline{\mathcal{H}}_t$, and the coefficients into vectors $\mathbf{A}^+$ and $\mathbf{A}^-$, the $z$-dependent transverse fields can be expressed as

$$\begin{bmatrix} \mathbf{E}_t(\mathbf{r}) \\ \mathbf{H}_t(\mathbf{r}) \end{bmatrix} = \begin{bmatrix} \overline{\mathcal{E}}_t(x,y) & \overline{\mathcal{E}}_t(x,y) \\ \overline{\mathcal{H}}_t(x,y) & -\overline{\mathcal{H}}_t(x,y) \end{bmatrix} \begin{bmatrix} \mathbf{A}^+(z) \\ \mathbf{A}^-(z) \end{bmatrix} \tag{A.3}$$

The wave propagation within the same layer can be expressed by $z$-dependent expansion coefficients,

$$\begin{bmatrix} \mathbf{A}^+(z) \\ \mathbf{A}^-(z) \end{bmatrix} = \begin{bmatrix} e^{i\overline{\beta}z} & \overline{\mathbf{0}} \\ \overline{\mathbf{0}} & e^{-i\overline{\beta}z} \end{bmatrix} \begin{bmatrix} \mathbf{A}^+(0) \\ \mathbf{A}^-(0) \end{bmatrix} = \overline{\varphi}_{2L \times 2L} \begin{bmatrix} \mathbf{A}^+(0) \\ \mathbf{A}^-(0) \end{bmatrix} \tag{A.4}$$

where $\overline{\varphi}$ is the propagation matrix within the same layer. Assume $z = z_0$ is the interface between two layers, the tangential fields need to satisfy the boundary conditions

$$\begin{bmatrix} \mathbf{E}_{t,\mathrm{I}}(z_0^-) \\ \mathbf{H}_{t,\mathrm{I}}(z_0^-) \end{bmatrix} = \begin{bmatrix} \mathbf{E}_{t,\mathrm{II}}(z_0^+) \\ \mathbf{H}_{t,\mathrm{II}}(z_0^+) \end{bmatrix} \tag{A.5}$$

Substituting Eq. (A.3) into Eq. (A.5), we obtain the matrix equation for wave propagation across the interface,

$$\begin{bmatrix} \overline{\mathcal{E}}_{t,\mathrm{I}} & \overline{\mathcal{E}}_{t,\mathrm{I}} \\ \overline{\mathcal{H}}_{t,\mathrm{I}} & -\overline{\mathcal{H}}_{t,\mathrm{I}} \end{bmatrix} \begin{bmatrix} \mathbf{A}^+(z_0^-) \\ \mathbf{A}^-(z_0^-) \end{bmatrix} = \begin{bmatrix} \overline{\mathcal{E}}_{t,\mathrm{II}} & \overline{\mathcal{E}}_{t,\mathrm{II}} \\ \overline{\mathcal{H}}_{t,\mathrm{II}} & -\overline{\mathcal{H}}_{t,\mathrm{II}} \end{bmatrix} \begin{bmatrix} \mathbf{A}^+(z_0^+) \\ \mathbf{A}^-(z_0^+) \end{bmatrix} \tag{A.6}$$

Solving the above matrix equation requires two steps. Firstly, we need to obtain the field distribution $\mathcal{E}_t^{(i)}(x,y)$ and $\mathcal{H}_t^{(i)}(x,y)$ of all the eigenmodes in both regions. For 1D periodic structures, these modes are expressed by 1D periodic scalar fields, which can be written in close-form as sinusoidal functions. This method is known as the analytical method [37]. However, for 2D periodic structures, there is no analytical expression for the vectorial eigenmodes without approximation. Secondly, we calculate all inner products (or overlap integrals) between modes on each side of the interface across the unit cell, and the matrix inversion can follow. For 1D periodic structures, the integrals can be calculated analytically as in [37]. However, for 2D periodic structures, this calculation can only be done numerically due to the lack of the close-form expressions of the modes, and the computation and storage costs are both high. For this reason, a more practical approach is to project the modes down to a lower-level basis, such as the finite-element basis, or the Bloch-wave basis, as shown in Table 2. This approach is also known as the RCWA [38, 45, 46], which particularly benefits the calculation of periodic structures. The Bloch-wave basis is a set of exponential functions $e^{i\mathbf{G} \cdot \mathbf{r}}$ in the Fourier domain, with $\mathbf{G} = \hat{x} G_x + \hat{y} G_y$ being the Bloch wave number based on the periodicities $\Lambda_x$ and $\Lambda_y$,

$$\begin{aligned} G_x &= \frac{2m\pi}{\Lambda_x}, m = 0, \pm 1, \pm 2, \ldots, \pm g_{x,max} \\ G_y &= \frac{2n\pi}{\Lambda_y}, n = 0, \pm 1, \pm 2, \ldots, \pm g_{y,max} \end{aligned} \tag{A.7}$$

For 1D periodic structures, we can set $g_{y,max} = 0$. Then the $i$-th eigenmode can be further expressed with state vectors $\tilde{\mathbf{E}}_t^{(i)}$ and $\tilde{\mathbf{H}}_t^{(i)}$ composed of the Bloch-mode expansion coefficients,

$$\mathcal{E}_t^{(i)} \to \tilde{\mathbf{E}}_t^{(i)} = \begin{bmatrix} \tilde{E}_{x,\mathbf{G}_1}^{(i)} \\ \vdots \\ \tilde{E}_{x,\mathbf{G}_N}^{(i)} \\ \tilde{E}_{y,\mathbf{G}_1}^{(i)} \\ \vdots \\ \tilde{E}_{y,\mathbf{G}_N}^{(i)} \end{bmatrix}_{2N \times 1} , \mathcal{H}_t^{(i)} \to \tilde{\mathbf{H}}_t^{(i)} = \begin{bmatrix} \tilde{H}_{x,\mathbf{G}_1}^{(i)} \\ \vdots \\ \tilde{H}_{x,\mathbf{G}_N}^{(i)} \\ \tilde{H}_{y,\mathbf{G}_1}^{(i)} \\ \vdots \\ \tilde{H}_{y,\mathbf{G}_N}^{(i)} \end{bmatrix}_{2N \times 1} \tag{A.8}$$

where $i = 1,2, \ldots, L$. In practical simulation, a finite number $N$ of Bloch modes are kept, where $N = (2g_{x,max} + 1)(2g_{y,max} + 1)$. Typical numerical eigensolver will return a square matrix, with its $L$ columns corresponding to eigenmodes in Eq. (A. 1) with $L = 2N$. Then the matrices $\overline{\mathcal{E}}_t$ and $\overline{\mathcal{H}}_t$ in Eq. (A. 6) can be expressed in the Fourier domain as square orthogonal matrices $\tilde{\overline{\mathbf{E}}}_t$ and $\tilde{\overline{\mathbf{H}}}_t$. And Eq. (A. 6) can be solved by matrix inversion,

$$\begin{bmatrix} A^+(z_0^+) \\ A^-(z_0^+) \end{bmatrix} = \begin{bmatrix} \tilde{\overline{\mathbf{E}}}_{t,II} & \tilde{\overline{\mathbf{E}}}_{t,II} \\ \tilde{\overline{\mathbf{H}}}_{t,II} & -\tilde{\overline{\mathbf{H}}}_{t,II} \end{bmatrix}^{-1}_{4N \times 4N} \begin{bmatrix} \tilde{\overline{\mathbf{E}}}_{t,I} & \tilde{\overline{\mathbf{E}}}_{t,I} \\ \tilde{\overline{\mathbf{H}}}_{t,I} & -\tilde{\overline{\mathbf{H}}}_{t,II} \end{bmatrix}_{4N \times 4N} \begin{bmatrix} A^+(z_0^-) \\ A^-(z_0^-) \end{bmatrix} = \overline{\mathcal{T}}_{4N \times 4N} \begin{bmatrix} A^+(z_0^-) \\ A^-(z_0^-) \end{bmatrix}$$
(A. 9)

where $\overline{\mathcal{T}}$ is the transfer matrix across the interface. With propagation and transfer matrices $\overline{\varphi}$ and $\overline{\mathcal{T}}$, we can obtain the vectors $A^+$ and $A^-$ at any $z$-distance, given that we know the forward incidence $A_{in}^+$ from the input port and the backward incidence from the output port being zero ($A_{out}^- = 0$). For instance, with an $x$-polarized normal-incident plane-wave in air as the input, the coefficients in the state vector $A_{in}^+$ correspond to plane waves propagating in different directions. Thus, only the first entry is non-zero,

$$A_{in}^+ = \begin{bmatrix} 1 \\ 0 \\ \vdots \\ 0 \end{bmatrix}_{2N \times 1}$$
(A. 10)

and the first entry corresponds to the first eigenmode, which is the normal incidence,

$$\tilde{\mathbf{E}}_t^{(1)} = \begin{bmatrix} 0 \\ \vdots \\ 1 \\ \vdots \\ 0 \\ 0 \\ \vdots \\ 0 \end{bmatrix}_{2N \times 1}$$
(A. 11)

where the only the $\left(\frac{N+1}{2}\right)$-th element is non-zero, corresponding to the Bloch mode of $G_x = G_y = 0$. Now the wave propagation through all layers is represented by matrix multiplications via Eq. (A. 4) and Eq. (A. 9). Eventually, the total scatter matrix $\overline{S}_{total}$ converted from the total transfer matrix $\overline{\mathcal{T}}_{total}$ produces the reflection and transmission of all modes,

$$\begin{bmatrix} A_{trans}^+ \\ A_{ref}^- \end{bmatrix} = \overline{S}_{total} \begin{bmatrix} A_{in}^+ \\ 0 \end{bmatrix}$$
(A. 12)

## Appendix B

In this appendix, we explain how eignemodes in the periodic region can be solved. The formulation of the matrix eigenequations using RCWA for solving $\tilde{\mathbf{E}}_t^{(i)}$ and $\tilde{\mathbf{H}}_t^{(i)}$ in Eq. (A. 9) is shown as follows.

The analytical mode-matching (AMM) and the rigorous coupled-wave analysis (RCWA) are the two common simulation approaches in the community of HCG/HCM. They both expand the field in each layer using the corresponding eigenmodes. The field propagation within a layer and field scattering at the layer interface are obtained via Eq. (A. 4) and Eq. (A. 6), respectively. With AMM, the eigenmodes are often expressed analytically [37], which is convenient for 1D HCMs. In cases where the HCMs are in 2D, or the building blocks have irregular shapes, RCWA is more favorable. With RCWA the eigenmodes $\mathcal{E}^{(i)}(x,y)e^{i\beta_i z}$ and their propagation constants $\beta_i$ are numerically solved by further expanding onto the Bloch-wave basis as

$$\mathbf{E}^{(i)}(\mathbf{r}) = \mathcal{E}^{(i)}(x,y)e^{i\beta_i z} = e^{i\beta_i z}\left(\sum_{\mathbf{G}} \tilde{\mathbf{E}}_{\mathbf{G}}^{(i)} e^{i\mathbf{G}\cdot\mathbf{r}}\right) e^{ik_{0x}x + ik_{0y}y}.$$

(B. 1)

A finite $N$-fold subset of the Bloch-wave basis with $N = (2g_{x,max} + 1)(2g_{y,max} + 1)$ is kept for numerical calculation. The vector wave equation for the $i$-th eigenmode is written as

$$\nabla \times \nabla \times \mathbf{E}^{(i)} = \frac{\omega^2}{c^2 \varepsilon_0} \mathbf{D}^{(i)} = \frac{\omega^2}{c^2} \varepsilon_r(\mathbf{r}) \mathbf{E}^{(i)}.$$

(B. 2)

Breaking into scalar equations we have

$$-\frac{\partial^2 E_x}{\partial z^2} + \frac{\partial^2 E_z}{\partial z \partial x} = \frac{\omega^2}{c^2 \varepsilon_0} D_x + \frac{\partial^2 E_x}{\partial y^2} - \frac{\partial^2 E_y}{\partial x \partial y},$$

$$-\frac{\partial^2 E_y}{\partial z^2} + \frac{\partial^2 E_z}{\partial z \partial y} = \frac{\omega^2}{c^2 \varepsilon_0} D_y + \frac{\partial^2 E_y}{\partial x^2} - \frac{\partial^2 E_x}{\partial y \partial x},$$

$$\left(\frac{\partial^2}{\partial x^2} + \frac{\partial^2}{\partial y^2}\right) E_z + \frac{\omega^2}{c^2 \varepsilon_0} D_z = \frac{\partial^2 E_x}{\partial z \partial x} + \frac{\partial^2 E_y}{\partial z \partial y}.$$

(B. 3)

Taking the Fourier transform of Eq. (**B. 2**), the displacement field in the Fourier domain can be written as

$$\tilde{\mathbf{D}}_{\mathbf{G}}^{(i)} = \sum_{\mathbf{G}'} \varepsilon_0 \varepsilon_{\mathbf{G}\mathbf{G}'} \tilde{\mathbf{E}}_{\mathbf{G}'}^{(i)}.$$

(B. 4)

And the $N \times N$ permittivity matrix is written as

$$\varepsilon_{\mathbf{G}\mathbf{G}'} = \frac{1}{S} \int \varepsilon_r(\mathbf{r}) e^{i(\mathbf{G}'-\mathbf{G})\cdot\mathbf{r}} dx dy.$$

(B. 5)

Substituting Eq. (**B. 1**), Eq. (**B. 4**), and Eq. (**B. 5**) into Eq. (**B. 3**), we obtain

$$\sum_{\mathbf{G}'} Q_{\mathbf{G}\mathbf{G}'} \tilde{E}_{z,\mathbf{G}'}^{(i)} = \beta_i \left[(k_{0x} + G_x)\tilde{E}_{x,\mathbf{G}}^{(i)} + (k_{0y} + G_y)\tilde{E}_{y,\mathbf{G}}^{(i)}\right].$$

(B. 6)

where $Q_{\mathbf{G}\mathbf{G}'}$ is an $N \times N$ matrix

$$Q_{\mathbf{G}\mathbf{G}'} = (k_{0x} + G_x)^2 (k_{0y} + G_y)^2 \delta_{\mathbf{G}\mathbf{G}'} + \frac{\omega^2}{c^2} \varepsilon_{\mathbf{G}\mathbf{G}'}.$$

(B. 7)

By matrix inversion we obtain the $z$-component of the electric field as

$$\tilde{E}_{z,\mathbf{G}}^{(i)} = \sum_{\mathbf{G}'} Q_{\mathbf{G}\mathbf{G}'}^{-1} \beta_i \left[(k_{ix} + G_x)\tilde{E}_{x,\mathbf{G}'}^{(i)} + (k_{iy} + G_y)\tilde{E}_{y,\mathbf{G}'}^{(i)}\right].$$

(B. 8)

Substituting Eq. (**B. 8**) into Eq. (**B. 3**) we can group the *x*- and *y*- components of the electric field into a $2N \times 2N$ matrix eigenequation

$$\begin{bmatrix} \bar{\mathbf{A}}_{11} & \bar{\mathbf{A}}_{12} \\ \bar{\mathbf{A}}_{21} & \bar{\mathbf{A}}_{22} \end{bmatrix} \begin{bmatrix} \widetilde{\mathrm{E}}^{(i)}_{x,G_1} \\ \vdots \\ \widetilde{\mathrm{E}}^{(i)}_{x,G_N} \\ \widetilde{\mathrm{E}}^{(i)}_{y,G_1} \\ \vdots \\ \widetilde{\mathrm{E}}^{(i)}_{y,G_N} \end{bmatrix} = \beta_i^2 \begin{bmatrix} \bar{\mathbf{B}}_{11} & \bar{\mathbf{B}}_{12} \\ \bar{\mathbf{B}}_{21} & \bar{\mathbf{B}}_{22} \end{bmatrix} \begin{bmatrix} \widetilde{\mathrm{E}}^{(i)}_{x,G_1} \\ \vdots \\ \widetilde{\mathrm{E}}^{(i)}_{x,G_N} \\ \widetilde{\mathrm{E}}^{(i)}_{y,G_1} \\ \vdots \\ \widetilde{\mathrm{E}}^{(i)}_{y,G_N} \end{bmatrix},$$

(B. 9)

where the $N \times N$ matrix elements are

$$(\bar{\mathbf{A}}_{11})_{GG'} = -(k_{iy} + G_y)^2 \delta_{GG'} + \frac{\omega^2}{c^2}\varepsilon_{GG'}$$

$$(\bar{\mathbf{A}}_{22})_{GG'} = -(k_{ix} + G_x)^2 \delta_{GG'} + \frac{\omega^2}{c^2}\varepsilon_{GG'}$$

$$(\bar{\mathbf{A}}_{12})_{GG'} = (\bar{\mathbf{A}}_{21})_{GG'} = (k_{ix} + G_x)(k_{iy} + G_y)\delta_{GG'}$$

$$(\bar{\mathbf{B}}_{11})_{GG'} = \delta_{GG'} - (k_{ix} + G_x)Q^{-1}_{GG'}(k_{ix} + G'_x)$$

$$(\bar{\mathbf{B}}_{22})_{GG'} = \delta_{GG'} - (k_{iy} + G_y)Q^{-1}_{GG'}(k_{iy} + G'_y)$$

$$(\bar{\mathbf{B}}_{12})_{GG'} = (\bar{\mathbf{B}}^T_{21})_{GG'} = -(k_{ix} + G_x)Q^{-1}_{GG'}(k_{iy} + G'_y)$$

(B. 10)

Numerical eigensolvers typically produce $L$ eigenvalues and eigenvectors with $L = 2N$. The eigenvectors representing the field profiles for all eigenmodes are grouped as

$$\bar{\bar{\mathbf{E}}}_t = \begin{bmatrix} \tilde{\mathbf{E}}^{(1)}_t & \cdots & \tilde{\mathbf{E}}^{(L)}_t \end{bmatrix} = \begin{bmatrix} \widetilde{\mathrm{E}}^{(1)}_{x,G_1} & \cdots & \widetilde{\mathrm{E}}^{(L)}_{x,G_1} \\ \vdots & & \\ \widetilde{\mathrm{E}}^{(1)}_{x,G_N} & \cdots & \widetilde{\mathrm{E}}^{(L)}_{x,G_1} \\ \widetilde{\mathrm{E}}^{(1)}_{y,G_1} & \cdots & \widetilde{\mathrm{E}}^{(L)}_{x,G_1} \\ \vdots & & \\ \widetilde{\mathrm{E}}^{(1)}_{y,G_N} & \cdots & \widetilde{\mathrm{E}}^{(L)}_{x,G_1} \end{bmatrix}_{2N \times L}$$

(B. 11)

and

$$\bar{\bar{\mathbf{H}}}_t = \frac{1}{\omega\mu_0} \begin{bmatrix} -\bar{\mathbf{B}}_{21} & -\bar{\mathbf{B}}_{22} \\ \bar{\mathbf{B}}_{11} & \bar{\mathbf{B}}_{12} \end{bmatrix} \bar{\bar{\mathbf{E}}}_t \cdot \bar{\mathbf{K}}_z$$

(B. 12)

which are used in Eq. (**A. 9**) for calculating the transfer matrices at interfaces. The permittivity matrix in Eq. (**B. 5**) can be obtained for building blocks with arbitrary shapes. Analytical expressions exist for certain shapes and on certain lattice structures. For 1D HCMs with rectangular cross-sections (bar width being $w$ and period being $\Lambda$),

$$\varepsilon_{GG'} = \varepsilon_a \delta_{GG'} + (\varepsilon_b - \varepsilon_a)\frac{\sin[(G-G')w/2]}{(G-G')\Lambda/2}.$$

(B. 13)

For 2D HCMs with cubical building blocks (widths $w_x$ and $w_y$) on a rectangular lattice (periods $\Lambda_x$ and $\Lambda_y$),

$$\varepsilon_{GG'} = \varepsilon_a \delta_{GG'} + (\varepsilon_b - \varepsilon_a)\frac{\sin[(G_x-G'_x)w_x/2]}{(G_x-G'_x)\Lambda_x/2}\frac{\sin[(G_y-G'_y)w_y/2]}{(G_y-G'_y)\Lambda_y/2}.$$

(B. 14)

For 2D HCMs with cylindrical building blocks (radius $r$) on a rectangular lattice,

$$\varepsilon_{GG'} = \begin{cases} \varepsilon_a + (\varepsilon_b - \varepsilon_a)\frac{\pi r^2}{\Lambda_x \Lambda_y} & \text{for } G = G' \\ 2(\varepsilon_b - \varepsilon_a)\frac{J_1(|G-G'|r)}{|G-G'|r}\frac{\pi r^2}{\Lambda_x \Lambda_y} & \text{for } G \neq G' \end{cases}. \tag{B.15}$$

For 2D HCMs with cylindrical building blocks (radius $r$) on a hexagonal lattice (periods $\Lambda_x = \Lambda$ and $\Lambda_y = \sqrt{3}\Lambda$),

$$\varepsilon_{GG'} = \begin{cases} \varepsilon_a + (\varepsilon_b - \varepsilon_a)\frac{2\pi r^2}{\sqrt{3}\Lambda^2} & \text{for } G = G' \\ 2(\varepsilon_b - \varepsilon_a)\frac{J_1(|G-G'|r)}{|G-G'|r}\frac{\pi r^2}{\sqrt{3}\Lambda^2}\left(1 + e^{i(G_x-G'_x)\Lambda/2}e^{i(G_y-G'_y)\sqrt{3}\Lambda/2}\right) & \text{for } G \neq G' \end{cases}.$$

(B.16)

Here, $J_1$ is the Bessel function of the first order, $\varepsilon_b$ and $\varepsilon_a$ are the relative permittivities of the building blocks and the surrounding medium, respectively.

## REFERENCES

1. E. Yablonovitch, "Inhibited Spontaneous Emission in Solid-State Physics and Electronics," Physical Review Letters **58**, 2059-2062 (1987).
2. S. John, "Strong localization of photons in certain disordered dielectric superlattices," Physical Review Letters **58**, 2486-2489 (1987).
3. J. D. Joannopoulos, S. G. Johnson, J. N. Winn, and R. D. Meade, *Photonic crystals: molding the flow of light* (Princeton university press, 2011).
4. S. Noda, A. Chutinan, and M. Imada, "Trapping and emission of photons by a single defect in a photonic bandgap structure," Nature **407**, 608-610 (2000).
5. T. A. Birks, J. C. Knight, and P. S. J. Russell, "Endlessly single-mode photonic crystal fiber," Opt. Lett. **22**, 961-963 (1997).
6. J. M. Dudley, G. Genty, and S. Coen, "Supercontinuum generation in photonic crystal fiber," Reviews of Modern Physics **78**, 1135-1184 (2006).
7. T. Baba, "Slow light in photonic crystals," Nat Photon **2**, 465-473 (2008).
8. C. Luo, S. G. Johnson, J. D. Joannopoulos, and J. B. Pendry, "All-angle negative refraction without negative effective index," Physical Review B **65**, 201104 (2002).
9. J. Valentine, J. Li, T. Zentgraf, G. Bartal, and X. Zhang, "An optical cloak made of dielectrics," Nat Mater **8**, 568-571 (2009).
10. C. F. R. Mateus, M. C. Y. Huang, C. Lu, C. J. Chang-Hasnain, and Y. Suzuki, "Broad-band mirror (1.12-1.62 um) using a subwavelength grating," IEEE Photonics Technology Letters **16**, 1676-1678 (2004).
11. M. C. Y. Huang, Y. Zhou, and C. J. Chang-Hasnain, "A nanoelectromechanical tunable laser," Nat Photon **2**, 180-184 (2008).
12. I. S. Chung, V. Iakovlev, A. Sirbu, A. Mereuta, A. Caliman, E. Kapon, and J. Mork, "Broadband MEMS-Tunable High-Index-Contrast Subwavelength Grating Long-Wavelength VCSEL," IEEE Journal of Quantum Electronics **46**, 1245-1253 (2010).
13. Y. Rao, W. Yang, C. Chase, M. C. Y. Huang, D. P. Worland, S. Khaleghi, M. R. Chitgarha, M. Ziyadi, A. E. Willner, and C. J. Chang-Hasnain, "Long-Wavelength VCSEL Using High-Contrast Grating," IEEE Journal of Selected Topics in Quantum Electronics **19**, 1701311-1701311 (2013).
14. E. Haglund, J. S. Gustavsson, J. Bengtsson, Å. Haglund, A. Larsson, D. Fattal, W. Sorin, and M. Tan, "Demonstration of post-growth wavelength setting of VCSELs using high-contrast gratings," Opt. Express **24**, 1999-2005 (2016).
15. K. Li, C. Chase, P. Qiao, and C. J. Chang-Hasnain, "Widely tunable 1060-nm VCSEL with high-contrast grating mirror," Opt. Express **25**, 11855-11866 (2017).
16. P. Qiao, K. T. Cook, K. Li, and C. J. Chang-Hasnain, "Wavelength-Swept VCSELs," IEEE Journal of Selected Topics in Quantum Electronics (2017).
17. Y. Zhou, M. Moewe, J. Kern, M. C. Y. Huang, and C. J. Chang-Hasnain, "Surface-normal emission of a high-Q resonator using a subwavelength high-contrast grating," Opt. Express **16**, 17282-17287 (2008).

18. T. Sun, S. Kan, G. Marriott, and C. Chang-Hasnain, "High-contrast grating resonators for label-free detection of disease biomarkers," Scientific Reports **6**, 27482 (2016).
19. W. Yang, J. Ferrara, K. Grutter, A. Yeh, C. Chase, Y. Yue, A. E. Willner, M. C. Wu, and C. J. Chang-Hasnain, "Low loss hollow-core waveguide on a silicon substrate," Nanophotonics **1**, 23 (2012).
20. B. Pesala, W. Yang, and C. J. Chang-Hasnain, "Compact On-Chip Optical Components Based on Multimode Interference Design Using High-Contrast Grating Hollow-Core Waveguides," IEEE Journal of Selected Topics in Quantum Electronics **22**, 279-287 (2016).
21. L. Zhu, V. Karagodsky, and C. Chang-Hasnain, "Novel high efficiency vertical to in-plane optical coupler," in *Proc. SPIE*, 2012), 82700L.
22. F. Lu, F. G. Sedgwick, V. Karagodsky, C. Chase, and C. J. Chang-Hasnain, "Planar high-numerical-aperture low-loss focusing reflectors and lenses using subwavelength high contrast gratings," Opt. Express **18**, 12606-12614 (2010).
23. D. Fattal, J. Li, Z. Peng, M. Fiorentino, and R. G. Beausoleil, "Flat dielectric grating reflectors with focusing abilities," Nat Photon **4**, 466-470 (2010).
24. B.-W. Yoo, M. Megens, T. Chan, T. Sun, W. Yang, C. J. Chang-Hasnain, D. A. Horsley, and M. C. Wu, "Optical phased array using high contrast gratings for two dimensional beamforming and beamsteering," Opt. Express **21**, 12238-12248 (2013).
25. W. Yang, T. Sun, Y. Rao, M. Megens, T. Chan, B.-W. Yoo, D. A. Horsley, M. C. Wu, and C. J. Chang-Hasnain, "High speed optical phased array using high contrast grating all-pass filters," Opt. Express **22**, 20038-20044 (2014).
26. T. Tran, V. Karagodsky, Y. Rao, W. Yang, R. Chen, C. Chase, L. C. Chuang, and C. J. Chang-Hasnain, "Surface-normal second harmonic emission from AlGaAs high-contrast gratings," Applied Physics Letters **102**, 021102 (2013).
27. L. Zhu, J. Kapraun, J. Ferrara, and C. J. Chang-Hasnain, "Flexible photonic metastructures for tunable coloration," Optica **2**, 255-258 (2015).
28. J. Ferrara, W. Yang, L. Zhu, P. Qiao, and C. J. Chang-Hasnain, "Heterogeneously integrated long-wavelength VCSEL using silicon high contrast grating on an SOI substrate," Opt. Express **23**, 2512-2523 (2015).
29. P. Qiao, K. Li, K. T. Cook, and C. J. Chang-Hasnain, "MEMS-tunable VCSELs using 2D high-contrast gratings," Opt. Lett. **42**, 823-826 (2017).
30. A. Arbabi, Y. Horie, M. Bagheri, and A. Faraon, "Dielectric metasurfaces for complete control of phase and polarization with subwavelength spatial resolution and high transmission," Nat Nano **10**, 937-943 (2015).
31. P. Qiao, L. Zhu, and C. J. Chang-Hasnain, "High-efficiency aperiodic two-dimensional high-contrast-grating hologram," in *Proc. SPIE*, 2016), 975708.
32. N. Yu and F. Capasso, "Flat optics with designer metasurfaces," Nat Mater **13**, 139-150 (2014).
33. P. Genevet, F. Capasso, F. Aieta, M. Khorasaninejad, and R. Devlin, "Recent advances in planar optics: from plasmonic to dielectric metasurfaces," Optica **4**, 139-152 (2017).
34. C. L. Holloway, E. F. Kuester, J. A. Gordon, J. O. Hara, J. Booth, and D. R. Smith, "An Overview of the Theory and Applications of Metasurfaces: The Two-Dimensional Equivalents of Metamaterials," IEEE Antennas and Propagation Magazine **54**, 10-35 (2012).
35. N. M. Estakhri and A. Alù, "Recent progress in gradient metasurfaces," J. Opt. Soc. Am. B **33**, A21-A30 (2016).
36. A. I. Kuznetsov, A. E. Miroshnichenko, M. L. Brongersma, Y. S. Kivshar, and B. Luk'yanchuk, "Optically resonant dielectric nanostructures," Science **354**(2016).
37. C. J. Chang-Hasnain and W. Yang, "High-contrast gratings for integrated optoelectronics," Adv. Opt. Photon. **4**, 379-440 (2012).
38. P. Qiao, L. Zhu, W. C. Chew, and C. J. Chang-Hasnain, "Theory and design of two-dimensional high-contrast-grating phased arrays," Opt. Express **23**, 24508-24524 (2015).
39. Z. Gang and Q. H. Liu, "The 2.5-D multidomain pseudospectral time-domain algorithm," IEEE Transactions on Antennas and Propagation **51**, 619-627 (2003).
40. A. Taflove and S. C. Hagness, *Computational electrodynamics* (Artech house, 2005).
41. W. C. Chew, E. Michielssen, J. M. Song, and J.-M. Jin, *Fast and efficient algorithms in computational electromagnetics* (Artech House, Inc., 2001).
42. J.-M. Jin, *The finite element method in electromagnetics* (John Wiley & Sons, 2015).
43. R. F. Harrington and J. L. Harrington, *Field computation by moment methods* (Oxford University Press, 1996).
44. R. E. Jorgenson and R. Mittra, "Efficient calculation of the free-space periodic Green's function," IEEE Transactions on Antennas and Propagation **38**, 633-642 (1990).
45. M. G. Moharam and T. K. Gaylord, "Rigorous coupled-wave analysis of planar-grating diffraction," J. Opt. Soc. Am. **71**, 811-818 (1981).

46. S. G. Tikhodeev, A. L. Yablonskii, E. A. Muljarov, N. A. Gippius, and T. Ishihara, "Quasiguided modes and optical properties of photonic crystal slabs," Physical Review B **66**, 045102 (2002).
47. V. Karagodsky, F. G. Sedgwick, and C. J. Chang-Hasnain, "Theoretical analysis of subwavelength high contrast grating reflectors," Opt. Express **18**, 16973-16988 (2010).
48. K. Li, Y. Rao, C. Chase, W. Yang, and C. J. Chang-Hasnain, "Beam-Shaping Single-Mode VCSEL With A High-Contrast Grating Mirror," in *Conference on Lasers and Electro-Optics*, OSA Technical Digest (2016) (Optical Society of America, 2016), SF1L.7.
49. T. Asano, B. S. Song, Y. Akahane, and S. Noda, "Ultrahigh-Q Nanocavities in Two-Dimensional Photonic Crystal Slabs," IEEE Journal of Selected Topics in Quantum Electronics **12**, 1123-1134 (2006).
50. S. David, A. Chelnikov, and J. M. Lourtioz, "Isotropic photonic structures: Archimedean-like tilings and quasi-crystals," IEEE Journal of Quantum Electronics **37**, 1427-1434 (2001).
51. N. Lawrence, J. Trevino, and L. D. Negro, "Aperiodic arrays of active nanopillars for radiation engineering," Journal of Applied Physics **111**, 113101 (2012).
52. S. Fan, J. N. Winn, A. Devenyi, J. C. Chen, R. D. Meade, and J. D. Joannopoulos, "Guided and defect modes in periodic dielectric waveguides," J. Opt. Soc. Am. B **12**, 1267-1272 (1995).
53. W. Yang, S. A. Gerke, K. W. Ng, Y. Rao, C. Chase, and C. J. Chang-Hasnain, "Laser optomechanics," Scientific Reports **5**, 13700 (2015).
54. R. Magnusson, "Wideband reflectors with zero-contrast gratings," Opt. Lett. **39**, 4337-4340 (2014).
55. A. Liu, W. Zheng, and D. Bimberg, "Comparison between high- and zero-contrast gratings as VCSEL mirrors," Optics Communications **389**, 35-41 (2017).
56. P. Lalanne and G. M. Morris, "Antireflection behavior of silicon subwavelength periodic structures for visible light," Nanotechnology **8**, 53 (1997).
57. Z. Bomzon, V. Kleiner, and E. Hasman, "Pancharatnam–Berry phase in space-variant polarization-state manipulations with subwavelength gratings," Opt. Lett. **26**, 1424-1426 (2001).
58. Z. Bomzon, G. Biener, V. Kleiner, and E. Hasman, "Radially and azimuthally polarized beams generated by space-variant dielectric subwavelength gratings," Opt. Lett. **27**, 285-287 (2002).
59. E. Hasman, V. Kleiner, G. Biener, and A. Niv, "Polarization dependent focusing lens by use of quantized Pancharatnam–Berry phase diffractive optics," Applied Physics Letters **82**, 328-330 (2003).
60. P. Viktorovitch, B. Ben Bakir, S. Boutami, J. L. Leclercq, X. Letartre, P. Rojo-Romeo, C. Seassal, M. Zussy, L. Di Cioccio, and J. M. Fedeli, "3D harnessing of light with 2.5D photonic crystals," Laser & Photonics Reviews **4**, 401-413 (2010).
61. Y. Zhou, M. C. Y. Huang, C. Chase, V. Karagodsky, M. Moewe, B. Pesala, F. G. Sedgwick, and C. J. Chang-Hasnain, "High-Index-Contrast Grating (HCG) and Its Applications in Optoelectronic Devices," IEEE Journal of Selected Topics in Quantum Electronics **15**, 1485-1499 (2009).
62. A. Liu, F. Fu, Y. Wang, B. Jiang, and W. Zheng, "Polarization-insensitive subwavelength grating reflector based on a semiconductor-insulator-metal structure," Opt. Express **20**, 14991-15000 (2012).
63. D. Zhao, H. Yang, Z. Ma, and W. Zhou, "Polarization independent broadband reflectors based on cross-stacked gratings," Opt. Express **19**, 9050-9055 (2011).
64. K. Ikeda, K. Takeuchi, K. Takayose, I.-S. Chung, J. Mørk, and H. Kawaguchi, "Polarization-independent high-index contrast grating and its fabrication tolerances," Appl. Opt. **52**, 1049-1053 (2013).
65. T. Mori, Y. Yamayoshi, and H. Kawaguchi, "Low-switching-energy and high-repetition-frequency all-optical flip-flop operations of a polarization bistable vertical-cavity surface-emitting laser," Applied Physics Letters **88**, 101102 (2006).
66. C. Sciancalepore, B. B. Bakir, X. Letartre, J. Harduin, N. Olivier, C. Seassal, J. M. Fedeli, and P. Viktorovitch, "CMOS-Compatible Ultra-Compact 1.55-um Emitting VCSELs Using Double Photonic Crystal Mirrors," IEEE Photonics Technology Letters **24**, 455-457 (2012).
67. H. Yang, D. Zhao, S. Chuwongin, J.-H. Seo, W. Yang, Y. Shuai, J. Berggren, M. Hammar, Z. Ma, and W. Zhou, "Transfer-printed stacked nanomembrane lasers on silicon," Nat Photon **6**, 615-620 (2012).
68. C. Chase, Y. Zhou, and C. J. Chang-Hasnain, "Size effect of high contrast gratings in VCSELs," Opt. Express **17**, 24002-24007 (2009).
69. T. D. Happ, A. Markard, M. Kamp, A. Forchel, S. Anand, J. L. Gentner, and N. Bouadma, "Nanofabrication of two-dimensional photonic crystal mirrors for 1.5 μm short cavity lasers," Journal of Vacuum Science & Technology B: Microelectronics and Nanometer Structures Processing, Measurement, and Phenomena **19**, 2775-2778 (2001).

70. M. Born and E. Wolf, *Principles of optics: electromagnetic theory of propagation, interference and diffraction of light* (Elsevier, 1980).
71. R. W. Wood, "On a Remarkable Case of Uneven Distribution of Light in a Diffraction Grating Spectrum," Proceedings of the Physical Society of London **18**, 269 (1902).
72. L. Rayleigh, "On the Dynamical Theory of Gratings," Proceedings of the Royal Society of London. Series A **79**, 399 (1907).
73. U. Fano, "Zur Theorie der Intensitätsanomalien der Beugung," Annalen der Physik **424**, 393-443 (1938).
74. A. Hessel and A. A. Oliner, "A New Theory of Wood's Anomalies on Optical Gratings," Appl. Opt. **4**, 1275-1297 (1965).
75. D. Maystre, "Theory of Wood's Anomalies," in *Plasmonics: From Basics to Advanced Topics*, S. Enoch and N. Bonod, eds. (Springer Berlin Heidelberg, Berlin, Heidelberg, 2012), pp. 39-83.
76. B. A. Munk, *Frequency selective surfaces theory and design* (John Wiley & Sons, 2000).
77. R. Mittra, C. H. Chan, and T. Cwik, "Techniques for analyzing frequency selective surfaces-a review," Proceedings of the IEEE **76**, 1593-1615 (1988).
78. V. Agrawal and W. Imbriale, "Design of a dichroic Cassegrain subreflector," IEEE Transactions on Antennas and Propagation **27**, 466-473 (1979).
79. H. L. Bertoni, L. h. S. Cheo, and T. Tamir, "Frequency-selective reflection and transmission by a periodic dielectric layer," IEEE Transactions on Antennas and Propagation **37**, 78-83 (1989).
80. S. S. Wang and R. Magnusson, "Theory and applications of guided-mode resonance filters," Appl. Opt. **32**, 2606-2613 (1993).
81. E. F. Kuester, M. A. Mohamed, M. Piket-May, and C. L. Holloway, "Averaged transition conditions for electromagnetic fields at a metafilm," IEEE Transactions on Antennas and Propagation **51**, 2641-2651 (2003).
82. R. A. Shelby, D. R. Smith, and S. Schultz, "Experimental Verification of a Negative Index of Refraction," Science **292**, 77 (2001).
83. N. Yu, P. Genevet, M. A. Kats, F. Aieta, J.-P. Tetienne, F. Capasso, and Z. Gaburro, "Light Propagation with Phase Discontinuities: Generalized Laws of Reflection and Refraction," Science **334**, 333 (2011).
84. C. L. Holloway, E. F. Kuester, and A. Dienstfrey, "A homogenization technique for obtaining generalized sheet transition conditions for an arbitrarily shaped coated wire grating," Radio Science **49**, 813-850 (2014).
85. S. Fan and J. D. Joannopoulos, "Analysis of guided resonances in photonic crystal slabs," Physical Review B **65**, 235112 (2002).
86. S. Sun, Q. He, S. Xiao, Q. Xu, X. Li, and L. Zhou, "Gradient-index meta-surfaces as a bridge linking propagating waves and surface waves," Nat Mater **11**, 426-431 (2012).
87. T. Zentgraf, T. P. Meyrath, A. Seidel, S. Kaiser, H. Giessen, C. Rockstuhl, and F. Lederer, "Babinet's principle for optical frequency metamaterials and nanoantennas," Physical Review B **76**, 033407 (2007).
88. B. Walther, C. Helgert, C. Rockstuhl, F. Setzpfandt, F. Eilenberger, E.-B. Kley, F. Lederer, A. Tünnermann, and T. Pertsch, "Spatial and Spectral Light Shaping with Metamaterials," Advanced Materials **24**, 6300-6304 (2012).
89. L. Huang, X. Chen, H. Mühlenbernd, G. Li, B. Bai, Q. Tan, G. Jin, T. Zentgraf, and S. Zhang, "Dispersionless Phase Discontinuities for Controlling Light Propagation," Nano Letters **12**, 5750-5755 (2012).
90. D. Lin, P. Fan, E. Hasman, and M. L. Brongersma, "Dielectric gradient metasurface optical elements," Science **345**, 298 (2014).
91. M. Khorasaninejad, F. Aieta, P. Kanhaiya, M. A. Kats, P. Genevet, D. Rousso, and F. Capasso, "Achromatic Metasurface Lens at Telecommunication Wavelengths," Nano Letters **15**, 5358-5362 (2015).
92. M. V. Berry, "The Adiabatic Phase and Pancharatnam's Phase for Polarized Light," Journal of Modern Optics **34**, 1401-1407 (1987).
93. E. Arbabi, A. Arbabi, S. M. Kamali, Y. Horie, and A. Faraon, "Multiwavelength polarization-insensitive lenses based on dielectric metasurfaces with meta-molecules," Optica **3**, 628-633 (2016).
94. A. Yariv and P. Yeh, *Photonics: optical electronics in modern communications* (oxford university press New York, 2007), Vol. 6.
95. S. Linden, C. Enkrich, M. Wegener, J. Zhou, T. Koschny, and C. M. Soukoulis, "Magnetic Response of Metamaterials at 100 Terahertz," Science **306**, 1351 (2004).
96. T. J. Yen, W. J. Padilla, N. Fang, D. C. Vier, D. R. Smith, J. B. Pendry, D. N. Basov, and X. Zhang, "Terahertz Magnetic Response from Artificial Materials," Science **303**, 1494 (2004).

97. V. M. Shalaev, W. Cai, U. K. Chettiar, H.-K. Yuan, A. K. Sarychev, V. P. Drachev, and A. V. Kildishev, "Negative index of refraction in optical metamaterials," Opt. Lett. **30**, 3356-3358 (2005).
98. Y. F. Yu, A. Y. Zhu, R. Paniagua-Domínguez, Y. H. Fu, B. Luk'yanchuk, and A. I. Kuznetsov, "High-transmission dielectric metasurface with 2π phase control at visible wavelengths," Laser & Photonics Reviews **9**, 412-418 (2015).
99. N. Shitrit, I. Bretner, Y. Gorodetski, V. Kleiner, and E. Hasman, "Optical Spin Hall Effects in Plasmonic Chains," Nano Letters **11**, 2038-2042 (2011).
100. E. Karimi, S. A. Schulz, I. De Leon, H. Qassim, J. Upham, and R. W. Boyd, "Generating optical orbital angular momentum at visible wavelengths using a plasmonic metasurface," Light Sci Appl **3**, e167 (2014).
101. A. Papakostas, A. Potts, D. M. Bagnall, S. L. Prosvirnin, H. J. Coles, and N. I. Zheludev, "Optical Manifestations of Planar Chirality," Physical Review Letters **90**, 107404 (2003).
102. U. Levy, H.-C. Kim, C.-H. Tsai, and Y. Fainman, "Near-infrared demonstration of computer-generated holograms implemented by using subwavelength gratings with space-variant orientation," Opt. Lett. **30**, 2089-2091 (2005).
103. D. Fattal, Z. Peng, T. Tran, S. Vo, M. Fiorentino, J. Brug, and R. G. Beausoleil, "A multi-directional backlight for a wide-angle, glasses-free three-dimensional display," Nature **495**, 348-351 (2013).
104. L. Huang, X. Chen, H. Mühlenbernd, H. Zhang, S. Chen, B. Bai, Q. Tan, G. Jin, K.-W. Cheah, C.-W. Qiu, J. Li, T. Zentgraf, and S. Zhang, "Three-dimensional optical holography using a plasmonic metasurface," Nature Communications **4**, 2808 (2013).
105. E. Plum, J. Zhou, J. Dong, V. A. Fedotov, T. Koschny, C. M. Soukoulis, and N. I. Zheludev, "Metamaterial with negative index due to chirality," Physical Review B **79**, 035407 (2009).
106. H. Zhang, L. Cao, and G. Jin, "Lighting effects rendering in three-dimensional computer-generated holographic display," Optics Communications **370**, 192-197 (2016).
107. T. Sun, W. Yang, V. Karagodsky, W. Zhou, and C. Chang-Hasnain, "Low-loss slow light inside high contrast grating waveguide," in 2012), 82700A-82700A-82707.
108. D. Taillaert, W. Bogaerts, P. Bienstman, T. F. Krauss, P. V. Daele, I. Moerman, S. Verstuyft, K. D. Mesel, and R. Baets, "An out-of-plane grating coupler for efficient butt-coupling between compact planar waveguides and single-mode fibers," IEEE Journal of Quantum Electronics **38**, 949-955 (2002).
109. B. Wang, J. Jiang, and G. P. Nordin, "Compact slanted grating couplers," Opt. Express **12**, 3313-3326 (2004).
110. C. J. Chang-Hasnain, K. Pei-cheng, K. Jungho, and C. Shun-lien, "Variable optical buffer using slow light in semiconductor nanostructures," Proceedings of the IEEE **91**, 1884-1897 (2003).
111. S. Kubo, D. Mori, and T. Baba, "Low-group-velocity and low-dispersion slow light in photonic crystal waveguides," Opt. Lett. **32**, 2981-2983 (2007).
112. D. M. Beggs, T. P. White, L. O'Faolain, and T. F. Krauss, "Ultracompact and low-power optical switch based on silicon photonic crystals," Opt. Lett. **33**, 147-149 (2008).
113. K. Nozaki, T. Tanabe, A. Shinya, S. Matsuo, T. Sato, H. Taniyama, and M. Notomi, "Sub-femtojoule all-optical switching using a photonic-crystal nanocavity," Nat Photon **4**, 477-483 (2010).
114. P. Russell, "Photonic Crystal Fibers," Science **299**, 358 (2003).
115. V. Karagodsky, C. Chase, and C. J. Chang-Hasnain, "Matrix Fabry–Perot resonance mechanism in high-contrast gratings," Opt. Lett. **36**, 1704-1706 (2011).
116. T. Sun, W. Yang, and C. Chang-Hasnain, "Surface-normal coupled four-wave mixing in a high contrast gratings resonator," Opt. Express **23**, 29565-29572 (2015).
117. T. Sun, J. Kim, J. M. Yuk, A. Zettl, F. Wang, and C. Chang-hasnain, "Surface-normal electro-optic spatial light modulator using graphene integrated on a high-contrast grating resonator," Opt. Express **24**, 26035-26043 (2016).
118. K. Hirose, Y. Liang, Y. Kurosaka, A. Watanabe, T. Sugiyama, and S. Noda, "Watt-class high-power, high-beam-quality photonic-crystal lasers," Nat Photon **8**, 406-411 (2014).
119. Y. Kurosaka, S. Iwahashi, Y. Liang, K. Sakai, E. Miyai, W. Kunishi, D. Ohnishi, and S. Noda, "On-chip beam-steering photonic-crystal lasers," Nat Photon **4**, 447-450 (2010).
120. B. Ellis, M. A. Mayer, G. Shambat, T. Sarmiento, J. Harris, E. E. Haller, and J. Vuckovic, "Ultralow-threshold electrically pumped quantum-dot photonic-crystal nanocavity laser," Nat Photon **5**, 297-300 (2011).
121. S. Chakravarty, W.-C. Lai, Y. Zou, H. A. Drabkin, R. M. Gemmill, G. R. Simon, S. H. Chin, and R. T. Chen, "Multiplexed Specific Label-Free Detection of NCI-H358 Lung Cancer Cell Line Lysates with Silicon Based Photonic Crystal Microcavity Biosensors," Biosensors & bioelectronics **43**, 50-55 (2013).

122. J.-Y. Ou, E. Plum, J. Zhang, and N. I. Zheludev, "Giant Nonlinearity of an Optically Reconfigurable Plasmonic Metamaterial," Advanced Materials **28**, 729-733 (2016).
123. J. Y. Ou, E. Plum, L. Jiang, and N. I. Zheludev, "Reconfigurable Photonic Metamaterials," Nano Letters **11**, 2142-2144 (2011).
124. J.-Y. Ou, E. Plum, J. Zhang, and N. I. Zheludev, "An electromechanically reconfigurable plasmonic metamaterial operating in the near-infrared," Nat Nano **8**, 252-255 (2013).
125. J. Valente, J.-Y. Ou, E. Plum, I. J. Youngs, and N. I. Zheludev, "A magneto-electro-optical effect in a plasmonic nanowire material," Nature Communications **6**, 7021 (2015).
126. Y.-W. Huang, H. W. H. Lee, R. Sokhoyan, R. A. Pala, K. Thyagarajan, S. Han, D. P. Tsai, and H. A. Atwater, "Gate-Tunable Conducting Oxide Metasurfaces," Nano Letters **16**, 5319-5325 (2016).
127. P. P. Iyer, M. Pendharkar, and J. A. Schuller, "Electrically Reconfigurable Metasurfaces Using Heterojunction Resonators," Advanced Optical Materials **4**, 1582-1588 (2016).
128. J. Sun, E. Timurdogan, A. Yaacobi, E. S. Hosseini, and M. R. Watts, "Large-scale nanophotonic phased array," Nature **493**, 195-199 (2013).
129. J. Rensberg, S. Zhang, Y. Zhou, A. S. McLeod, C. Schwarz, M. Goldflam, M. Liu, J. Kerbusch, R. Nawrodt, S. Ramanathan, D. N. Basov, F. Capasso, C. Ronning, and M. A. Kats, "Active Optical Metasurfaces Based on Defect-Engineered Phase-Transition Materials," Nano Letters **16**, 1050-1055 (2016).
130. H.-S. Ee and R. Agarwal, "Tunable Metasurface and Flat Optical Zoom Lens on a Stretchable Substrate," Nano Letters **16**, 2818-2823 (2016).
131. S. M. Kamali, E. Arbabi, A. Arbabi, Y. Horie, and A. Faraon, "Highly tunable elastic dielectric metasurface lenses," Laser & Photonics Reviews **10**, 1002-1008 (2016).
132. M. Imada, A. Fujimori, and Y. Tokura, "Metal-insulator transitions," Reviews of Modern Physics **70**, 1039-1263 (1998).